\documentclass[authoryear,twocolumn,5p,final,times,12pt]{elsarticle}

\usepackage{amssymb}
\usepackage{lipsum}
\usepackage[fleqn]{amsmath}
\usepackage{booktabs}
\usepackage{multicol}

\journal{International Journal of Multiohase Flow}

\begin{document}

\begin{frontmatter}


 \title{Length-scale cascade and spread rate of atomizing planar liquid jets}
 \author[aff1]{Arash Zandian\corref{an}}
 \cortext[an]{Corresponding author}
 \ead{azandian@uci.edu}
  \author[aff1]{William A. Sirignano}
 \address[aff1]{Department of Mechanical and Aerospace Engineering, University of California, Irvine, CA 92697, USA}
 \author[aff2]{Fazle Hussain}
 \address[aff2]{Department of Mechanical Engineering, Texas Tech University, Lubbock, TX 79409, USA}

\begin{abstract}
The primary breakup of a planar liquid jet is explored via direct numerical simulation (DNS) of the incompressible Navier-Stokes equation with level-set and volume-of-fluid interface capturing methods. PDFs of the local radius of curvature and the local cross-flow displacement of the liquid-gas interface are evaluated over wide ranges of the Reynolds number ($Re$), Weber number ($We$), density ratio and viscosity ratio. The temporal cascade of liquid-structure length scales and the spread rate of the liquid jet during primary atomization are analyzed. 
The formation rate of different surface structures, e.g.~lobes, ligaments and droplets, are compared for different flow conditions and are explained in terms of the vortex dynamics in each atomization domain that we identified recently. With increasing $We$, the average radius of curvature of the surface decreases, the number of small droplets increases, and the cascade and the surface area growth occur at faster rates. The spray angle is mainly affected by $Re$ and density ratio, and is larger at higher $We$, at higher density ratios, and also at lower $Re$. The change in the spray spread rate versus $Re$ is attributed to the angle of ligaments stretching from the jet core, which increases as $Re$ decreases. Gas viscosity has negligible effect on both the droplet-size distribution and the spray angle. Increasing the wavelength-to-sheet-thickness ratio, however, increases the spray angle and the structure cascade rate, while decreasing the droplet size. The smallest length scale is determined more by surface tension and liquid inertia than by the liquid viscosity, while gas inertia and liquid surface tension are the key parameters in determining the spray angle. 
\end{abstract}


\end{frontmatter}


\section{Introduction}\label{Intro}

Atomization, the process by which a liquid stream disintegrates into droplets, has numerous industrial, automotive, environmental, as well as aerospace applications. In the field of engineering, atomization of liquid fuels in energy conversion devices is important because it governs the spread rate of the liquid jet (i.e.~spray angle), size of fuel droplets, and droplet evaporation rate. The spray angle also measures the liquid sheet instability growth, which controls the rate and intensity of the atomization.

The common purpose of breaking a liquid stream into spray is to increase the liquid surface area so that subsequent heat and mass transfer can be increased or a coating can be obtained. The spatial distribution, or dispersion, of the droplets is important in combustion systems because it affects the mixing of the fuel with the oxidant, hence the flame length and thickness. The size, velocity, volume flux, and number density of droplets in sprays critically affect the heat, mass, and momentum transport processes, which, in turn, affect the flame stability and ignition characteristics. Smaller drop size leads to higher volumetric heat release rates, wider burning ranges, higher combustion efficiency, lower fuel consumption and lower pollutant emission. In some other applications however, small droplets must be avoided because their settling velocity is low and under certain meteorological conditions, they can drift too far downwind \citep{Negeed}.

Most previous research has attempted to assess the manner by which the final droplet-size distribution -- after the atomization is fully developed -- is affected by the gas and liquid properties and by the nozzle geometry \citep{Dombrowski, Senecal, Mansour, Samuelsen, Lozano, Carvalho, Varga, Marmottant, Negeed}. General conclusions are that the Sauter Mean Diameter (SMD $=\Sigma N_i d_i^3/\Sigma N_i d_i^2$, where $N_i$ is the number of droplets per unit volume in size class $i$, and $d_i$ is the droplet diameter) decreases with increasing relative gas-liquid velocity, increasing liquid density, and decreasing surface tension, while viscosity is found to have little effect \citep{Samuelsen}. The main focus has been on the final droplet-size distribution and the spatial growth of the spray (spray angle), but little emphasis has been placed on the temporal cascade of the surface length scales -- growth of Kelvin-Helmholtz (KH) waves and their cascade into smaller structures leading to final breakup into ligaments and then droplets -- and the spread rate of the spray (spray angle) in primary atomization. In order to better control and optimize the atomization efficiency, the transient process from the point of injection to the fully-developed state needs to be better understood. Hence the spray angle and the rate at which the length scales cascade at different flow conditions must be analyzed. This is the main focus of the current study. The data presented herein will be crucial in the analysis and design of atomizers.


In recent years, more computational studies have addressed the liquid-jet breakup length scales and spray width. Most of these, however, have qualitatively investigated the effects of fluid properties and flow parameters on the final droplet size distribution or the spray angle. Only a few studies quantified the spatial variation of the droplet/ligament size along or across the spray axis. However, there are no detailed study of the temporal variation of the liquid-structure size distribution and the spray width (spray angle) during primary atomization. Most recent computational studies analyze the spatially developing instability leading to breakup of liquid streams; however, all of them are at relatively low values of the Weber number ($We < 2000$). That is, although some of those works are described as ``atomization" studies, they all fit better under the classical ``wind-induced capillary instabilities" defined by \citet{Ohnesorge} and \citet{Reitz}. Most of these studies are linear \citep{Otto}, two-dimensional (2D) inviscid \citep{Matas}, two-dimensional \citep{Fuster,Agbaglah}, or three-dimensional (3D) large-eddy simulations \citep{Agbaglah}. Of course, these did not resolve the smaller structures that form during the cascade process of the breakup. An analysis with spatial development offers some advantage with practical realism over temporal analysis. At the same time, the additional constraints imposed by the boundary conditions remove generality in the delineation of the important relevant physics. For these reasons, we follow the path of temporal-instability analysis in the classical atomization (high $We$ range) provided by \citet{Jarrahbashi1}. The goal is to reveal and interpret the physics in the cascade process known as atomization. Note that some spatial development is provided when the temporal analysis covers a domain that is several wavelengths in size. Relations between spatially developing and temporal results are discussed for single-phase and two-phase flows by \cite{Gaster} and \cite{Fuster}, respectively.

\cite{Jarrahbashi2} studied wide ranges of density ratio ($0.05$--$0.5$), Reynolds number \mbox{($320<Re<8000$)}, and $We$ of $O$($10^4$--$10^5$) in round liquid jet breakup. They defined the radial scale of the two-phase mixture as the outermost radial location of the continuous liquid and showed that the radial spray growth increases with increasing gas-to-liquid density ratio. They showed that droplets are larger for higher gas densities and lower $We$ values, and form at earlier times for higher density ratios. However, the effects of $Re$ and $We$ were not fully studied there.

\begin{figure*}[!t]
	\centering\includegraphics[width=0.75\linewidth]{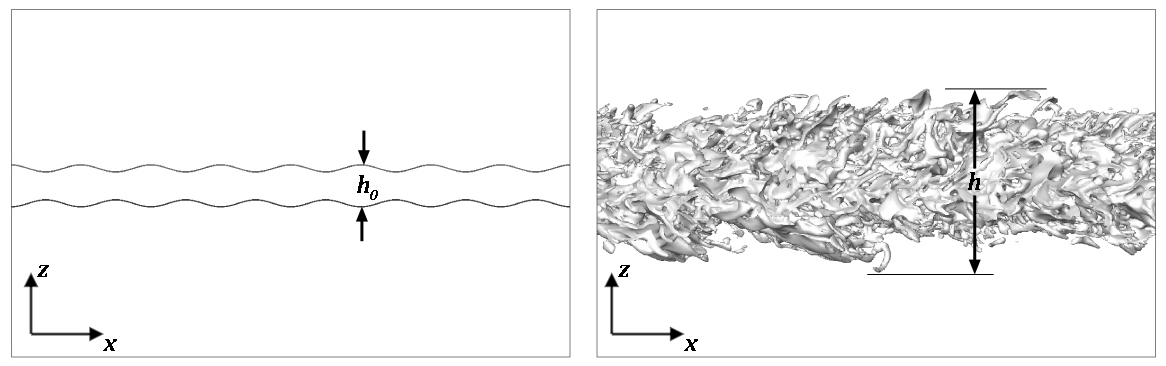}
	\caption{Schematic showing the spray width basis as defined by \protect\cite{Arash1}.}
	\label{fig:expansion}
\end{figure*}

In a similar study, \cite{Arash1} explored the effects of $We$ on the droplet size as well as the liquid sheet expansion rate. They also showed the variation in the size and number density of droplets for $We$ in the range $3000$--$72,000$; however, both these quantities were obtained from visual post-processing, which incur non-negligible errors. Their qualitative comparison showed that droplet and ligament sizes decrease with increasing $We$, while the number of droplets increases. Their results applied for a limited time after the injection and did not show how fast the length scales cascade. \cite{Arash1} similar to \cite{Jarrahbashi2} defined the distance of the farthest point on the continuous liquid sheet surface from the jet centerplane as the cross-flow width of the spray, as schematically depicted in Fig.~\ref{fig:expansion}. Even though this definition provides a simple description of the spray growth, it lacks statistical information about the number density of liquid structures at different cross-flow distances from the jet centerplane. Thus, this definition is not optimal for the spray width analysis and failed to show the effects of $We$ on the spray expansion rate.

\begin{figure}[!t]
	\centering\includegraphics[width=1.0\linewidth]{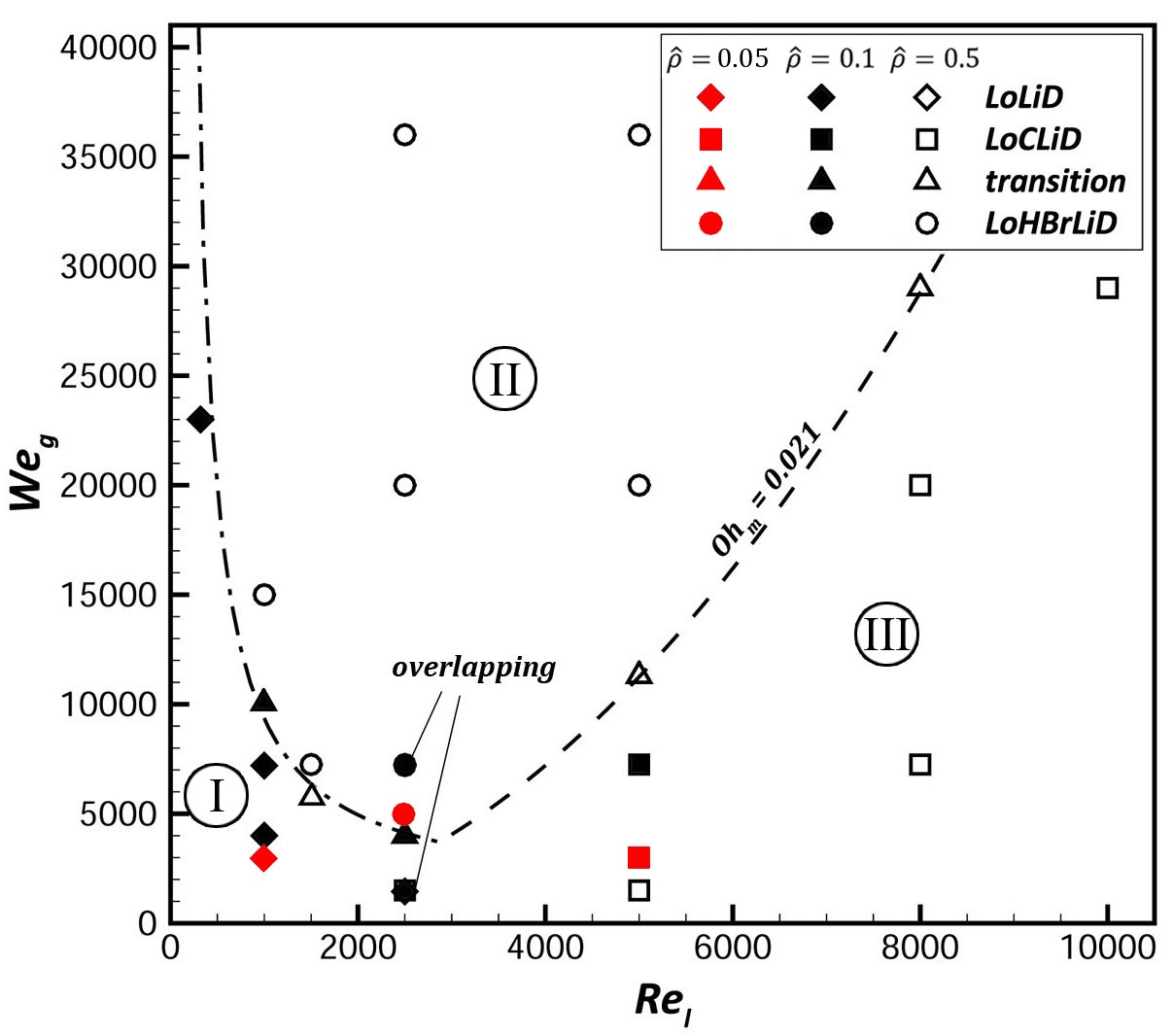}
	\caption{The breakup characteristics based on $We_g$ and $Re_l$, showing the $LoLiD$ mechanism (Atomization Domain I) denoted by diamonds, the $LoHBrLiD$ mechanism (Atomization Domain II) denoted by circles, the $LoCLiD$ mechanism (Atomization Domain III) denoted by squares, and the transitional region denoted by triangles. The cases with density ratio of $0.1$ ($\hat{\rho}=0.1$) are shaded. The $\hat{\rho}=0.1$ and $\hat{\rho}=0.5$ cases that overlap at the same point on this diagram are noted. --\,$\cdot$\,--\,$\cdot$\,--, transitional boundary at low $Re_l$; and --~--~--, transitional boundary at high $Re_l$ \protect\citep{Arash2}. The red symbols denote the cases for lower $\hat{\rho} = 0.05$, added to the original diagram by \protect\cite{Arash3}.}
	\label{fig:domains_gas_based}
\end{figure}

\begin{figure*}[!t]
	\centering\includegraphics[width=0.75\linewidth]{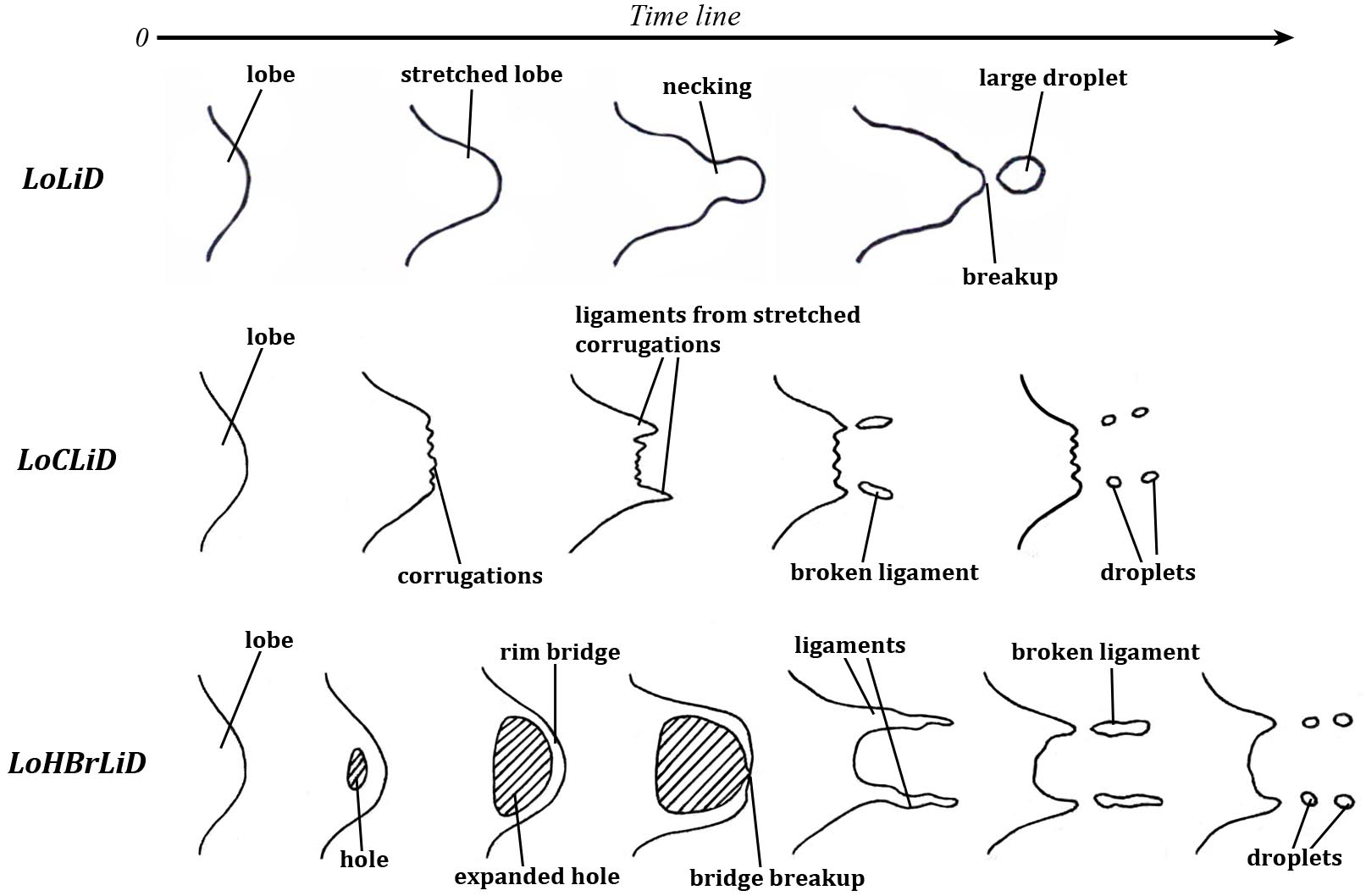}
	\caption{Sketch showing the cascade of structures on a liquid lobe from a top view, for the $LoLiD$ (top), $LoCLiD$ (center), and $LoHBrLiD$ (bottom) processes. The gas flows on top of these structures from left to right, and time increases to the right \protect\citep{Arash2}.}
	\label{fig:sketches}
\end{figure*}

More recently, \cite{Arash2} identified three atomization mechanisms and defined their domains of dominance on a gas Weber number ($We_g$) versus liquid Reynolds number ($Re_l$) diagram, shown in Fig.~\ref{fig:domains_gas_based}. The liquid structures seen in each of these atomization domains are sketched in Fig.~\ref{fig:sketches}, where the evolution of a liquid lobe is shown from a top view. These domain classifications were shown to apply to both planar and round jets. At high $Re_l$, the liquid sheet breakup characteristics change based on a modified Ohnesorge number, \mbox{$Oh_m \equiv \sqrt{We_g}/Re_l$}, as follows: (i) at high $Oh_m$ and high $We_g$, lobes become thin and puncture, creating holes and bridges. Bridges break as perforations expand and create ligaments, which then stretch and break into droplets by capillary action. This domain is indicated as Atomization \mbox{Domain II} in Fig.~\ref{fig:domains_gas_based}, and its process is called \mbox{\textit{LoHBrLiD}} based on the cascade of structures in this domain (\mbox{$Lo \equiv$ Lobe}, $H \equiv$ Hole, $Br \equiv$ Bridge, \mbox{$Li \equiv$ Ligament}, and \mbox{$D \equiv$ Droplet}); \mbox{(ii) at} low $Oh_m$ and high $Re_l$, holes are not seen at early times; instead, many corrugations form on the lobe front edge and stretch into ligaments. This process is called \textit{LoCLiD} ($C \equiv$ Corrugation) and occurs in Atomization Domain III (see Fig.~\ref{fig:domains_gas_based}), and results in ligaments and droplets without having the hole and bridge formation steps. The third is a \textit{LoLiD} process that occurs at low $Re_l$ and low $We_g$ (Atomization Domain I in Fig.~\ref{fig:domains_gas_based}), but with some difference in details from the $LoCLiD$ process. 

The main difference between the two processes (Domains I and III) at low and high $Re_l$ is that, at higher $Re_l$ the lobes become corrugated before stretching into ligaments. Hence, each lobe may produce multiple ligaments, which are typically thinner and shorter than those at lower $Re_l$. At low $Re_l$, on the other hand, because of the higher viscosity, the entire lobe stretches into one thick, usually long ligament. There is also a transitional region between the atomization domains, shown by the dashed-lines in Fig.~\ref{fig:domains_gas_based}. As seen in Fig.~\ref{fig:sketches}, the primary atomization follows a cascade process, where larger liquid structures, e.g.~waves and lobes, cascade into smaller and smaller structures; e.g.~bridges, corrugations, ligaments, and droplets. The time scales of the structure formations were also determined and were correlated to $Re_l$ and $We_g$. However, the length scale of the structures observed in each domain and their cascade rates were not discussed. The main focus of this study is on quantifying the cascade rate of different surface structures at the three atomization domains that were identified above. Moreover, the growth rate of the spray width in primary atomization is also quantified for the first time and is compared for different atomization domains. In the remainder of this section, some of the older methods used for quantification of the mean droplet size and spray angle are introduced and their pertinent results are presented.

\cite{Desjardins} defined the half-width of the planar jets as the distance from the jet centerline to the point at which the mean streamwise velocity excess is half of the centerline velocity. They showed that high $We$ jets grow faster, which suggests that surface tension stabilizes the jets. They also demonstrated that, regardless of $We$, droplets are generated through the creation and stretching of liquid ligaments (pertinent to Domain I in Fig.~\ref{fig:domains_gas_based}). Ligaments are longer, thinner, and more numerous as $We$ is increased. The corrugation length scales appear larger for lower $Re_l$ -- attributed to lesser energy contained in small eddies at lower $Re_l$. Consequently, earlier deformation on the smaller scales of the interface is more likely to take place on relatively larger length scales as $Re_l$ is reduced. Their work was limited to $We$ of $O$($10^2$--$10^3$), which were low compared to our range of interest for common liquid fuels and high-pressure operations. The effects of density ratio were also not studied in their work.

\cite{Dombrowski} developed theoretical expressions for the size of drops produced from fan spray sheets, and showed that under certain operating conditions in super atmospheric pressures, the drop size increases with ambient density. Verified experimentally, the drop size initially decreases, passing through a minimum with further increase of ambient density. They also showed that, for relatively thin sheets ($h/\lambda<1.25$; $h$ is sheet thickness and $\lambda$ is the unstable wavelength), the drop size increases with the surface tension and the liquid density, but depends inversely on the liquid injection pressure and the gas density. For relatively thick sheets ($h/\lambda>1.5$), however, the drop size is independent of surface tension or injection pressure and increases with increasing gas density. In another linear stability analysis, \cite{Senecal} derived expressions for the ligament diameter for short and long waves. They showed that the ligament diameter is directly proportional to the sheet thickness, but inversely to the square root of $We_g$. They also analytically related the final droplet size to the ligament diameter and the Ohnesorge number ($Oh=\sqrt{We}/Re$), and showed that the SMD decreases with time. The gas viscosity and the nonlinear physics of the problem, however, were neglected.

Effects of mean drop size and velocity on liquid sheets were measured using a phase Doppler particle analyzer (PDPA) by \cite{Mansour}. The spray angle decreased notably with increasing liquid flow rate while maintaining the same air pressure. They related this behavior to the reduction in the specific energy of air per unit volume of liquid leaving the nozzle. Increasing the air pressure for a fixed liquid flow rate resulted in an increase in the spray angle. \cite{Mansour} also measured the SMD along and across the spray axis, and showed a significant reduction in droplet size by increasing the air-to-liquid mass-flux ratio.


\cite{Carvalho} performed detailed measurements of the spray angle versus gas and liquid velocities in a planar liquid film surrounded by two air streams in a range of $Re$ from $500$ to $5000$. For low gas-to-liquid momentum ratio, the dilatational wave dominates the liquid film disintegration mode, and the atomization quality is rather poor, with a narrow spray angle (pertinent to Domain III). For higher gas-to-liquid momentum ratios, sinusoidal waves dominate and the atomization quality is considerably improved, as the spray angle increases significantly (Domain I). Regardless of the gas velocity, a region of maximum spray angle occurs, followed by a sharp decrease for higher values of the liquid velocity, and the maximum value of the spray angle decreases with the gas flow rate (corresponding to a transition from Domain I to III).

At present, it is very challenging to predict the droplet-diameter distribution as a function of injection conditions. For combustion applications, many empirical correlations relate the droplet size to the injection parameters \citep{Lefebvre}; however, detailed studies of fundamental breakup mechanisms -- especially during the initial period -- are needed to construct predictive models. A summary of several of these expressions for airblast atomization was compiled by \cite{Lefebvre}. The dependence of the primary droplet size ($d$) on the atomizing gas velocity ($U_g$) is most often expressed as a power law, $d \propto U_g^{-n}$, where $0.7\leq n\leq 1.5$. Physical explanations for particular values of the exponent $n$ are generally lacking. \cite{Varga} found that the mean droplet size is not very sensitive to the liquid jet diameter -- actually opposite to intuition; the droplet size is observed to increase slightly with decreasing nozzle diameter. This effect was attributed to the slightly longer gas boundary-layer attachment length. \citet{Varga} also observed a clear reduction in droplet size with lowering surface tension (transition from Domain I to II). They suggested the scaling $d\propto We_g^{-1/2}$, where $We_g$ is the Weber number based on gas properties and jet diameter. Their study was limited to very low $Oh$ of $O(10^{-3})$.

\cite{Marmottant} presented probability density functions (PDF) of the ligament size and the droplet size in their experimental study of the round liquid spray formation. The ligament size $d_0$ was found to be distributed around the mean in a nearly Gaussian distribution, but the droplet diameter $d$ was more broadly distributed and skewed. Rescaled by $d_0$, which depends on the gas velocity $U_g$, the size distribution keeps roughly the same shape for various gas flows. The average droplet size after ligament breakup was found to be $d\simeq0.4d_0$, with its distribution $P(d)$ having an exponential fall-off at large diameters parameterized by the average ligament size $\left\langle d_0 \right\rangle$, namely $P(d) \sim \exp(-nd/\left\langle d_0 \right\rangle)$. The parameter $n\approx 3.5$ slowly increased with $U_g$. The mean droplet size in the spray decreases like $U_g^{-1}$. Most of their experiments were at low $Re$ and low $We$ ranges (Domain I).

More recently, \cite{Negeed} analytically and experimentally studied the effects of nozzle shape and spray pressure on the liquid sheet characteristics. They showed that the droplet mass mean diameter decreases by increasing the $Re_l$ or by increasing the water sheet $We$ (transition from Domain I to Domain III), since the inertia force increases by increasing both parameters. The spray pressure difference was also shown to have a similar effect on the mean droplet size as $Re_l$. Their study was limited to very high $Re_l$ values  ($10,000$--$36,000$) and very low $We_g$ values of $O(10^1)$; i.e.~Domain III.

The scarcity of studies in Domain II is especially notable since the trend towards much higher operating pressures has started in engine designs. The studies introduced above indicate that there is no proper method in the literature for quantification of the cascade rate and spray spread rate at early liquid-jet breakup. Therefore, implementation of a new methodology -- as will be introduced here -- is essential for evaluation of these quantities. Quantification of these parameters is very important for understanding and identifying the most significant causes, which are helpful in controlling the atomization process.

\subsection{Objectives}

Our objectives are to (i) study the effects of the key non-dimensional parameters, i.e.~$Re_l$, $We_g$, gas-to-liquid density ratio and viscosity ratio, and also the wavelength-to-sheet-thickness ratio, on the temporal variation of the spray width and the liquid-structures length scale; (ii) establish a new model and definition for measurement of the liquid surface length-scale distribution and spray width for the early period of spray formation; and (iii) explain the roles different breakup regimes play in the cascade of length scales and spray development during the early atomization period. The results are separated by atomization domains to clarify the effects of each atomization mechanism on the cascade process and spray expansion. Moreover, the time portions of the behaviors are related to the various structures formed during the primary atomization period.

To address these objectives, two PDFs are obtained from the numerical results, for a wide range of liquid-structure size and transverse location of the liquid-gas interface at different times to give a broader understanding of the temporal variation of the length-scale distribution and the spray width. The temporal evolution of the distribution functions is given rather than just showing the fully-developed asymptotic length scales, as is well explored and examined in the literature. The first PDF indicates the size of the small liquid structures through the local radius of curvature of the gas-liquid interface. The novelty of this model is that we do not address only the droplets that are formed, nor present only the size of the droplets as the length scale of the atomization problem -- as what the SMD presents. Instead, the length-scale distribution here comprises the size of all the liquid structures -- from the initial surface waves, to the lobes, bridges, ligaments, and finally droplets. Thus, this analysis reveals the change in the overall length scale of the jet surface even before the droplets are formed -- not examined or quantified before. The second PDF indicates the location of the liquid-gas interface, giving more than just the outermost displacement of the spray, as was typically measured in most past computational studies. Rather, we present the ``density" of the liquid surface at any transverse location, which provides a more meaningful and realistic presentation of the spray width. ``Density" here means the relative liquid surface area at a given control volume at any distance from the jet midplane.

In Section \ref{Modeling}, the numerical methods that have been used and the most important flow parameters involved in this study are presented along with the post-processing methods used to obtain the PDFs. Section \ref{Results} is devoted to the analysis of the effects of $We$ (Section~\ref{Weber effects}), $Re$ (Section \ref{Reynolds effects}), density ratio (Section \ref{Density ratio}), viscosity ratio (Section \ref{Viscosity ratio}), and sheet thickness (Section \ref{Thickness}) on the spray width and the liquid-structure size. Conclusions are given in Section \ref{Conclusion}, accompanied by a short summary of our most important findings.

\section{Numerical Modeling} \label{Modeling}

The 3D Navier-Stokes equations with level-set and volume-of-fluid interface capturing methods yield computational results for the liquid segment which captures the liquid-gas interface deformations with time.

The incompressible continuity and Navier-Stokes equations, including the viscous diffusion and surface tension forces and neglecting the gravitational force are as follows:
\begin{equation}
\nabla \cdot \textbf{u}=0,
\label{eqn:incomp continuity}
\end{equation}
\begin{equation}
\begin{aligned}
&\frac{\partial (\rho \textbf{u})}{\partial t}+ \nabla \cdot (\rho \textbf{uu})= \\ 
& - \nabla p+\nabla \cdot (2\mu \textbf{D})- \sigma \kappa \delta (d) \textbf{n},
\label{eqn:momentum}
\end{aligned}
\end{equation}
where $\textbf{D}$ is the rate of deformation tensor,
\begin{equation}
\textbf{D} = \frac{1}{2} \left[ (\nabla \textbf{u})+(\nabla \textbf{u})^T \right].
\end{equation}

$\textbf{u}$ is the velocity vector; $p$, $\rho$, and $\mu$ are the pressure, density and dynamic viscosity of the fluid, respectively. The last term in Eq.~(\ref{eqn:momentum}) is the surface tension force per unit volume, where $\sigma$ is the surface tension coefficient, $\kappa$ is the surface curvature, $\delta (d)$ is the Dirac delta function, $d$ is the distance from the interface, and $\textbf{n}$ is the unit vector normal to the liquid/gas interface.

Direct numerical simulation is done with an unsteady 3D finite-volume solver for the incompressible Navier-Stokes equations describing the planar liquid sheet segment (initially stagnant), which is subject to instabilities due to a gas stream that flows past it on both sides. A uniform staggered grid is used with the mesh size of $\Delta=2.5~\mu$m and a time step of $5~$ns -- finer grid resolution of $1.25~\mu$m is used for the cases with higher $We_g$ \mbox{($\geq 36,000$)} and higher $Re_l$ ($=5000$). Spatial discretization and time marching are given by the third-order accurate QUICK scheme and the Crank-Nicolson scheme, respectively. The continuity and momentum equations are coupled through the SIMPLE algorithm.

The level-set method developed by Osher and his coworkers \citep{Zhao, Sussman, Osher} captures the liquid-gas interface. The level set $\phi$ is a distance function with zero value at the liquid-gas interface, positive values in the gas phase and negative values in the liquid phase. All the fluid properties for both phases in the Navier-Stokes equations are defined based on the $\phi$ value and the equations are solved for both phases simultaneously. Properties such as density and viscosity vary continuously but with a very large gradient near the liquid-gas interface. The level-set function $\phi$ is also advected by the velocity field;
\begin{equation}
\frac{\partial \phi}{\partial t} + \textbf{u} \cdot \nabla \phi = 0.
\label{eqn:level set}
\end{equation}

The interface curvature needed to calculate the surface tension force can also be obtained from the level-set function ($\kappa = \nabla . \frac{\nabla \phi}{|\nabla \phi|}$), which for a 3D domain results in the following equation.
\begin{eqnarray}
\kappa&=&\frac{\phi_{xx}(\phi_y^2+\phi_z^2) + \phi_{yy}(\phi_x^2+\phi_z^2) + \phi_{zz}(\phi_x^2+\phi_y^2)}{{(\phi_x^2 +\phi_y^2+\phi_z^2)}^{3/2}} \nonumber \\ 
&&-\frac{2\phi_x\phi_y\phi_{xy} + 2\phi_y\phi_z\phi_{yz} + 2\phi_z\phi_x\phi_{xz}}{{(\phi_x^2 +\phi_y^2+\phi_z^2)}^{3/2}}.
\label{eqn:curvature}
\end{eqnarray}

For detailed descriptions of this interface capturing see \cite{Sussman}.

At low density ratios, a transport equation similar to Eq.~(\ref{eqn:level set}) is used for the volume fraction $f$, also called the Volume-of-Fluid (VoF) variable, in order to describe the temporal and spatial evolution of the two-phase flow \citep{Hirt}. 
\begin{equation}
\frac{\partial f}{\partial t} + \textbf{u} \cdot \nabla f = 0.
\label{eqn:VoF}
\end{equation}
where the VoF-variable $f$ represents the volume of (liquid phase) fluid fraction at each cell.

The fully conservative momentum convection and volume fraction transport, the momentum diffusion, and the surface tension are treated explicitly. To ensure a sharp interface of all flow discontinuities and to suppress numerical dissipation of the liquid phase, the interface is reconstructed at each time step by the PLIC (piecewise linear interface calculation) method of \cite{Rider}. The normal direction of the interface is found by considering the volume fractions in a neighborhood of the cell considered (similar to Eq.~\ref{eqn:curvature}), where $f$ changes most rapidly. A least-square method introduced by \cite{Puckett} is used for normal reconstruction, where the interface is approximated by a straight line in the cell block. Once interface reconstruction has been performed, direction-split geometrical fluxes are computed for VoF advection \citep{Popinet}. This advection scheme preserves sharp interfaces and is close to second-order accurate for practical atomization applications. We use the paraboloid fitting technique of \cite{Popinet} to compute an accurate estimate of the curvature from the discrete volume fraction values. The capillary effects in the momentum equations are represented by a capillary tensor introduced by \cite{Zaleski}.

\subsection{Flow Configuration}

\begin{figure*}[t!]
	\centering\includegraphics[width=0.95\linewidth]{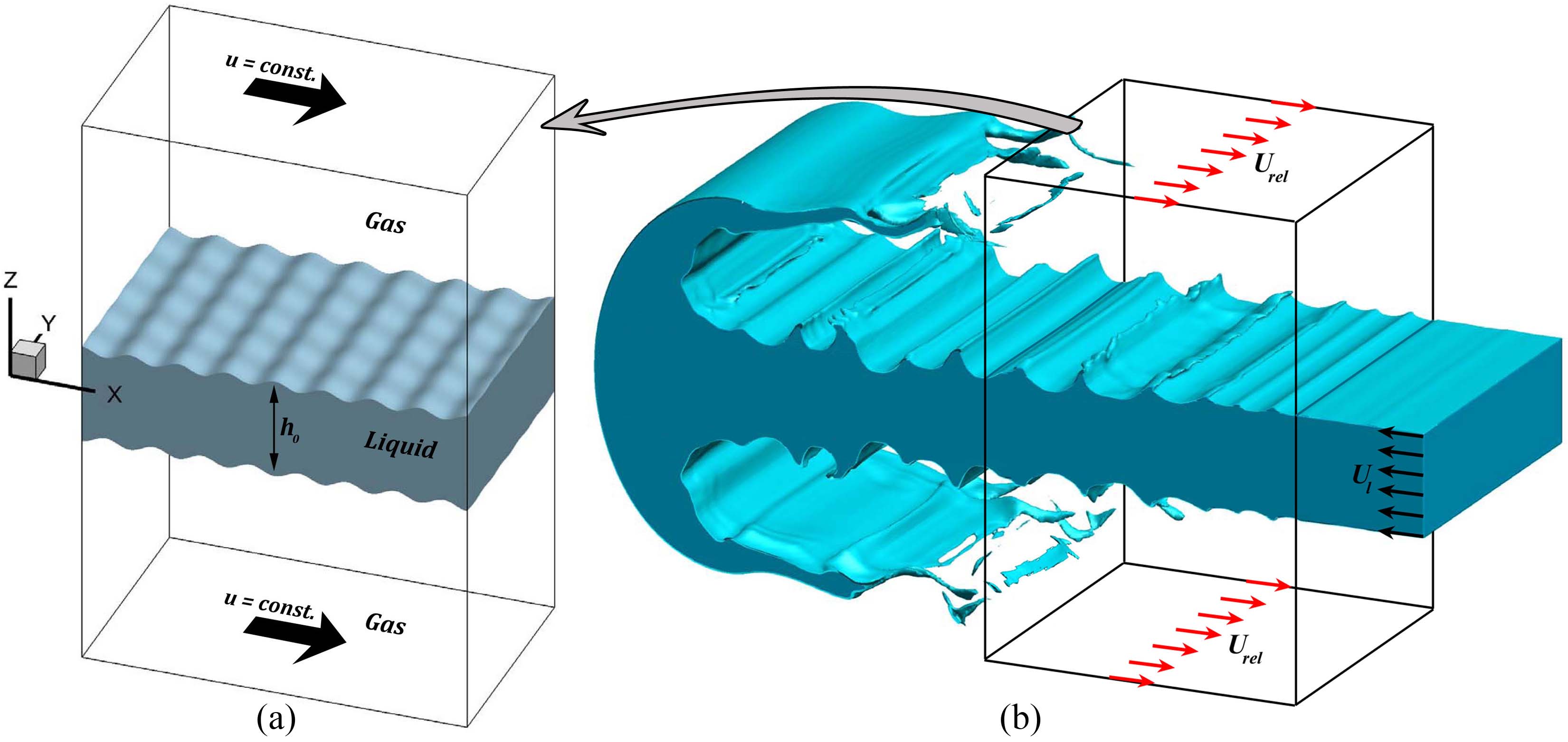}
	\caption{The computational domain with the initial liquid and gas zones (a), and a spatially developing full liquid jet (b).}
	\label{fig:geo}
\end{figure*}

The computational domain, shown in Fig.~\ref{fig:geo}(a), consists of a cube, which is discretized into uniform-sized cells. The liquid segment, which is a sheet of thickness $h_0$ ($h_0=50~\mu$m for the thin sheet and $200~\mu$m for the thick sheet), is located at the center of the box and is stationary in the beginning. The domain size in terms of the sheet thickness is \mbox{$16h_0\times10h_0\times10h_0$}, in the $x$, $y$ and $z$ directions, respectively, for the thin sheet, and \mbox{$4h_0\times4h_0\times8h_0$} for the thick sheet. The liquid segment is surrounded by the gas zones on top and bottom. The gas moves in the positive $x$- (streamwise) direction with a constant velocity ($U=100~$m/s) at top and bottom boundaries, and its velocity diminishes to the interface velocity with a boundary layer thickness obtained from 2D full-jet simulations. In the liquid, the velocity decays to zero at the center of the sheet with a hyperbolic tangent profile. For more detailed description of the initial velocity profile and boundary layer thickness, see figure 12 of \cite{Arash1}. The normal gradient ($\partial/\partial z$) is set to zero for the normal and spanwise velocity components ($w$ and $v$) at the top and bottom boundaries, so that the gas would be free to flow into the domain following the entrainment by the jet. Periodic boundary conditions for all components of velocity as well as the level-set/VoF variable are imposed on the four sides of the computational domain; i.e.~the $x$- and $y$-planes.

The computational box resembles a segment of the liquid sheet far upstream of the jet cap, as shown in Fig.~\ref{fig:geo}(b). This sub-figure shows the starting behavior of a spatially developing full liquid jet injected with constant velocity $U_l$ into quiescent gas. A Galilean transformation of the velocity field shows that the gas stream flows upstream with respect to the liquid jet with a relative velocity $U_{rel}=U_l$ (denoted by the red arrows) in the box shown in Fig.~\ref{fig:geo}(b). Our temporal study focuses on this region of the jet stream away from the jet cap, where previous studies indicate that surface deformations are periodic \citep{Shinjo}. 

This study involves a temporal computational analysis with a relative velocity between the two phases. Due to friction, the relative velocity decreases with time. Furthermore, the domain is several wavelengths long in the streamwise direction so that some spatial development occurs. In order to reduce the dependence of the results on details of the boundary conditions, specific configurations (e.g.,~air-assist or air-blast atomization) are avoided. However, calculations are made with the critical non-dimensional parameters in the ranges of practical applications.

The liquid-gas interface is initially perturbed symmetrically on both sides with a sinusoidal profile and predefined wavelength and amplitude obtained from the 2D full-jet simulations \citep[see][]{Arash1}. Two analyses without forced or initial surface perturbations -- the full-jet 2D simulations \citep[figure 11 of][]{Arash1} and the initially non-perturbed 3D simulations \citep[figure 17 of][]{Arash2} -- show KH wavelengths in the moderate range of $80$--$125~\mu$m over a wide range of $Re_l$, $We_g$ and $\hat{\rho}$ studied here. In order to expedite the appearance and growth of the KH waves, initial perturbations with wavelength of $100~\mu$m are imposed on the interface with a small amplitude of $4~\mu$m for the thick sheet and $2~\mu$m for the thin sheet. This amplitude is small enough that subharmonics would have a chance to form and grow. Similar to \cite{Jarrahbashi1}, our results show that at higher $Re_l$ and lower $\hat{\rho}$, smaller waves appear superimposed on the initial perturbations. The waves also merge to create larger waves at lower $We_g$. Both streamwise ($x$-direction) and spanwise ($y$-direction) perturbations are considered in this study.

The most important dimensionless groups in this study are the Reynolds number ($Re$), the Weber number ($We$), as well as the gas-to-liquid density ratio ($\hat{\rho}$) and viscosity ratio ($\hat{\mu}$), as defined below. The initial wavelength-to-sheet-thickness ratio ($\Lambda$) is also an important parameter.
\begin{subequations}\label{parameters}
	\begin{equation}
	Re = \frac{\rho Uh_0}{\mu} \; , \quad
	We = \frac{\rho U^2 h_0}{\sigma} \; ,
	\end{equation}
	\begin{equation}
	\hat{\rho} = \frac{\rho_g}{\rho_l} \; , \quad
	\hat{\mu} = \frac{\mu_g}{\mu_l} \; , \quad
	\Lambda = \frac{\lambda_0}{h_0}.
	\end{equation}
\end{subequations}

\begin{table*}[!t]
	\caption{\label{tab:parameter range}Range of dimensionless parameters.}
	\begin{center}
		\begin{tabular}{lccccc}
			\hline \toprule
			Parameter  & $Re_l$   &   $We_g$ & $\hat{\rho}$ & $\hat{\mu}$ & $\Lambda$ \\[3pt] \midrule 
			Range   & ~1000--5000~ & ~1500--36,000~ & ~0.05--0.9~ & ~0.0005--0.05~ & 0.5--2.0\\
			\bottomrule \hline
		\end{tabular}
	\end{center}
\end{table*}

The initial sheet thickness $h_0$ is considered as the characteristic length, and the relative gas--liquid velocity $U$ as the characteristic velocity. The subscripts $l$ and $g$ refer to the liquid and gas, respectively. Theoretically, if the flow field is infinite in the streamwise direction (as in our study), a Galiliean transformation shows that only the relative velocity between the two streams is consequential. Spatially developing flow fields, however, are at most semi-infinite so that both velocities at the flow-domain entry and their ratio (or their momentum-flux ratio) are important. A wide range of $Re$ and $We$ at high and low density and viscosity ratios is covered in this research. Kerosene with density of $800$~kg~m$^{-3}$ is used as the liquid. The liquid viscosity and surface tension coefficient are obtained from the given Reynolds and Weber numbers, respectively, and the gas properties are calculated from the desired density and viscosity ratio. For a typical case at $Re_l=2500$ and $We_g=5000$, $\mu_l\approx0.005$~Pa~s and $\sigma\approx0.03$~kg~s$^{-2}$ are obtained. The range of dimensionless parameters analyzed in this study are given in Table~\ref{tab:parameter range}. The range of parameters considered here covers the more practical ranges usually seen in most atomization applications; however, higher $\hat{\rho}$ (high pressures) and higher $We_g$ values are also studied to fully explore and portray their effects. Even though current practical applications usually perform at density ratios lower than $0.05$, high pressure (high density) applications are currently being discussed for future gas turbines/engines. The analysis performed in this study is in line with this trend of future injection systems towards higher density ratios.

The grid independency tests were performed in detail by \cite{Arash3}. They showed that the errors in the size and location of ligaments and droplets, and the magnitude of the velocity and boundary layer thickness computed using different mesh resolutions were within an acceptable range. The effects of mesh resolution on the most important flow parameters, e.g. surface structures, velocity and vorticity profiles were studied in detail by \cite{Arash3} and the mass conservation of the LS method was confirmed. The domain-size independency were also checked in both streamwise and spanwise directions to ensure that the resolved wavelengths were not affected by the domain length and width. The transverse (cross-flow) dimension of the domain was chosen such that the top and bottom boundaries remain far from the interface at all times, so that the surface deformation is not directly affected by the boundary conditions. \cite{Arash3} showed that, as the KH waves amplify, the convective velocity of the interface at the base of the waves was in very good agreement with the Dimotakis speed \citep{Dimotakis} defined as $U_D = (U_l+\sqrt{\hat{\rho}}U_g)/(1+\sqrt{\hat{\rho}})$. The effects of the fuzzy zone thickness -- where properties have large gradients to approximate the discontinuities between the two phases -- have been previously addressed by \cite{Jarrahbashi1}. The effect of mesh resolution on the PDFs and length scale measurements is detailed in Section \ref{mesh test}.

\subsection{Data analysis} \label{Data analysis}

Twice the inverse of the liquid surface curvature ($\kappa=1/R_1+1/R_2$, where $R_1$ and $R_2$ are the two radii of curvature of the surface in a 3D domain) has been defined as the local length scale in this study. This length scale represents the local radius of curvature of the interface, and is a proper quantity enabling us to monitor the overall size of the surface structures. It would eventually asymptote to the droplet radius after the entire jet is broken into approximate spherical droplets. Based on this definition, a length scale is obtained at each computational cell in the fuzzy zone at the interface. The curvature of each cell containing the interface is computed at every time step; then, the structure length scale ($L$) is defined as
\begin{equation}
L_{ijk} = \frac{2}{|\kappa_{ijk}|} ,
\end{equation}
where $\kappa$ is the curvature, and the $ijk$ indices denote the coordinates of the cell in a 3D mesh. The length scale $L_{ijk}$ of each cell is used to create a PDF of the length scales. The $\kappa_{ijk}$ value is measured per computational cell and weighted by the interface area in that cell to obtain the PDFs. Only cells containing the interface are included in this analysis. The bin size used for the length-scale analysis is $\mathrm{d}L = \Delta = 2.5~\mu$m. The probability of the length scale in the interval $(L,L+\mathrm{d}L)$ is obtained by multiplying the PDF value at that length scale, $f(L)$, by the bin size:
\begin{equation}
prob(L\leq L' \leq L+\mathrm{d}L)\equiv P(L) = f(L)\mathrm{d}L .
\label{eqn:prob}
\end{equation}

This is an operational definition of the PDF. Since the probability is unitless, $f(L)$ has units of the inverse of the length scale; i.e.~m$^{-1}$. However, in our study, the length scale is nondimensionalized by the initial wavelength. Thus, $f(L/\lambda_0)$ becomes unitless. The relation between the length-scale PDF and its probability could be derived from Eq.~(\ref{eqn:prob}); 
\begin{equation}
f(L/\lambda_0) = \frac{P(L/\lambda_0)}{\mathrm{d}L/\lambda_0} = 40 P(L/\lambda_0) ,
\label{eqn:PDF}
\end{equation}
where $\mathrm{d}L$ is the bin size and $\lambda_0=100~\mu$m is the initial KH wavelength. The probability of having the length scale in the finite interval $[a,b]$ can be determined by integrating the PDF;
\begin{eqnarray}
prob(a\leq \frac{L}{\lambda_0} \leq b)&\equiv& P(a \leq \frac{L}{\lambda_0} \leq b) \nonumber\\
& = &\int_a^b f(\frac{L'}{\lambda_0})\frac{\mathrm{d}L'}{\lambda_0} .
\end{eqnarray}

In order to obtain the average length scale at each time step, the length scales are integrated along the liquid-gas interface and divided by the total interface area. The average length scale $\delta$ is nondimensionalized using the initial perturbation wavelength, $\lambda_0$. The average length scale is defined as
\begin{equation}
\delta = \frac{1}{\lambda_0}\frac{\int L\mathrm{d}s}{S} \approx \frac{1}{\lambda_0}\frac{\sum L_is_i}{\sum s_i} ,
\end{equation}
where $S$ is the total surface area of the interface, $s_i$ is the interface area in cell $i$ and $L_i$ is the length scale of that particular cell.

Since $L$ has a wide range from a few microns to infinity (if the curvature is zero at a cell), we neglect the length scales that are very large, i.e.~$L>4\lambda_0$, so that $\delta$ would not be biased towards large scales due to those off values. $\delta$ can show the overall change in the size of the structures on the liquid surface; therefore, one can track the stretching of the surface -- if $\delta$ grows -- or its cascade into smaller structures and appearance of subharmonic instabilities -- as the mean decays.

Similar to the length scale, a PDF is obtained for the transverse distance of the interface from the centerplane. This is related to the interface density model introduced by \citet{Chesnel}, where the interface density was defined as the ratio of the interface area within the considered control volume. In our study, however, the interface density is measured at different transverse locations to form the PDFs. This PDF also has units of m$^{-1}$, as discussed before; however, it is normalized by the initial sheet thickness. This gives a better statistical data about the distribution of the transverse location of the spray interface rather than just presenting the outermost location of the liquid surface. Thereby, the concentration, or density, of the liquid surface at any transverse plane is measured. This quantity also shows the breakup of surface structures and demonstrates how uniformly the spray spreads; i.e.~the quality of spray development. In fact, this PDF is a more generalized version of the average liquid volume fraction distribution. To account for both sides of the liquid sheet in this analysis, the cross-flow distance $h$ of each local point at the interface is defined as the absolute value of its $z$-coordinate ($z=0$ at the centerplane);
\begin{equation}
h_{ijk} = |z_{ijk}| .
\end{equation}

The probability of the spray width is obtained from an equation similar to Eq.~(\ref{eqn:prob}), where $L$ is replaced by $h$. The relation between the spray-width PDF, $f(h/h_0)$, and its probability, $P(h/h_0)$, is
\begin{equation}
f(h/h_0) = \frac{P(h/h_0)}{\mathrm{d}h/h_0} ,
\end{equation}
where $\mathrm{d}h$ is the bin size for the spray-width PDF, taken to be equal to the mesh size, i.e.~$\mathrm{d}h=\Delta=2.5~\mu$m, and $h_0$ is the initial sheet thickness; $h_0=50~\mu$m for the thin sheet, and $200~\mu$m for the thick sheet.

The mean spray width (sheet thickness) $\zeta$ is also obtained by integrating $h$ along the interface, and dividing it by the total interface area. The mean spray width is nondimensionalized by the initial sheet thickness $h_0$;
\begin{equation}
\zeta = \frac{1}{h_0}\frac{\int h \mathrm{d}s}{S} \approx \frac{1}{h_0}\frac{\sum h_is_i}{\sum s_i} ,
\end{equation}
where $h_i$ is the cross-flow distance of the interface in cell $i$ from the midplane. This definition represents how dense the surface is at any distance from the jet centerplane. Therefore, it gives a more realistic representation of the jet growth in a way that is more useful for many applications such as combustion and coating. A simple height function for the spray would be a single-valued function of downstream distance and time; our PDF accounts for the multi-valued nature of the real situation.

\section{Results and discussion} \label{Results}

\subsection{Data analysis verification} \label{mesh test}

\begin{figure}[!b]
	\centering\includegraphics[width=1.0\linewidth]{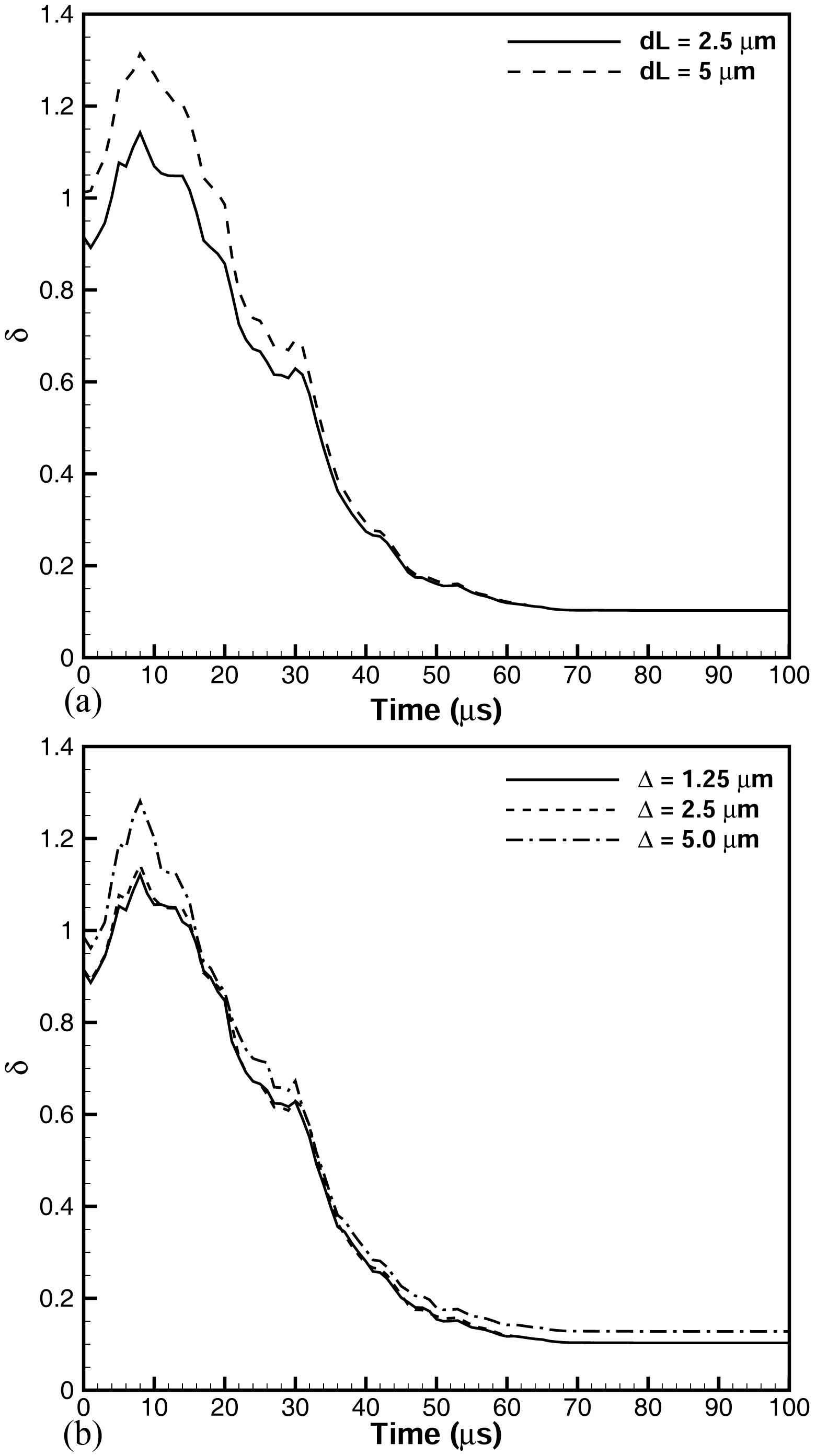}
	\caption{Effect of bin size (a) and mesh resolution (b) on the temporal evolution of $\delta$; $Re_l=2500$, $We_g=7250$, \mbox{$\hat{\rho}=0.5$}, $\hat{\mu}=0.0066$, and $\Lambda = 2.0$.}
	\label{fig:tests}
\end{figure}

\begin{figure}[!b]
	\centering\includegraphics[width=1.0\linewidth]{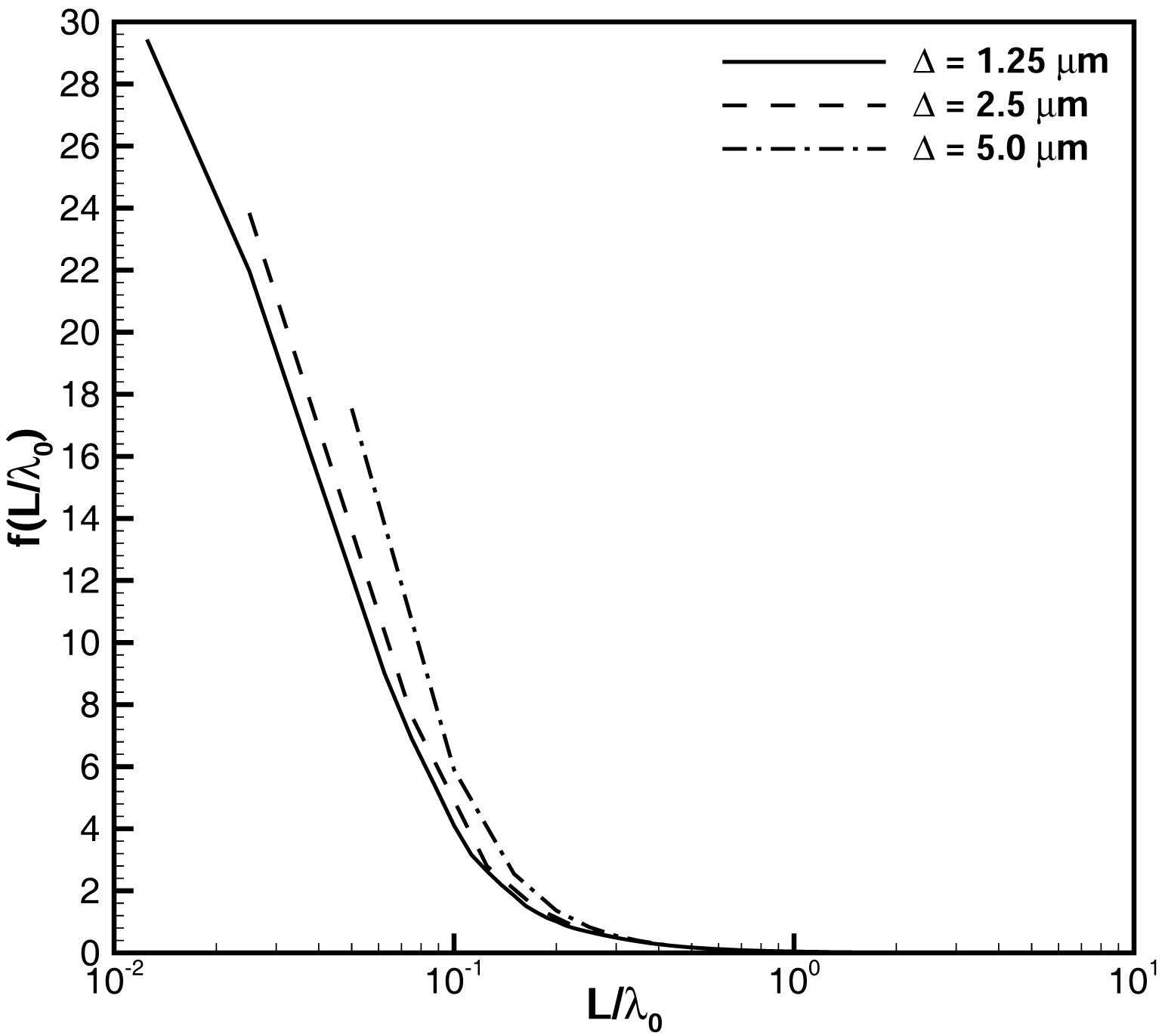}
	\caption{Effect of mesh resolution on the length scale PDF, $f(L/\lambda_0)$, at $t=70~\mu$s; $Re_l=2500$, $We_g=7250$, \mbox{$\hat{\rho}=0.5$}, $\hat{\mu}=0.0066$, and $\Lambda = 2.0$.}
	\label{fig:mesh PDF}
\end{figure}

\begin{figure}[!t]
	\centering\includegraphics[width=1.0\linewidth]{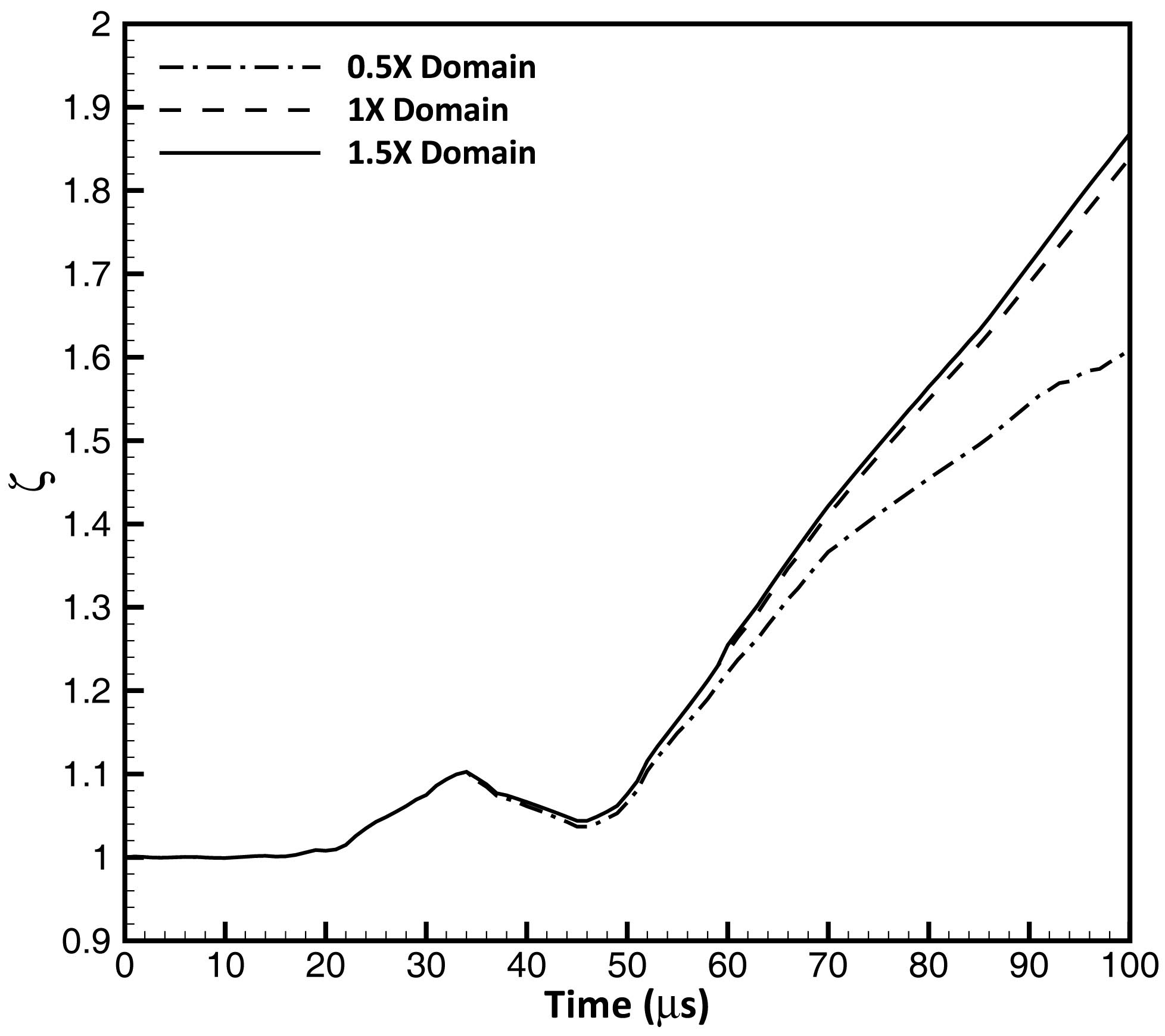}
	\caption{Effect of computational domain size on the temporal evolution of $\zeta$; $Re_l=2500$, $We_g=7250$, \mbox{$\hat{\rho}=0.5$}, $\hat{\mu}=0.0066$, and $\Lambda = 2.0$.}
	\label{fig:domain size}
\end{figure}

The choice of bin size and the sensitivity of the measurements against the mesh resolution and computational-domain size are tested and verified in this section. The results of these tests are given in Figs.~\ref{fig:tests}--\ref{fig:domain size}. Fig.~\ref{fig:tests}(a) compares the temporal variation of $\delta$ with two different bin sizes used for measurement of this parameter. The solid line is the result obtained by the actual bin size used in our analysis ($\mathrm{d}L=\Delta=2.5~\mu$m), and the dashed line denotes the variation of $\delta$ with time using a bin size twice as big; i.e.~$\mathrm{d}L=2\Delta=5~\mu$m. Both cases converge at about $40~\mu$s and are in good agreement thereafter. Both test cases result in the same asymptotic length scale at the end of the computation; however, in the early stages, the larger bin size produces slightly larger scales, especially around the maximum point ($t\approx10~\mu$s). The maximum error in the larger bin size is around $13\%$. The error gradually decreases after $10~\mu$s and becomes less than $1\%$ at $40~\mu$s. The temporal trend of the length scales, i.e.~when the scales are growing or declining, predicted by both test cases, match very well. The only difference is that the larger bin overpredicts the length scales at early stages. Since the magnitude in the early stage of the length scale growth (around the time when the maximum errors occur) is not the main goal of our research, it can be concluded that choosing $\mathrm{d}L=2.5~\mu$m as the bin size is acceptable for the purposes of this study. Smaller bin size is not recommended in this analysis since the smallest length scale that the simulations are able to capture is as small as the mesh size; thus, choosing a bin smaller than the mesh resolution would be meaningless.

Fig.~\ref{fig:tests}(b) shows the temporal evolution of $\delta$ for three different grid resolutions. The result for the grid size used in this study ($\Delta=2.5~\mu$m) is denoted by dashed line in this plot. Two other grid resolutions -- one twice as big ($\Delta=5~\mu$m) and the other one half of the current grid size ($\Delta=1.25~\mu$m) -- are also plotted in dash-dotted and solid lines, respectively. The larger grid is unable to predict the asymptotic length scale and has about $5\%$ error near the asymptote. The maximum error of this large grid is about $10\%$ and occurs near the maximum scale. The result of the finest grid however, matches perfectly with the current grid after $15~\mu$s. The maximum difference between the results of the two finer grids is less than $1\%$. To further delineate the effects of mesh resolution on the length scales population, the PDF of length scales, $f(L/\lambda_0)$, obtained by these three meshes are compared in Fig.~\ref{fig:mesh PDF} at $70~\mu$s. For each of these cases, the bin size is equal to their mesh resolution. The PDF results of the coarsest grid are quite overpredicted, with more than $20\%$ error near the smallest scale. The difference between the two finer grids however, is much smaller. The largest error at the smallest $L$ is about $5\%$. The error in mass distribution would be even much smaller since the smallest scales have much smaller volumes too. The difference in the PDFs also does not impact the reported results for $\delta$. Therefore, this comparison verifies that the $2.5~\mu$m grid resolution is justified for our study.

The size of the computational domain (especially in the transverse direction) has a major influence on the evaluation of the mean spray width. Fig.~\ref{fig:domain size} compares the temporal growth of the $\zeta$ for three domain sizes -- the original domain used in this study (1X), a domain half the original size (0.5X), and another $1.5$ times larger (1.5X). The largest domain predicts a very similar result compared to the original case, with very slight difference in $\zeta$ after $70~\mu$s. However, the difference in the results of the two larger domains never exceeds $1.5\%$, which verifies the appropriateness of our chosen domain size. However, a larger domain is definitely required if one wants to study the process further in time. The predicted spray width of the 0.5X Domain is acceptable until $55~\mu$s, but it departs from the correct trend henceforth and becomes noticeably underpredicted. The error of the smallest domain reaches about $14\%$ by the end of the simulation. This clearly shows that the boundary conditions have a major impact on the predicted results for the small domain.

\subsection{Weber number effects} \label{Weber effects}

\begin{figure}
	\centering\includegraphics[width=1.0\linewidth]{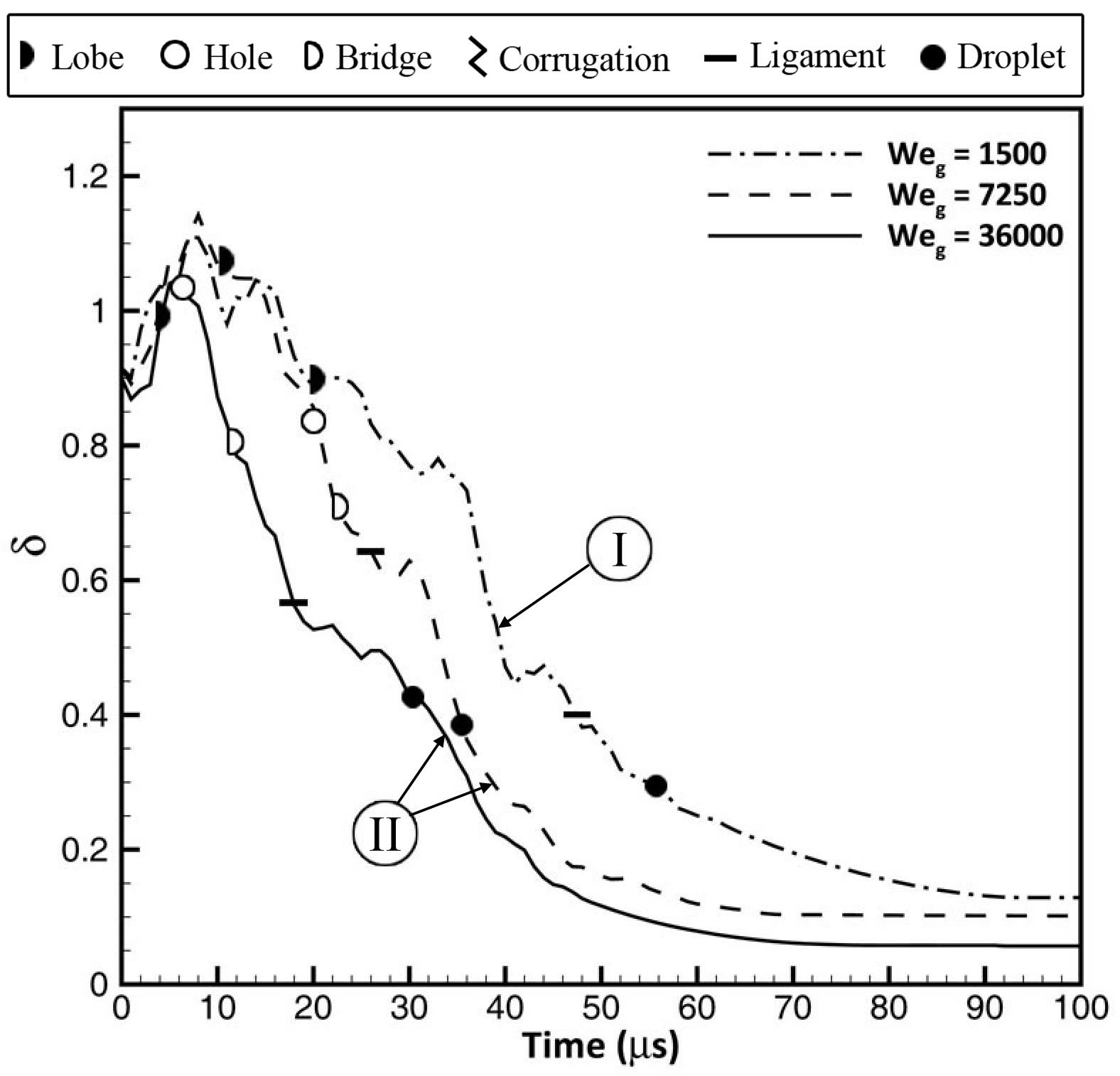}
	\caption{Effect of $We_g$ on the temporal variation of $\delta$; $Re_l=2500$, \mbox{$\hat{\rho}=0.5$}, $\hat{\mu}=0.0066$, and $\Lambda = 2.0$. The symbols indicate the first time when different liquid structures form.}
	\label{fig:average scale}
\end{figure}

The temporal variation of $\delta$ for low, medium, and high $We_g$ and moderate $Re_l$ are illustrated in Fig.~\ref{fig:average scale}. The density ratio is kept the same ($0.5$) among all these cases; hence, $We_g$ is only changed through the surface tension. The effects of $\hat{\rho}$ are analyzed in Section~\ref{Density ratio}, and the combined effects of $\hat{\rho}$ and $We$ are revealed there. The symbols on the plot denote the first instant at which different liquid structures form during the atomization process. The definition of each symbol is introduced above the plot in Fig.~\ref{fig:average scale} and will be used hereafter in the proceeding plots. The symbols help us compare the rate of formation of each structure, say ligament or droplet, at different flow conditions. The relation between the formation of each structure and the behavior of the length scale or spray width can also be better understood using these symbols. The atomization domain for each process is also denoted on the plot. Note that the zigzag symbol denoting the ``corrugation" formation does not appear in Fig.~\ref{fig:average scale} because this structure forms only in Domain III (a Domain III result will be discussed in the next sub-section). All computations are stopped at $100~\mu$s, which is sufficient for the quantities of interest to reach a steady state. It will be shown later that further continuation of the computations is not justified because the interface gets too close to the top and bottom computational boundaries, so that the length scale and sheet width get affected by the boundary conditions.

The lowest $We_g$ falls in the ligament stretching ($LoLiD$) category in Atomization Domain I, while the two higher $We_g$ cases follow the hole-formation mechanism ($LoHBrLiD$) in Domain II; see Fig.~\ref{fig:domains_gas_based}. $\delta$ decreases with time for all cases, to be expected for the cascade of structures presented in Fig.~\ref{fig:sketches} -- lobes to holes and bridges, to ligaments and then to droplets. $\delta$ becomes smaller as $We_g$ increases, and its cascade is hastened by increasing $We_g$. At $40~\mu$s, the average length scale for $We_g=36,000$ is almost $0.22\lambda_0 = 22~\mu$m, while it increases to $0.3\lambda_0$ for $We_g = 7250$, and to $0.5\lambda_0$ at the lowest $We_g$. This shows the clear influence of surface tension on the length-scale cascade and on the size of the ligaments and droplets. An increase in surface tension suppresses the instabilities and increases the structure size. This trend is consistent with both analytical \citep{Negeed} and numerical \citep{Desjardins} results. The droplets and ligaments formed in Domain II are generally smaller than in Domain I.

The length scale starts from $0.9\lambda_0$ in all three cases due to the initial perturbations, and $\delta$ increases for the first $10~\mu$s, until it reaches a maximum. $We_g$, and hence surface tension, does not notably affect the initial length scale growth. This growth involves the initial stretching of the waves, which creates flat regions near the braids -- they have low curvatures, hence large length scales. As lobes and ligaments form later, $\delta$ decreases because (i) the radius of curvature of these structures is much smaller than the initial waves, and (ii) the total interface area increases by the formation of lobes, bridges, ligaments, and droplets, smearing out the influence of the large length scales. Fig.~\ref{fig:average scale} also shows that all the structures -- especially ligaments and droplets -- form sooner with increasing $We_g$. Moreover, ligaments and droplets form slower in Domain I than in Domain II; that is, the $LoHBrLiD$ mechanism is more efficient than the $LoLiD$ process in terms of cascade rate at the same $Re_l$ range.

\cite{Arash2} showed that two distinct characteristic times exist for the formation of holes and the stretching of lobes and ligaments. At a given $Re_l$, as surface tension increases (i.e.~decreasing $We_g$), the characteristic time for hole formation increases, thereby delaying the hole formation. Thus, for lower $We_g$, most of the earlier ligaments are formed by direct stretching of the lobes and/or corrugations, while the hole formation is inhibited. On the other hand, at relatively large $Re_l$ \mbox{($>3000$)}, as liquid viscosity is increased (i.e.~decreasing $Re_l$), at the same $We_g$, the ligament-stretching time gets larger. In this case, hole formation prevails over the ligament stretching mechanism, resulting in more holes on the lobes. As $We_g$ increases, the time at which the first hole forms decreases. This indicates that the hole formation time should be inversely proportional to $We_g$. 

\begin{figure}
	\centering\includegraphics[width=0.97\linewidth]{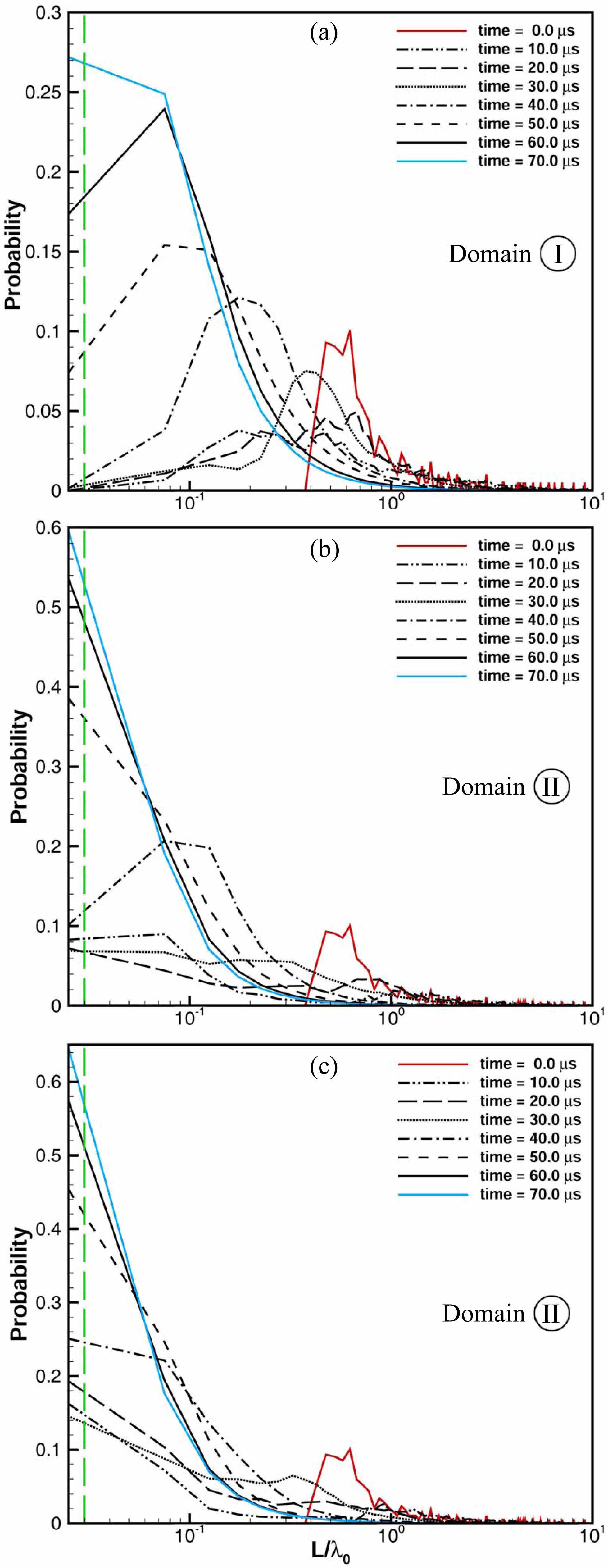}
	\caption{PDF of the normalized length scales at different times for $We_g=1500$ (a), $We_g = 7250$ (b), and $We_g=36,000$ (c); $Re_l=2500$, $\hat{\rho}=0.5$, $\hat{\mu}=0.0066$, and $\Lambda = 2.0$. The broken green lines indicate the cell size.}
	\label{fig:scale PDF}
\end{figure}

At low $Re_l$ ($<3000$), the liquid viscosity has an opposite effect on the hole formation and ligament stretching \citep{Arash2}. As shown in Fig.~\ref{fig:domains_gas_based}, near the left transitional boundary, the time scale of the stretching becomes relatively smaller than the hole-formation time scale as $Re_l$ is reduced at a constant $We_g$. Therefore, there is a reversal to ligament stretching as $Re_l$ is decreased at a fixed $We_g$. Keeping all these effects in mind, the following two nondimensional characteristic times were proposed by \cite{Arash2};
\begin{subequations}\label{eqn:thole}
\begin{equation} \label{eqn:tau h}
\frac{U\tau_h}{h_0} \hspace{1pt} \propto \hspace{1pt} \frac{1}{We_g}\left( 1+\frac{k}{Re_l} \right)
\end{equation}
\begin{equation}
\frac{U\tau_s}{h_0} \hspace{1pt} \propto \hspace{1pt} \hspace{1pt} \frac{1}{Re_l},
\end{equation}
\end{subequations}
where $\tau_h$ and $\tau_s$ are the dimensional characteristic times for hole formation and ligament stretching, respectively, and $k$ is a dimensionless constant. The results in Fig.~\ref{fig:average scale} are consistent with Eq.~(\ref{eqn:thole}), which suggests the hole formation time scale to be inversely proportional to $We_g$. Therefore, the hole formation time for $We_g = 36,000$ should be nearly $5$ times smaller than for $We_g=7250$ since $Re_l$ is the same for both cases. From Fig.~\ref{fig:average scale}, the first instant when a hole is formed is almost $5~\mu$s for $We_g = 36,000$ (solid line), and $22~\mu$s for $We_g=7250$ (dashed line). Thus, the ratio of $\tau_h$ for these two cases is about $4.4$, in good agreement with the result obtained from Eq.~(\ref{eqn:tau h}).

Even though $\delta$ gives a good insight into the temporal variation of the overall scale of the liquid structures, it does not show the distribution of the scales, for which the length-scale PDFs are needed. Fig.~\ref{fig:scale PDF} compares the PDFs of the length scales of the three cases of Fig.~\ref{fig:average scale} at different times on a log-scale.

\begin{figure*}[!t]
	\centering\includegraphics[width=0.85\linewidth]{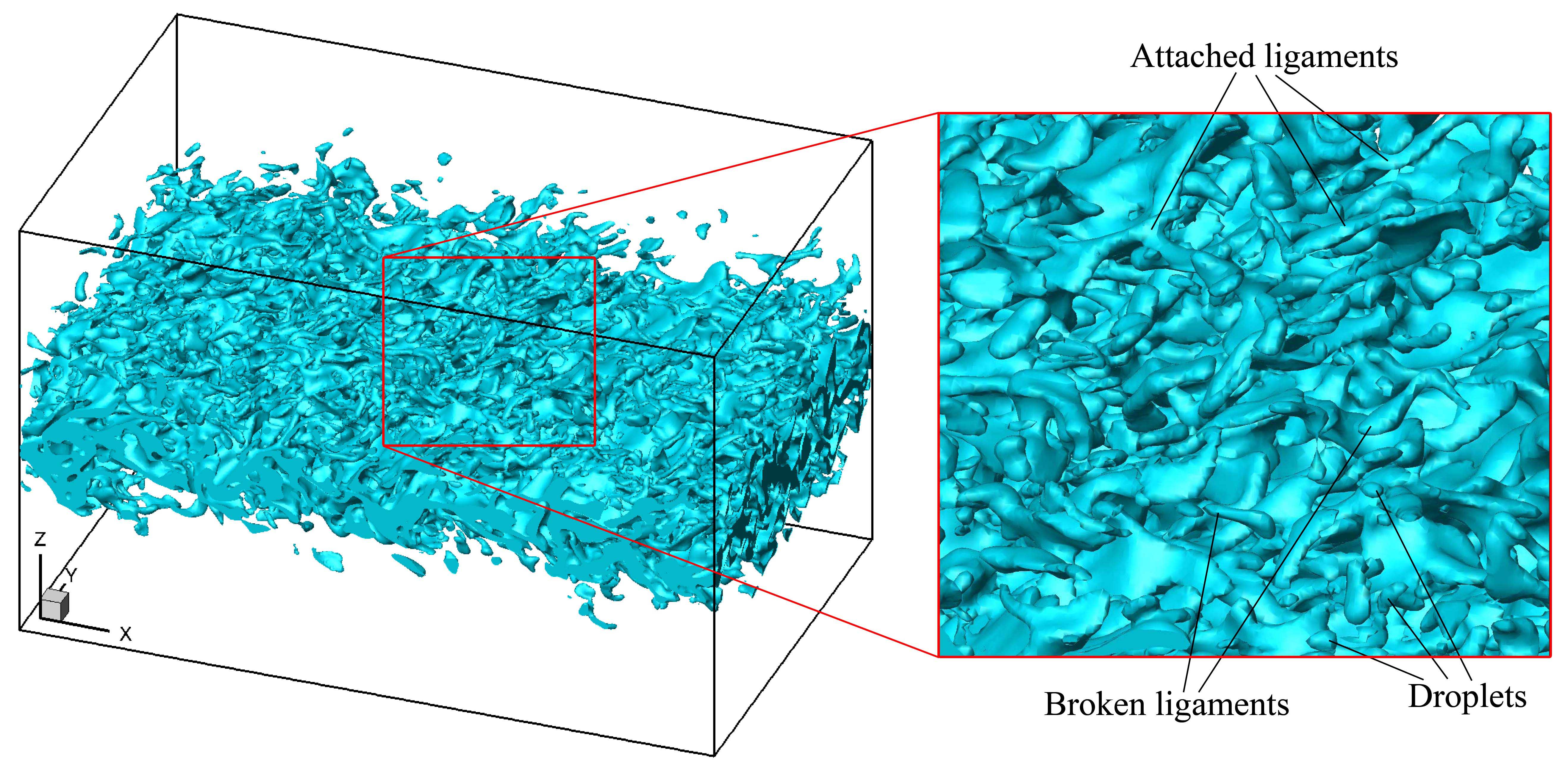}
	\caption{Liquid-jet surface at $70~\mu$s for $We_g=7250$; $Re_l=2500$, $\hat{\rho}=0.5$, $\hat{\mu}=0.0066$, and $\Lambda = 2.0$.}
	\label{fig:Surface_vs_Weber}
\end{figure*}

All cases start from the same initial distribution indicated by the red line. The length scales in the range $0.5\lambda_0$--$0.6\lambda_0$ have the maximum probability of approximately $10\%$. Later, the most probable length scale becomes smaller, while the probability of the dominant scale increases. The transition towards smaller scales is faster as $We_g$ increases since lowering surface tension reduces the resistance of the liquid surface against deformations. At $50~\mu$s, the most probable length scale becomes $0.1\lambda_0$ for \mbox{$We_g=1500$} (dashed-line in Fig.~\ref{fig:scale PDF}a). Higher $We_g$ cases reach the same most probable length scale at $40~\mu$s (dash-dotted line in Fig.~\ref{fig:scale PDF}b) for $We_g=7250$, and at less than $10~\mu$s (not shown) for $We_g=36,000$.

The PDFs also show an increase in the probability of smallest scales at higher $We_g$. At $t=70~\mu$s, for example, the lowest $We_g$ has a probability of about $27\%$ for the smallest computed length scale of $2.5~\mu$m; see where the blue curve intersects the vertical axis in Fig.~\ref{fig:scale PDF}($a$). That probability at the same time increases to $59\%$ as $We_g$ increases to $7250$. At even higher $We_g=36,000$, the probability of $2.5~\mu$m or lower is slightly more than $64\%$ at $70~\mu$s. Since the smaller length scale also implies smaller volume, the conclusion is that the number density of the small droplets also increases with $We_g$. The green broken lines in Fig.~\ref{fig:scale PDF} indicate the normalized cell size. In all cases and at all times, more than $98\%$ of the computed length scales lie to the right of this line and are larger than the mesh size, which justifies the sufficient resolution of the current grid. The effects of grid resolution on the liquid structures scale was demonstrated in detail by \citet{Arash3} and it was shown that the grid resolution used for the current analysis is fine enough to capture the smallest radii of curvature.

Fig.~\ref{fig:Surface_vs_Weber} shows the liquid surface for the moderate $We_g=7250$ case at $70\mu$s. As shown in the magnified image, at this time the liquid surface is mainly comprised of ligaments (either broken or still attached) and droplets, while little or almost no lobes with very large scales are present. Therefore, the evolution of the surface is summarized mainly in stretching and breakup of the ligaments henceforth. 

\begin{figure}[!t]
	\centering\includegraphics[width=1.0\linewidth]{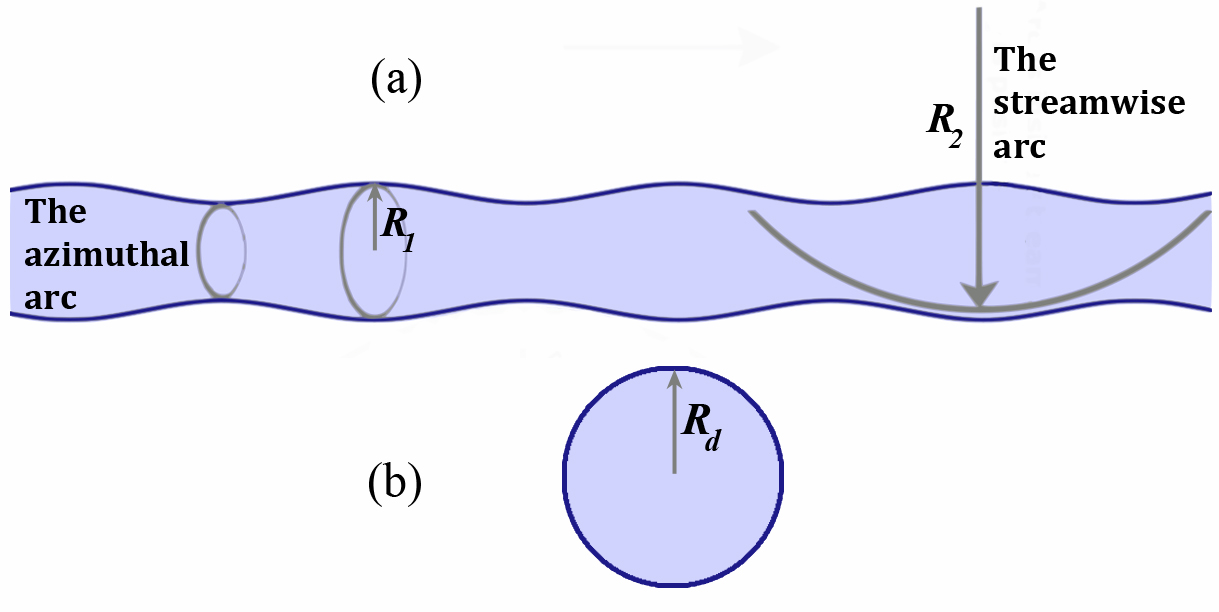}
	\caption{Schematic of the radii of curvatures on a typical ligament (a), and its resulting droplet (b).}
	\label{fig:ligament arc}
\end{figure}

As shown in Fig.~\ref{fig:ligament arc}, there are two radii of curvature in a typical perturbed ligament. $R_1$ is the smaller azimuthal radius which is initially equal to the radius of the cylindrical ligament. $R_2$, the radius of curvature of the streamwise arc of the ligament after it undergoes Rayleigh-Plateau (RP) instability, is much larger than $R_1$. Theoretical analyses of \cite{Rayleigh} show that for a cylindrical liquid segment of radius $R_1$, unstable components are only those where the product of the wave number with the initial radius is less than unity; i.e.~$kR_1<1$. Thus, the minimum unstable RP wavelength for a ligament of radius $R_1$ is $\lambda_{RP}=2\pi R_1$. The volume of a cylinder of radius $R_1$ and length $\lambda_{RP}$ is $V=2\pi^2R_1^3$. If this segment of the cylinder (ligament) breaks into a droplet, a simple mass balance shows that the resulting droplet radius ($R_d$ in Fig.~\ref{fig:ligament arc}b) would be $R_d\approx1.67R_1$. $R_2$ is never smaller than $R_1$; $R_2\approx R_1$ based on an approximate sinusoidal surface shape at the instant of ligament breakup, and $R_2=\infty$ in case of unperturbed ligament; i.e.~$R_1\leq R_2<\infty$. Considering these limits and using the relation between $R_1$ and $R_d$, the extents of ligament length scale $L_{l}=2/(1/R_1 + 1/R_2)$ as a function of $R_d$ follow
\begin{subequations}\label{ligament scale}
	\begin{equation}
	\text{if}~R_2=\infty \rightarrow L_l=\frac{2}{1/R_1}=2R_1\approx 1.2R_d ,
	\end{equation}
	\begin{equation}
	\text{if}~R_2=R_1 \rightarrow L_l=\frac{2}{1/R_1 + 1/R_1}=R_1\approx 0.6R_d
	\end{equation}
\end{subequations}
thus, $0.6R_d<L_l<1.2R_d$. The droplet length scale is $L_d=R_d$. Therefore, the time-averaged ligament length scale is approximately equal to the droplet length scale during the short period of RP instability growth and ligament breakup. This simple analysis explains why the average length scale becomes almost constant after $70\mu$s (see Fig.~\ref{fig:average scale}) while the ligament breakup is still occurring and the number of droplets is increasing. Since the ligament formation is delayed (about $30~\mu$s) at lower $We_g$, the asymptotic length scale is also expected to occur later for $We_g=1500$. This is consistent with Fig.~\ref{fig:average scale}, where the length scale asymptotes about $27~\mu$s later for $We_g=1500$ compared to $We_g=7250$.

\begin{figure}[!t]
	\centering\includegraphics[width=1.0\linewidth]{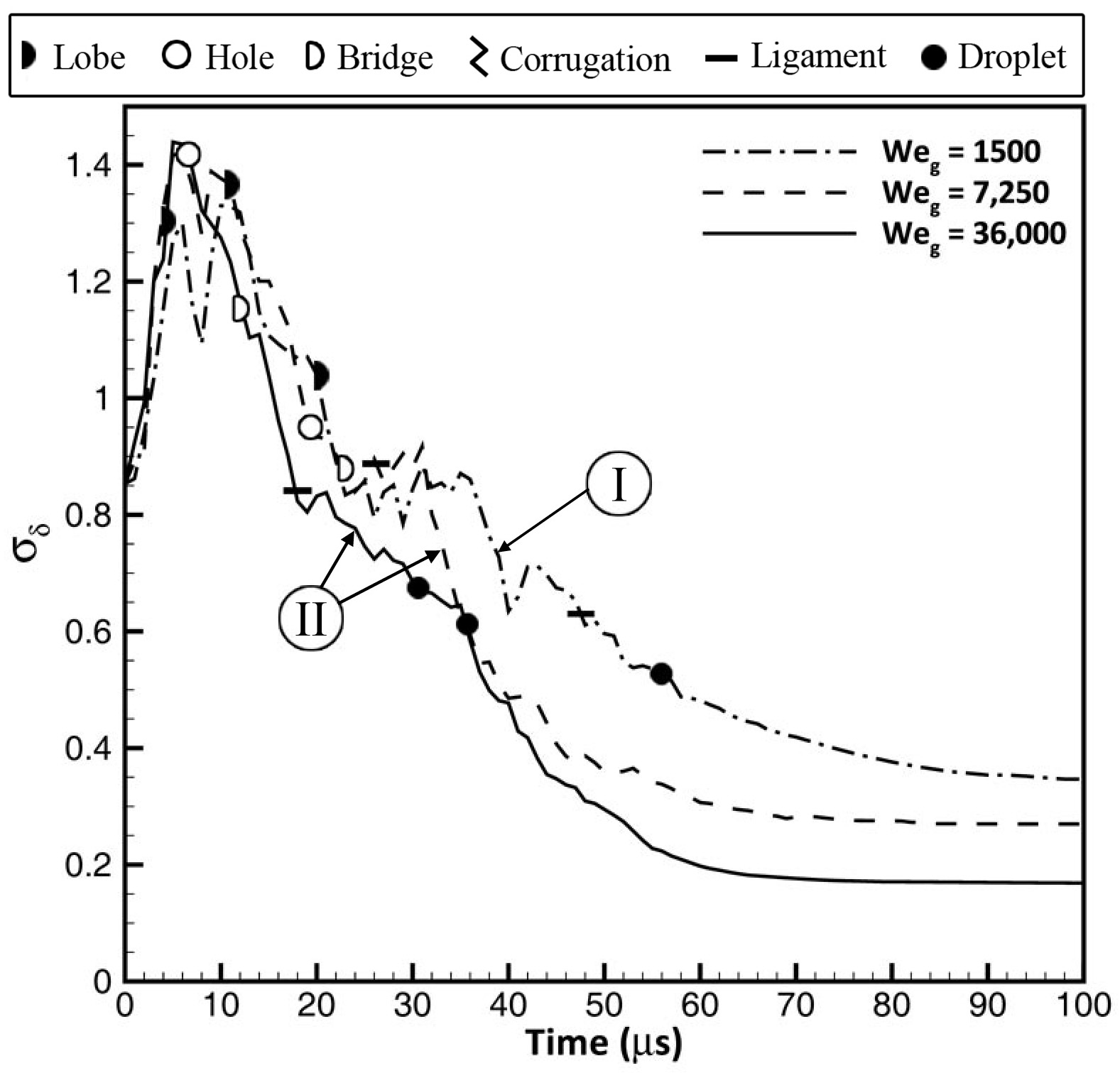}
	\caption{Temporal variation of the standard deviation of the dimensionless length-scale PDFs of Fig.~\ref{fig:scale PDF} for $We_g=1500$, $7250$, and $36,000$; $Re_l=2500$, $\hat{\rho}=0.5$, $\hat{\mu}=0.0066$, and $\Lambda = 2.0$.}
	\label{fig:scale std}
\end{figure}

The standard deviation of the PDFs of Fig.~\ref{fig:scale PDF}, illustrated in Fig.~\ref{fig:scale std}, provides a more accurate quantitative measure of the length-scale distribution. The length scales range from a few microns, i.e.~a small fraction of the initial wavelength, to several hundred microns, e.g.~four times the initial wavelength. Therefore, the standard deviation of $\delta$ ($\sigma_\delta$) is very large, even in the beginning. $\delta$ is about $0.9\lambda_0=90~\mu$m at the start of the computations (see Fig.~\ref{fig:average scale}), but $\sigma_\delta$ is about $0.85\lambda_0=85~\mu$m at this time.

At the early stage, $\sigma_\delta$ increases as larger scales become more probable following the flattening and stretching of the waves. Later, the flow field gets filled with more small ligaments and droplets and more curved surfaces, which reduce both the mean and the standard deviation. However, even at the end of the process, $\sigma_\delta$ is still around $0.2$--$0.4\lambda_0=20$--$40~\mu$m; so, a wide range of length scales is still present in the flow. $\sigma_\delta$ decreases with increasing $We_g$, as the smaller capillary force allows the larger scales to deform easily and cascade more quickly into smaller scales with higher curvatures; this reduces the deviation of the scales. $\sigma_\delta$ becomes almost constant when $\delta$ asymptotes to its ultimate value.

The length-scale PDFs show that: (i) the asymptotic length scale (ligament and droplet size) decreases with increasing $We_g$; (ii) the number of small droplets increases with increasing $We_g$; and (iii) the cascade of length scales occurs faster at higher $We_g$. The last item is also implied by the temporal evolution of $\delta$. The first two items are consistent with the literature, but the third item is a new finding.

\begin{figure}[!b]
	\centering\includegraphics[width=1.0\linewidth]{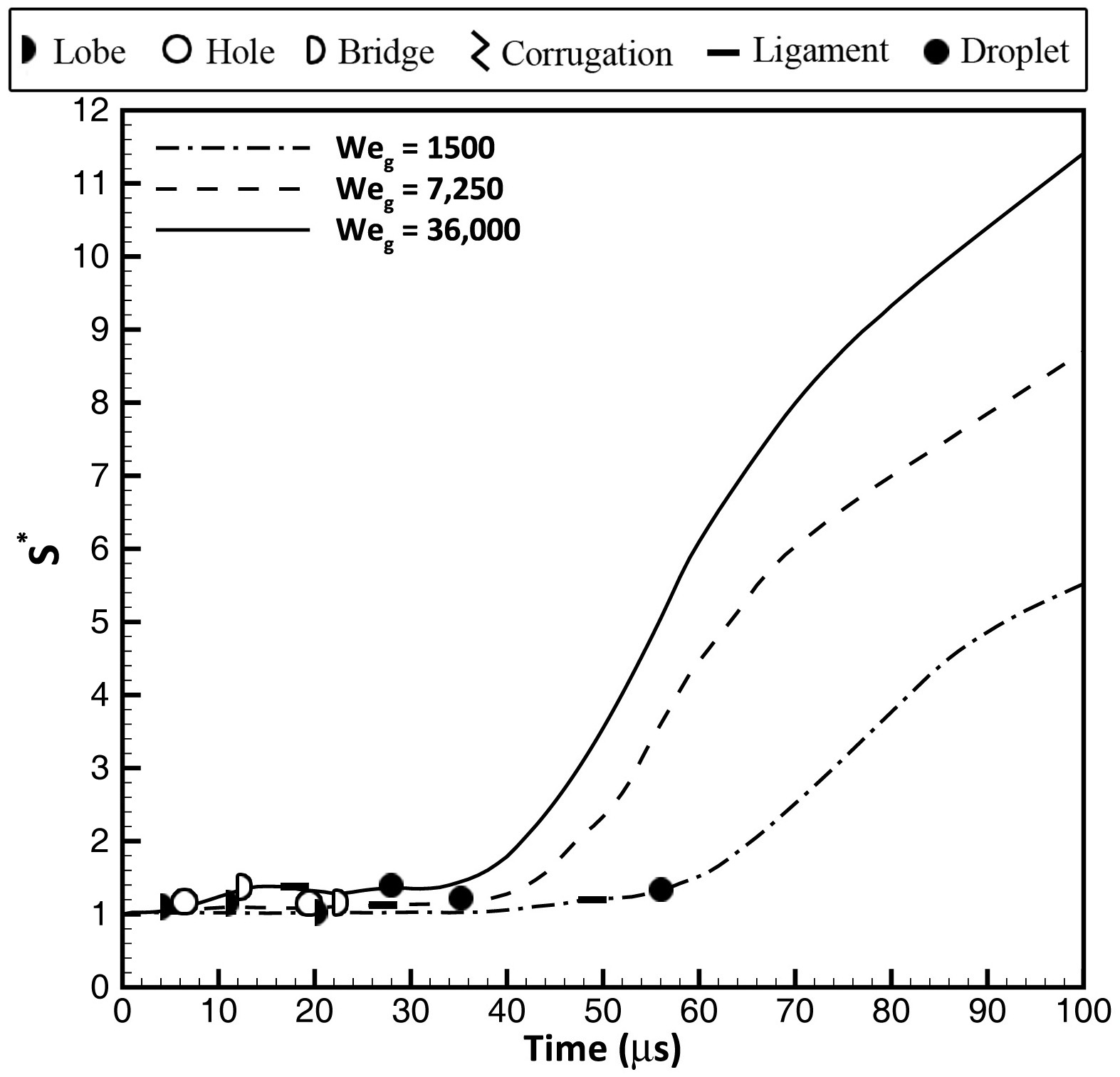}
	\caption{Temporal growth of the non-dimensional liquid surface area ($S^*$) for $We_g=1500$, $7250$, and $36,000$; $Re_l=2500$, $\hat{\rho}=0.5$, $\hat{\mu}=0.0066$, and $\Lambda = 2.0$.}
	\label{fig:area growth}
\end{figure}

Temporal growth of the non-dimensional liquid surface area ($S^*$) for the three $We_g$ cases is plotted in Fig.~\ref{fig:area growth}. The surface area ($S$) is non-dimensionalized by the initial liquid surface area; i.e.~$S^*=S/S_0$. Therefore, $S^*$ grows monotonically in time from 1 at $t=0$; \mbox{$S_0\approx0.8$~mm$^2$}.

\begin{figure*}[!t]
	\centering\includegraphics[width=1.0\linewidth]{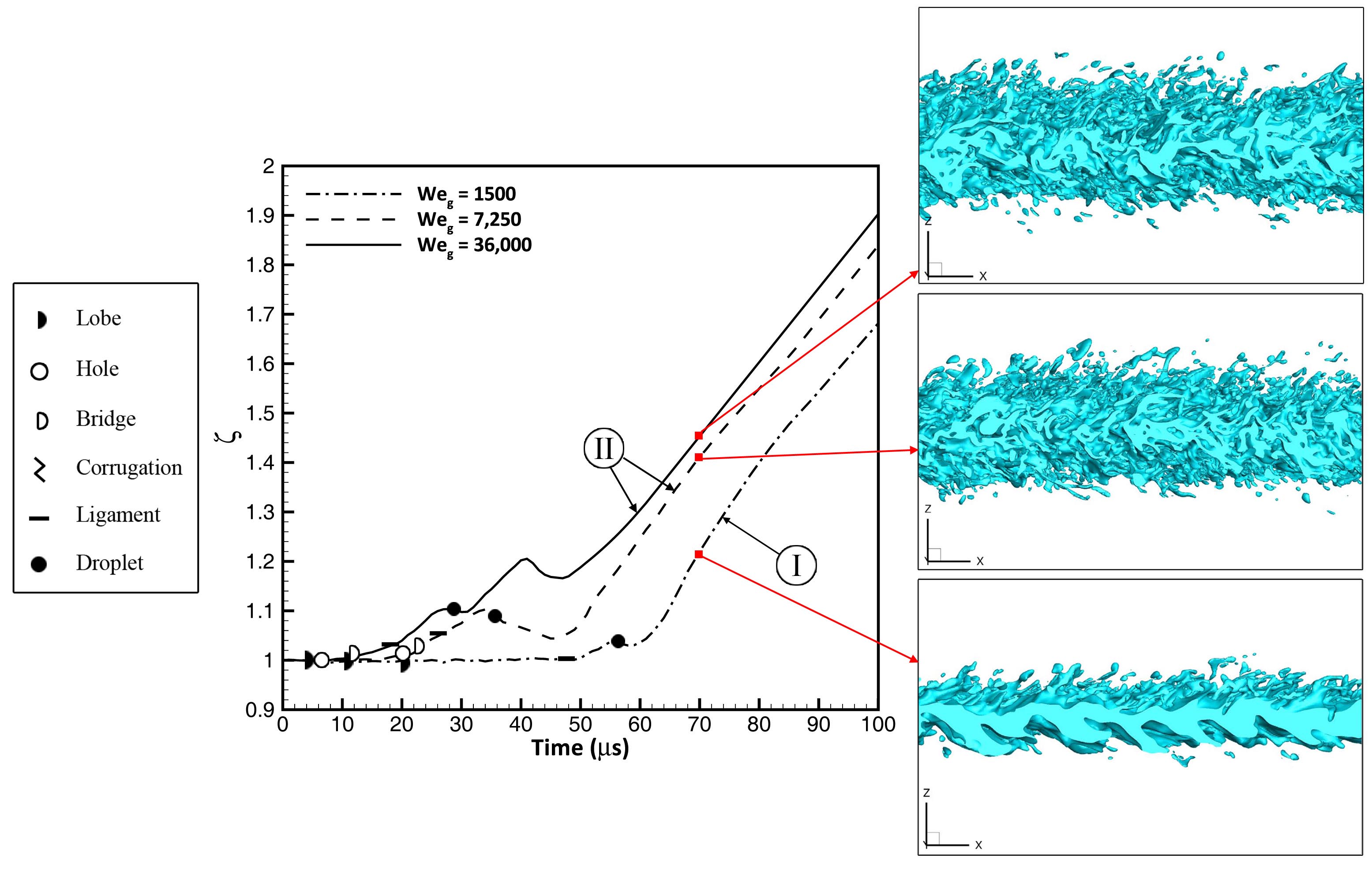}
	\caption{Effect of $We_g$ on the temporal variation of $\zeta$; $Re_l=2500$, $\hat{\rho}=0.5$, $\hat{\mu}=0.0066$, and $\Lambda = 2.0$. The liquid jet surface at $70~\mu$s is shown for each process.}
	\label{fig:average expansion}
\end{figure*}

As expected, the surface area growth rate is higher for higher $We_g$, where the surface deformations occur and grow faster and ligaments and droplets form earlier. $S^*$ grows very gradually in the first $30~\mu$s, but experiences a sudden increase in the growth rate after the formation of first ligaments and droplets. This abrupt growth in surface area occurs sooner and at a higher rate for higher $We_g$ following the earlier formation of ligaments in those cases. The $S^*$ growth rate decreases slightly towards the end of the computations, while the area still keeps growing. The cause of this gradual decrease in growth rate is speculated to be mainly due to the breakup of ligaments into droplets. Surface tension minimizes the surface area after ligament breakup. The surface deformation at later times mainly consists of formation of ligaments and their breakup into droplets. Following the simplified ligament and droplet radii model shown in Fig.~\ref{fig:ligament arc}, a simple calculation reveals that the surface area of the resulting droplet is $10\%$ smaller than the surface area of its mother ligament; i.e.~$S_d=0.9S_l$. Therefore, the ligament breakup decreases the $S^*$ growth rate associated with mere stretching of the ligaments. Even though the mean length scale has reached an asymptote at the final stage ($t>70~\mu$s), the increase in $S^*$ indicates that the surface dynamics are still in progress and ligament and droplet formation still continues. The rate of growth of $S^*$ becomes almost constant in the asymptotic phase. At $100~\mu$s, the surface area has grown more than 11 times for $We_g=36,000$, while for $We_g=1500$, the surface has less than $6$ times its initial area.

As mentioned earlier, defining the jet width (thickness) as the distance to the farthest liquid point from the centerplane might render results which are prone to misinterpretation. This definition does not take into account the interface location distribution and only considers the farthest liquid location. For this purpose, the PDFs of the transverse interface location are used to describe the expansion of the liquid sheet in a more meaningful form.

The effect of $We_g$ on the temporal evolution of the average spray width ($\zeta$) is shown in Fig.~\ref{fig:average expansion} along with the liquid-jet interface picture at $70~\mu$s of each process. The expansion rate of the liquid jet can be distinguished better with the new definition of the jet width \citep[compare this plot with figure~40 of][]{Arash1}. The jet images at $70~\mu$s show that the two higher $We_g$ jets would have a very similar width if the widths were presented as the farthest liquid points from the centerplane. However, our method clearly shows the difference between these two cases, and indicates that a larger portion of the liquid surface is in fact located at a farther transverse distance from the jet center for the higher $We_g$. 

The spray expands faster at higher $We_g$ and results in a wider spray at the end -- a result which is in agreement with numerical results of \citet{Desjardins}. $\zeta$ remains close to $1.0$ for the first $50$ microseconds for $We_g=1500$ since the high surface tension suppresses instability waves, lobe stretching, and ligament breakup. The jet expands much sooner at higher $We_g$; for example, at about $20~\mu$s for $We_g=7250$, and at $7~\mu$s for $We_g=36,000$. The structures stretch much more quickly at higher $We_g$ and are less suppressed by the surface tension forces; thus, they can expand more freely and are carried around more easily by the gas flow, after breakup. The expansion of the jet at the lowest $We_g$ (dash-dotted line in Fig.~\ref{fig:average expansion}) coincides with the formation of the first ligament (at $50~\mu$s). Therefore, the lobes are much less amplified at such a low $We_g$, and the ligament formation and stretching are primarily in the normal direction in Domain I. Generally, the spray angle is larger in Domain II than in Domain I at a comparable $Re_l$ range.

\begin{figure}
	\centering\includegraphics[width=0.97\linewidth]{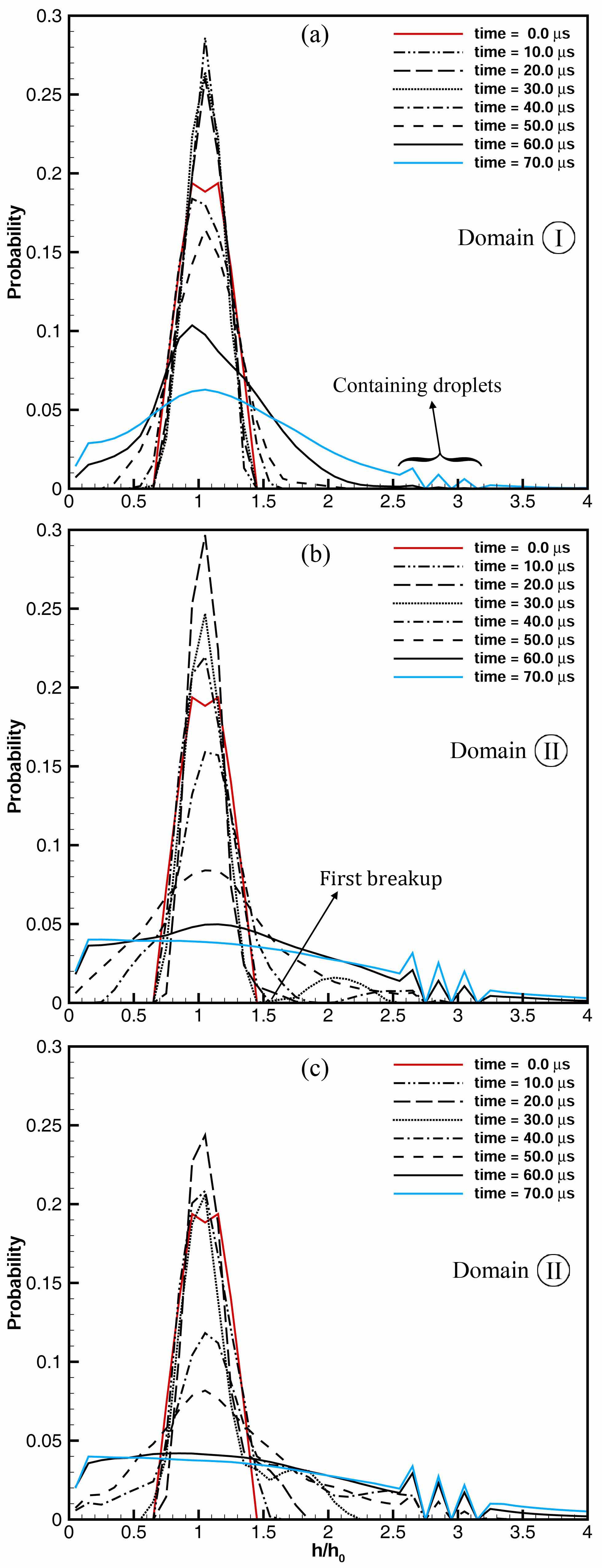}
	\caption{Nondimensional spray-width PDF at different times for $We_g=1500$ (a), $We_g = 7250$ (b), and $We_g=36,000$ (c); $Re_l=2500$, $\hat{\rho}=0.5$, $\hat{\mu}=0.0066$, and $\Lambda = 2.0$.}
	\label{fig:expansion PDF}
\end{figure}

The spikes and oscillations in $\zeta$ are caused by the detachment of a liquid structure, e.g.~bridges, ligaments or droplets, from the jet core. The first decay in $\zeta$ coincides with the formation of the first droplets; i.e.~the black circles in Fig.~\ref{fig:average expansion}. The spray-width PDFs, given in Fig.~\ref{fig:expansion PDF}, support this view.

All three cases in Fig.~\ref{fig:expansion PDF} start from a bell-shaped distribution around $h=h_0$, denoted by the red solid lines. The two peaks on the two sides of $h/h_0 = 1$ are due to the initial perturbations amplitude of $5~\mu$m imposed on the surface of the sheet. Since there are more computational cells near the peak and trough of the perturbations compared to the neutral plane, i.e.~$h/h_0=1$, the probability of those sizes are slightly higher. The probability of the initial thickness value increases in all cases during the first $20~\mu$s since the wave amplitude decays as the waves get stretched in the flow direction, in the initial stage.

\begin{figure*}[!t]
	\centering\includegraphics[width=1.0\linewidth]{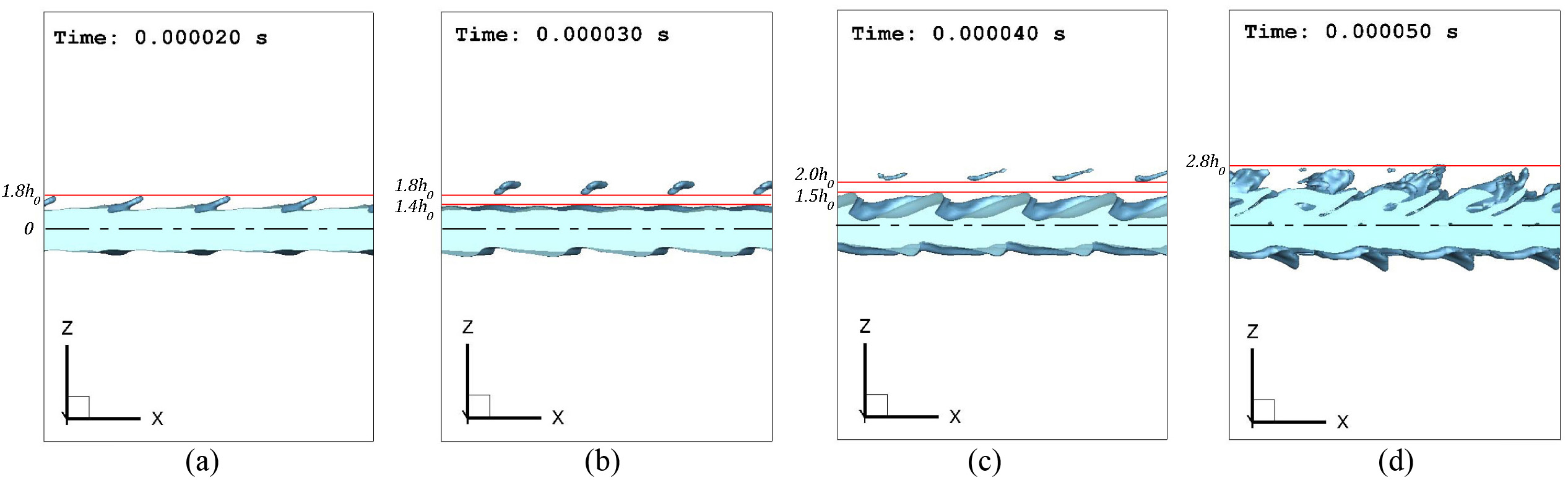}
	\caption{Side view of the liquid sheet surface at $t=20~\mu$s (a), $30~\mu$s (b), $40~\mu$s (c), and \mbox{$50~\mu$s (d)}; $Re_l=2500$, $We_g=7250$, $\hat{\rho}=0.5$, $\hat{\mu}=0.0066$, and $\Lambda = 2.0$. Gas flows from left to right.}
	\label{fig:expansion We14k}
\end{figure*}

Later, when the waves start to grow due to the KH instability and roll-up of the lobes over the primary KH vortices, the jet width increases and the distribution becomes wider and skews towards larger values on the right. Meanwhile, smaller lengths are also observed. With time, the peak of the initial distribution curve decreases while the distribution broadens. This decline occurs faster at higher $We_g$, since the spray grows faster for lower surface tension. At $t=50~\mu$s, only $8\%$ of the surface, i.e.~computational cells at the interface, lie near the initial thickness for $We_g=36,000$ (dashed-line in Fig.~\ref{fig:expansion PDF}c), and the spray has grown upto three times the initial sheet thickness; i.e.~$h/h_0 \approx 3$. At the same time, the farthest transverse distance reached by the liquid is about $2.8h_0$ from the centerplane for $We_g=7250$; see where the dashed-line in Fig.~\ref{fig:expansion PDF}(b) meets the horizontal axis. At still lower $We_g$, the maximum spray width is just slightly more than $2h_0$ at $50~\mu$s, and still more than $16\%$ of the surface lies around the initial sheet thickness; see Fig.~\ref{fig:expansion PDF}(a).

Fig.~\ref{fig:expansion PDF}(b) shows a missing section (having zero probability) around \mbox{$1.4<h/h_0<1.8$} at $t=30~\mu$s. The same missing section moves to the right (to $1.5<h/h_0<2.0$) at $t=40~\mu$s, and finally vanishes at $50~\mu$s. These sections -- marked as the ``first breakup" -- coincide with the places where the sudden decline in $\zeta$ was seen in Fig.~\ref{fig:average expansion} for $We_g=7250$, explaining the oscillations in the average spray width. The missing section appears since some part of the liquid jet (ligament, bridge or droplet) detaches from the jet core into the gas flow, leaving behind a vertical gap empty of any liquid surface at the breakup location, as shown in the sequential liquid surface images in Fig.~\ref{fig:expansion We14k}.

The lobes form and stretch until about $25~\mu$s, resulting in an increase in $\zeta$. At $30~\mu$s (Fig.~\ref{fig:expansion We14k}b), the bridge breaks from the lobe and creates a gap near the detachment location, where there are no liquid elements. Therefore, $\zeta$ suddenly drops at that instant, even though the distance of the farthest liquid element from the centerplane is still growing; i.e.~the intersection of the PDF curves with the horizontal axis in Fig.~\ref{fig:expansion PDF}(b) moves to the right. While the detached liquid blob moves away from the interface, the jet surface stretches outward again due to KH instability; the missing zone moves outward following this motion (Fig.~\ref{fig:expansion We14k}c). $\zeta$ grows again when the new lobes and ligaments stretch enough to compensate for the broken (missing) section. At this time, the lobes and ligaments fill in the missing gap while the broken liquid structures advect farther from the interface, as shown in Fig.~\ref{fig:expansion We14k}(d) at $50~\mu$s. Fig.~\ref{fig:expansion We14k} also shows that the instabilities start from a symmetric distribution but gradually move towards an antisymmetric mode (Fig.~\ref{fig:expansion We14k}d). The transition towards antisymmetry is seen in all cases studied here and is explained in detail via vortex dynamics analysis by \citet{Arash3}. It is shown that transition towards antisymmetry is faster for thin liquid sheets due to the higher induction of the KH vortices on the opposite sides of the liquid surface. In practical atomization conditions, the antisymmetric mode has a higher growth rate and thus eventually dominates the symmetric mode.

The PDFs and the average spray-width plots indicate the first instance of ligament/bridge breakup, when $\zeta$ starts to decline. Since the lower $We_g$ has less stretching and fewer ligament detachments at early time -- due to the high surface tension -- its spike is less intense and also appears much later (about $55~\mu$s). The higher $We_g$, however, breaks much sooner, at about $27~\mu$s.

\begin{figure}[!t]
	\centering\includegraphics[width=1.0\linewidth]{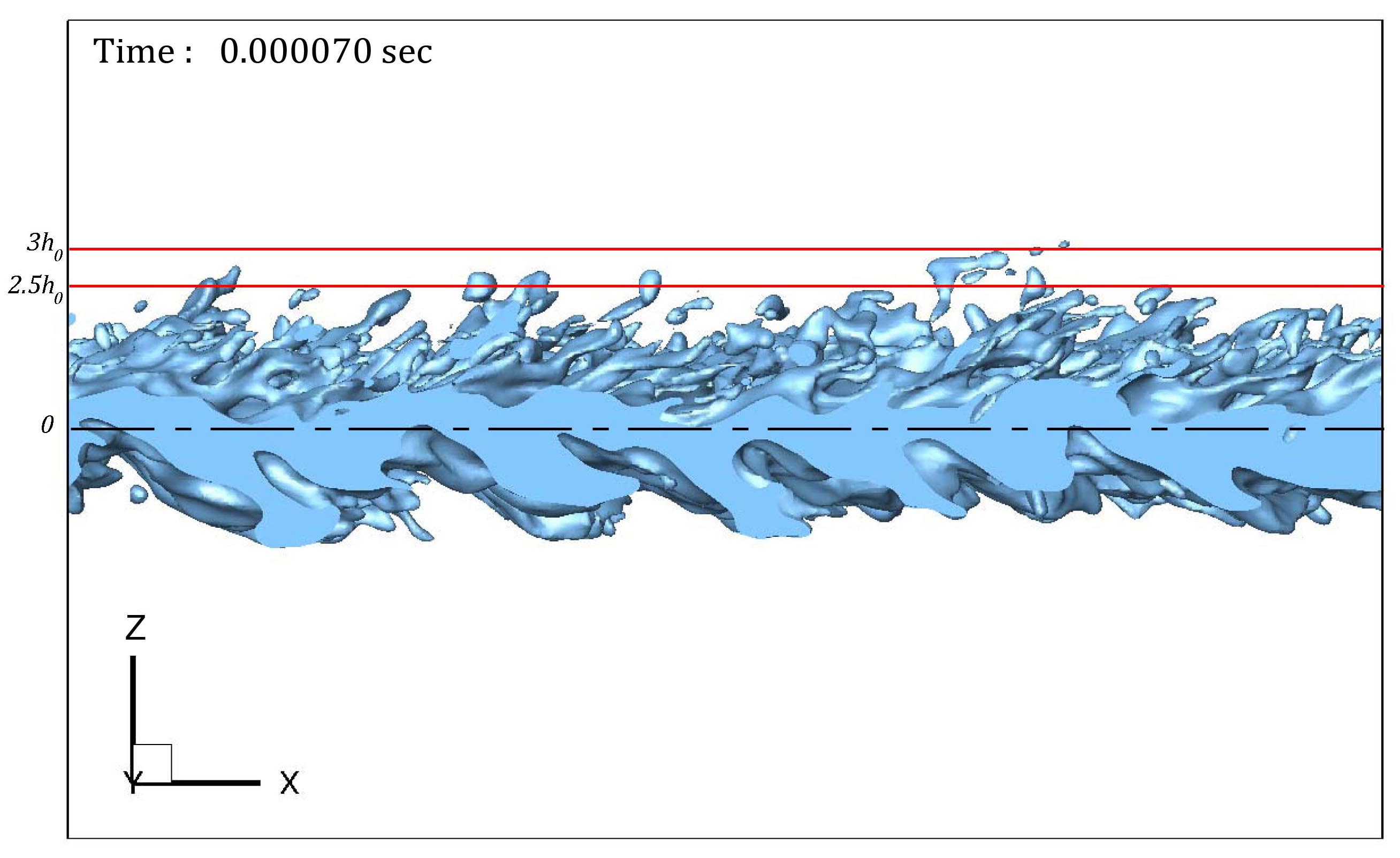}
	\caption{Side-view of the liquid surface at $t=70~\mu$s; $We_g=1500$, $Re_l=2500$, $\hat{\rho}=0.5$, $\hat{\mu}=0.0066$, and $\Lambda = 2.0$. Gas flows from left to right.}
	\label{fig:expansion t70}
\end{figure}

There are also some later oscillations near the maximum spray width of the PDF plots (see Fig.~\ref{fig:expansion PDF}), for all three cases. The reason for these spikes is explained using the liquid isosurface at $70~\mu$s for $We_g=1500$, illustrated in Fig.~\ref{fig:expansion t70}. The planes $2.5h_0$ and $3.0h_0$ away from the centerline are marked with the red lines. Undulations in the PDF curve occur in this range; see the blue line in Fig.~\ref{fig:expansion PDF}(a) marked as ``containing droplets". This range is mostly empty of liquids, i.e.~filled with gas, except for rare cells which contain the occasional droplets or detached ligaments. In the spray-width PDF plot, the empty spaces have zero or almost negligible probability, while other transverse heights containing liquid droplets and broken ligaments have greater probability. Thus, waviness is seen in the spray-width PDF at later times and at greater distances from the centerplane. Those spikes represent the droplets that are thrown outward from the jet core, as the spray spreads.

The reason for having non-zero probability for zero width in the PDF plots of Fig.~\ref{fig:expansion PDF} can also be assessed in Fig.~\ref{fig:expansion t70}. Since the antisymmetric mode is dominant in the considered range of $Re_l$ and $We_g$, the trough of the surface wave reaches the centerplane (indicated by the broken black line in Fig.~\ref{fig:expansion t70}) and even crosses it. Thus, after some time, non-zero probabilities occur for the surfaces that intersect the centerplane.

\begin{figure}[!b]
	\centering\includegraphics[width=1.0\linewidth]{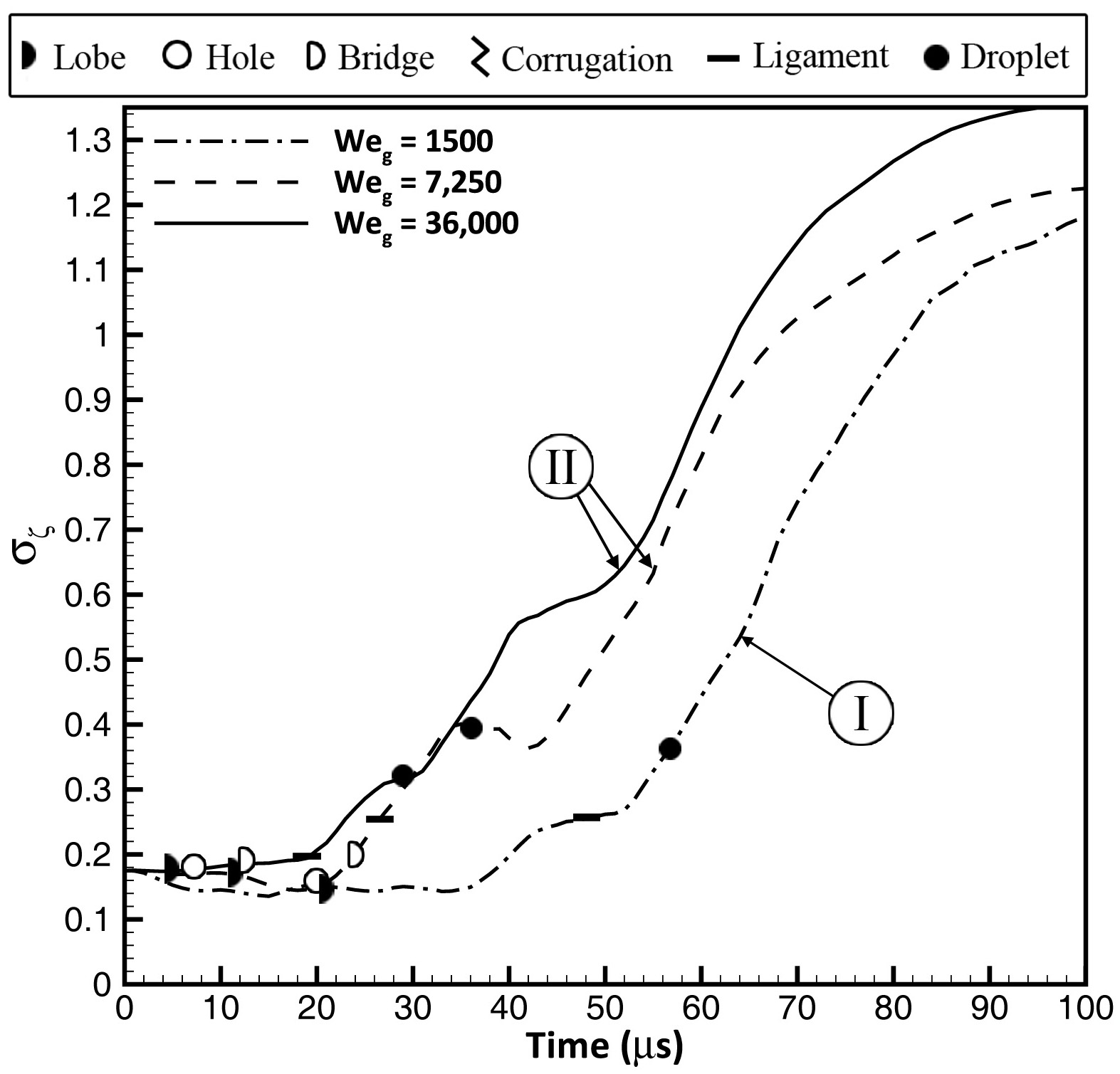}
	\caption{Temporal variation of the standard deviation of the dimensionless spray-width PDFs of Fig.~\ref{fig:expansion PDF} for $We_g=1500$, $7250$, and $36,000$; $Re_l=2500$, $\hat{\rho}=0.5$, $\hat{\mu}=0.0066$, and $\Lambda = 2.0$.}
	\label{fig:expansion std}
\end{figure}

Fig.~\ref{fig:expansion std} shows the standard deviation of the spray-width ($\sigma_\zeta$) PDFs of Fig.~\ref{fig:expansion PDF}. Initially, $\sigma_\zeta$ is slightly below $0.2h_0=10~\mu$m, which is exactly equal to the peak-to-peak amplitude of the initial perturbations. At early times, as the waves stretch in the streamwise direction, their amplitude decreases; hence, a larger portion of the interface gets closer to the initial sheet surface. This reduces $\sigma_\zeta$. This reduction is greater for lower $We_g$ because of the stabilizing role of surface tension. Later, as the spray expands, the distance between the outermost and the innermost liquid surface grows, and the spray-width PDF gets wider; see Fig.~\ref{fig:expansion PDF}. $\sigma_\zeta$ increases with increasing $We_g$. The rate of increase of $\sigma_\zeta$ decreases at about $70~\mu$s for the highest $We_g$ and at $90~\mu$s for the lowest one. The reason for this change in pace is that, beyond this point, some kind of saturation occurs in the computational box by the broken liquid blobs that get too close to the top and bottom boundaries; notice that the liquid particles cannot leave the box from the normal boundaries. This could clearly influence the dynamics of atomization. Thus, the computations are not continued beyond $100~\mu$s. A larger computational box is required if one would want to continue the analysis in time, but this is not plausible for this study due to its computational cost.

\subsection{Reynolds number effects} \label{Reynolds effects}

Practically, $Re_l$ should have a major effect on both the final droplet size and the spray angle as well as the cascade rate, since both the inertia and viscous effects are involved in these quantities. These effects are studied quantitatively here. Three different $Re_l$ values are compared in this study; $Re_l=1000$, $2500$, and $5000$ -- each representing one of the domains in the $We_g$ vs.~$Re_l$ plot with a particular breakup characteristic, as defined in Figs.~\ref{fig:domains_gas_based} and \ref{fig:sketches}. $Re_l=5000$ (in Domain III) and $Re_l=1000$ (in Domain I) have the stretching characteristics during the primary breakup, with and without the corrugation formation, respectively. The $Re_l=2500$ case follows the hole/bridge formation mechanism in Domain II.    

\begin{figure}[!t]
	\centering\includegraphics[width=1.0\linewidth]{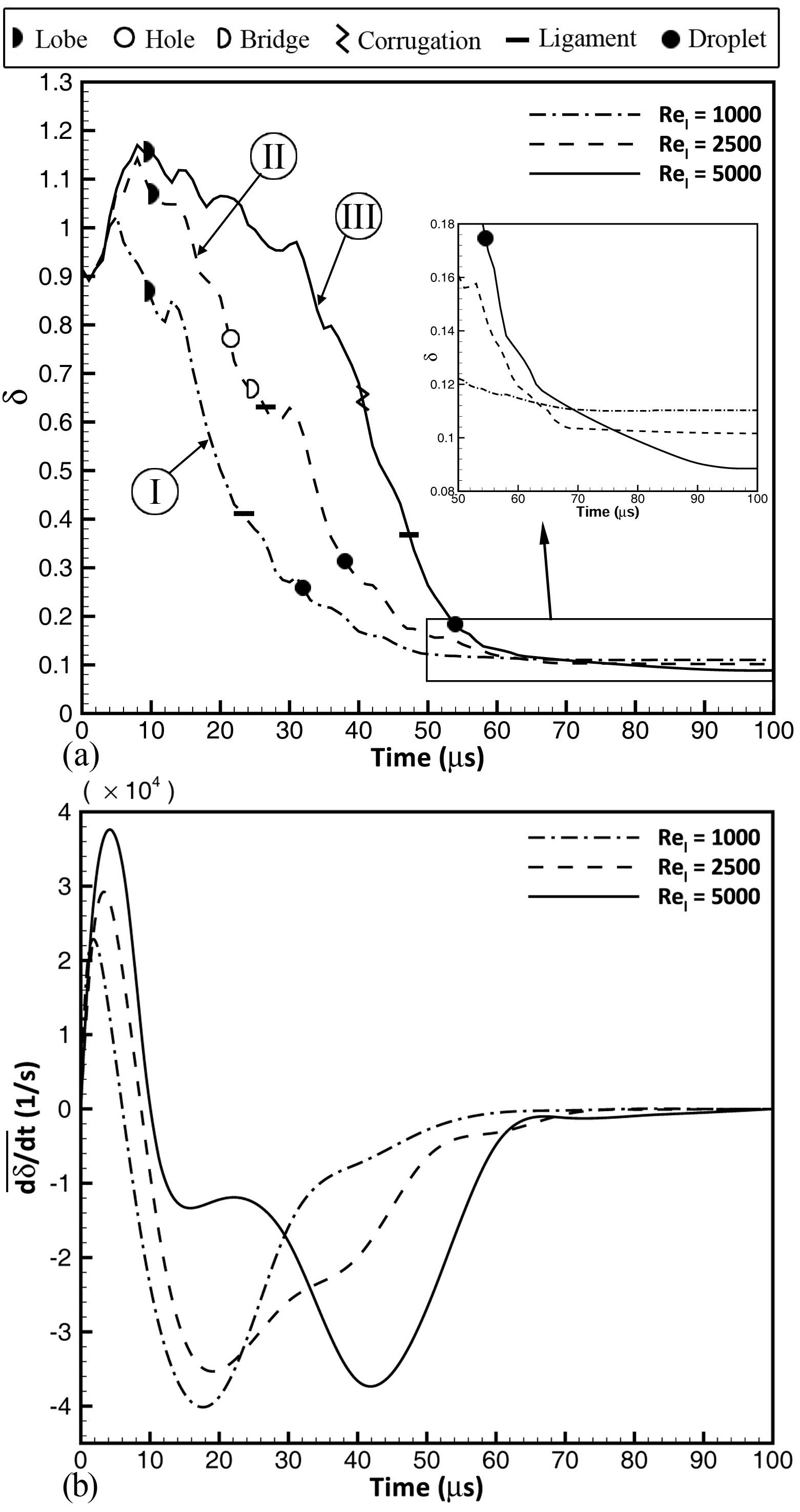}
	\caption{Effect of $Re_l$ on the temporal variation of $\delta$ (a), and the average cascade rate (b); $We_g=7250$, $\hat{\rho}=0.5$, $\hat{\mu}=0.0066$, and $\Lambda = 2.0$.}
	\label{fig:scale Reynolds}
\end{figure}

Figs.~\ref{fig:scale Reynolds}(a) and \ref{fig:scale Reynolds}(b) respectively demonstrate the effects of $Re_l$ on the average length scale ($\delta$) and the average cascade rate ($\overline{\mathrm{d}\delta/\mathrm{d}t}$) over time. The average cascade rate is obtained by calculating the average slope of the curvatures in Fig.~\ref{fig:scale Reynolds}(a) in $5~\mu$s and $10~\mu$s time spans. As expected, the liquid surface is stretched more in the streamwise direction with increasing $Re_l$, creating flatter surfaces early on. Thus, as $Re_l$ increases, relatively larger length scales occur at early times. For higher $Re_l$, the lobes stretch for a longer time before breakup, yielding maximum length scale at a later time. The breakup process and cascade of length scales occur later for higher $Re_l$ (see Fig.~\ref{fig:scale Reynolds}b).

After the early injection period, the rate of cascade of the large liquid structures, e.g.~lobes and bridges, into smaller structures, e.g.~corrugations, ligaments and droplets, is greater at higher $Re_l$, as observed from the mean slopes in Fig.~\ref{fig:scale Reynolds}(a) in the downfall portion of the plot. As shown in Fig.~\ref{fig:scale Reynolds}(b), only one maximum cascade rate is observed in Domain I at about $18~\mu$s, with an average rate of $-4~\mu$m/$\mu$s, which corresponds to the stretching of lobes into ligaments. In Domain II (dashed line), two major rates are observed -- a larger rate at $\approx 20~\mu$s, which corresponds to the stretching of lobes, and another lower rate at $\approx35~\mu$s, corresponding to the formation of holes and ligaments. In Domain III (solid line), there are two local minimums. There is a slower rate ($-1.3~\mu$m/$\mu$s) which spans from $10~\mu$s to about $25~\mu$s, and a larger rate of $\approx-3.8~\mu$m/$\mu$s, which occurs between $25~\mu$s and $50~\mu$s. The first rate corresponds to the stretching of the lobe, and the second (larger) rate corresponds to the formation of corrugations on the lobe edges. Since corrugations result in much smaller scales compared to the lobes, this transition results in a sudden growth in the cascade rate. The average cascade rate goes to zero with the formation of droplets, as was discussed before.

By decreasing the liquid viscosity, i.e.~increasing $Re_l$, with the other properties held constant, the breakup occurs faster but later. This can be attributed to the stabilizing effects of viscosity, which damps the small scale instabilities at low $Re_l$. Fig.~\ref{fig:scale Reynolds}(a) also shows that $Re_l$ affects the ultimate $\delta$ value. Even though the effect of $Re_l$ on the final length scale is not as significant as $We_g$, the magnified subplot of Fig.~\ref{fig:scale Reynolds}(a) shows that the asymptotic scale decreases from $0.11\lambda_0$ to $0.09\lambda_0$ as $Re_l$ increases from $1000$ to $5000$. Even though the rate of cascade of length scales is larger at higher $Re_l$, the asymptotic length scale is achieved later for higher $Re_l$ since the largest length scales are also larger for higher $Re_l$. $\delta$ for the highest $Re_l$ case just becomes smaller than the two lower $Re_l$ cases at about $75~\mu$s. \cite{Negeed} and \cite{Desjardins} also qualitatively showed that the final droplet size decreases with increasing $Re_l$; however, they did not quantify the droplet size nor its cascade rate.

Even though the formation of lobes is not affected much by $Re_l$, the ligaments and droplets form notably later as $Re_l$ increases. This counter-intuitive fact is related to the process of ligament and droplet formation. At a constant $We_g$, the ligament formation in the $LoCLiD$ process (Domain III) is slower than in the $LoHBrLiD$ process (Domain II), and both are slower than that in the $LoLiD$ process (Domain I). All these cascade processes are explained via vortex dynamics by \cite{Arash3}. They show that the formation of corrugations at higher $Re_l$ takes longer and requires downstream advection of the split KH vortex by a distance of one wavelength ($\approx100~\mu$m), until the hairpin vortices get undulated and induce the corrugations. The $LoHBrLiD$ process, on the other hand, requires only stretching and overlapping of the hairpins over the KH vortices, which occur faster. The $LoLiD$ mechanism involves cross-flow advection of the KH vortices, which occurs slightly more quickly. However, the droplets formed in the $LoCLiD$ process are smaller than in the other two processes, and the $LoLiD$ process results in the thickest ligaments and the largest droplets.

\begin{figure}[!t]
	\centering\includegraphics[width=1.0\linewidth]{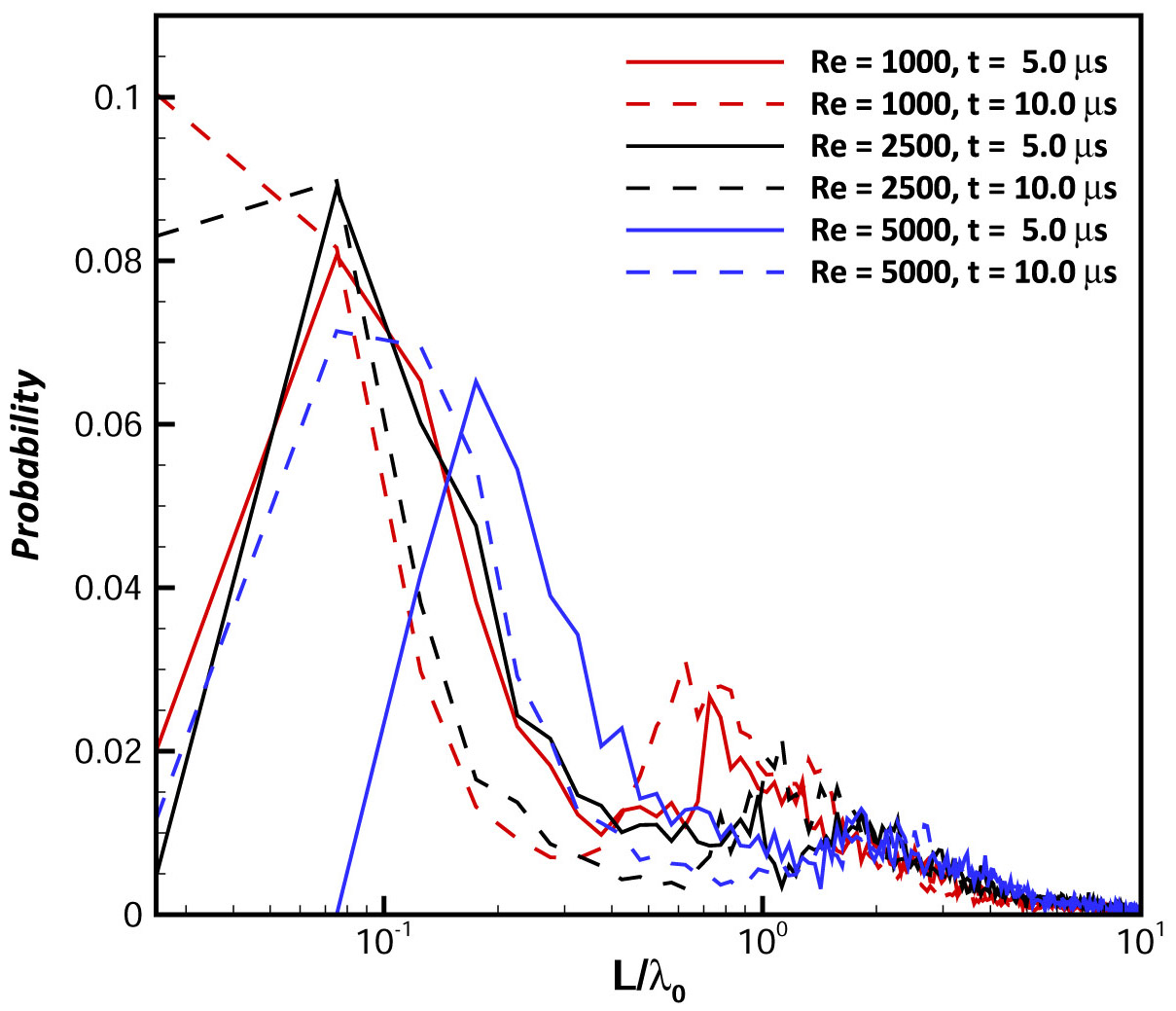}
	\caption{PDF of the normalized length scales at $t = 5~\mu$s (solid lines) and $10~\mu$s (dashed-lines) for $Re_l=1000$ (red line), $Re_l=2500$ (black line), and $Re_l = 5000$ (blue line); $We_g=7250$, $\hat{\rho}=0.5$, $\hat{\mu}=0.0066$, and $\Lambda = 2.0$.}
	\label{fig:scale PDF Reynolds}
\end{figure}

To better understand the difference in $\delta$ for the three $Re_l$ cases at early injection stage, the length scale PDFs for these cases are plotted at $5~\mu$s and $10~\mu$s in Fig.~\ref{fig:scale PDF Reynolds}. Both times are within the initial injection period when the maximum $\delta$ occurs; see Fig.~\ref{fig:scale Reynolds}(a). The $Re_l=1000$, $2500$, and $5000$ cases are shown by red, black, and blue lines in Fig.~\ref{fig:scale PDF Reynolds}, respectively. The PDFs are denoted by solid lines at $5~\mu$s and by dashed lines at $10~\mu$s.

The highest $Re_l$ (blue line) has the highest probability of larger length scales ($L/\lambda_0>2.0$) at both $5~\mu$s and $10~\mu$s, confirming the earlier claim that higher $Re_l$ causes more stretched surfaces and larger scales (less curved surfaces) early on. Even though the maximum distribution of the length scales moves to smaller scales from $5~\mu$s to $10~\mu$s for $Re_l=5000$, $\delta$ still grows, as shown in Fig.~\ref{fig:scale Reynolds}(a). In particular, a large population of the cells contains surfaces with very large length scales. As $Re_l$ decreases, transition towards smaller scales occurs faster at these early times. Therefore, $\delta$ decreases sooner for lower $Re_l$. Clearly, there are two factors in determining the mean length scale: the sizes of the smallest and largest length scales, and the population of those scales. At an early stage, i.e.~$t<10~\mu$s, the smallest as well as the largest length scales are the same for all cases; however, it is the population of those small scales compared to the large ones that controls $\delta$. Since there are more large scales at higher $Re_l$, more time is needed for those structures to cascade to smaller scales; thereby, $\delta$ keeps growing for a longer period at higher $Re_l$ (Fig.~\ref{fig:scale Reynolds}b). After this initial period, however, the cascade is faster for the higher $Re_l$ because of the lower viscous resistance against surface deformation; therefore, the smallest bin gets populated at a higher rate.

\begin{figure}[t!]
	\centering\includegraphics[width=1.0\linewidth]{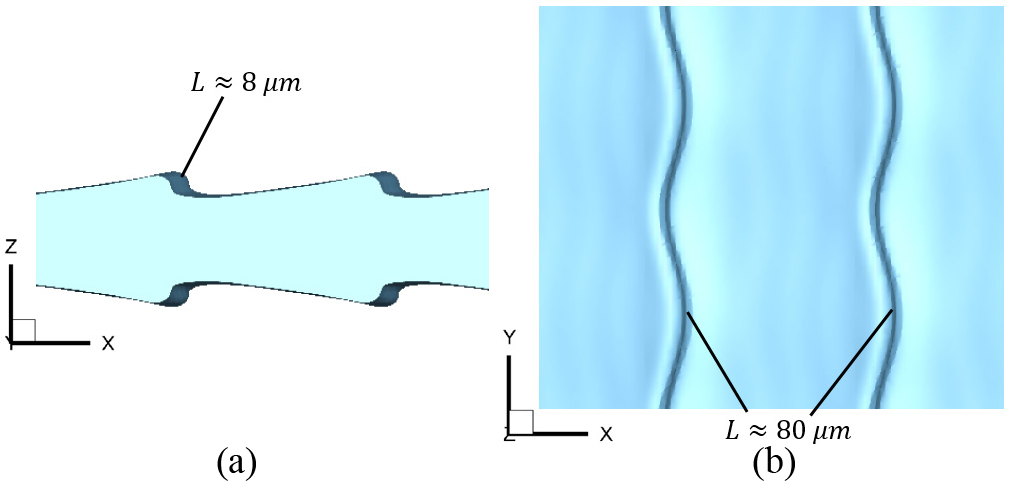}
	\caption{Liquid surface at $t=5~\mu$s from a side view (a), and top view (b); $Re_l=1000$, $We_g=7250$, $\hat{\rho}=0.5$, $\hat{\mu}=0.0066$, and $\Lambda = 2.0$.}
	\label{fig:t5 surface}
\end{figure}

There are two distinct spikes in the length-scale PDFs of Fig.~\ref{fig:scale PDF Reynolds} at $5~\mu$s -- one at a low scale of approximately $L=0.08\lambda_0=8~\mu$m, and another with lower probability but larger scale of  $L=0.8\lambda_0=80~\mu$m for $Re_l=1000$. Similar two spikes are seen for other $Re_l$ cases at an early time. The cause of the two spikes is shown in Fig.~\ref{fig:t5 surface}. The smaller scale represents the radius of curvature of the streamwise KH wave crest, which has the largest probability because this length scale exists for many cells along the spanwise direction on the front edge of the waves. This is shown in the side-view of Fig.~\ref{fig:t5 surface}(a). The other spike with the larger length scale but smaller probability applies to the radius of curvature of the spanwise waves. These points are indicated on the top-view of Fig.~\ref{fig:t5 surface}(b). This length scale occurs for all the cells near the tip of the protruded liquid lobe.

The conclusion from the above results is that there are two stages in the liquid-sheet distortion: (i) the initial stage of distortion when the lengths grow, and (ii) the final asymptote in time. Viscosity but not surface tension is dominant in the first stage, which is inertially driven, while surface tension more than viscosity affects the mean scale at the latter stage.

\begin{figure}[!t]
	\centering\includegraphics[width=1.0\linewidth]{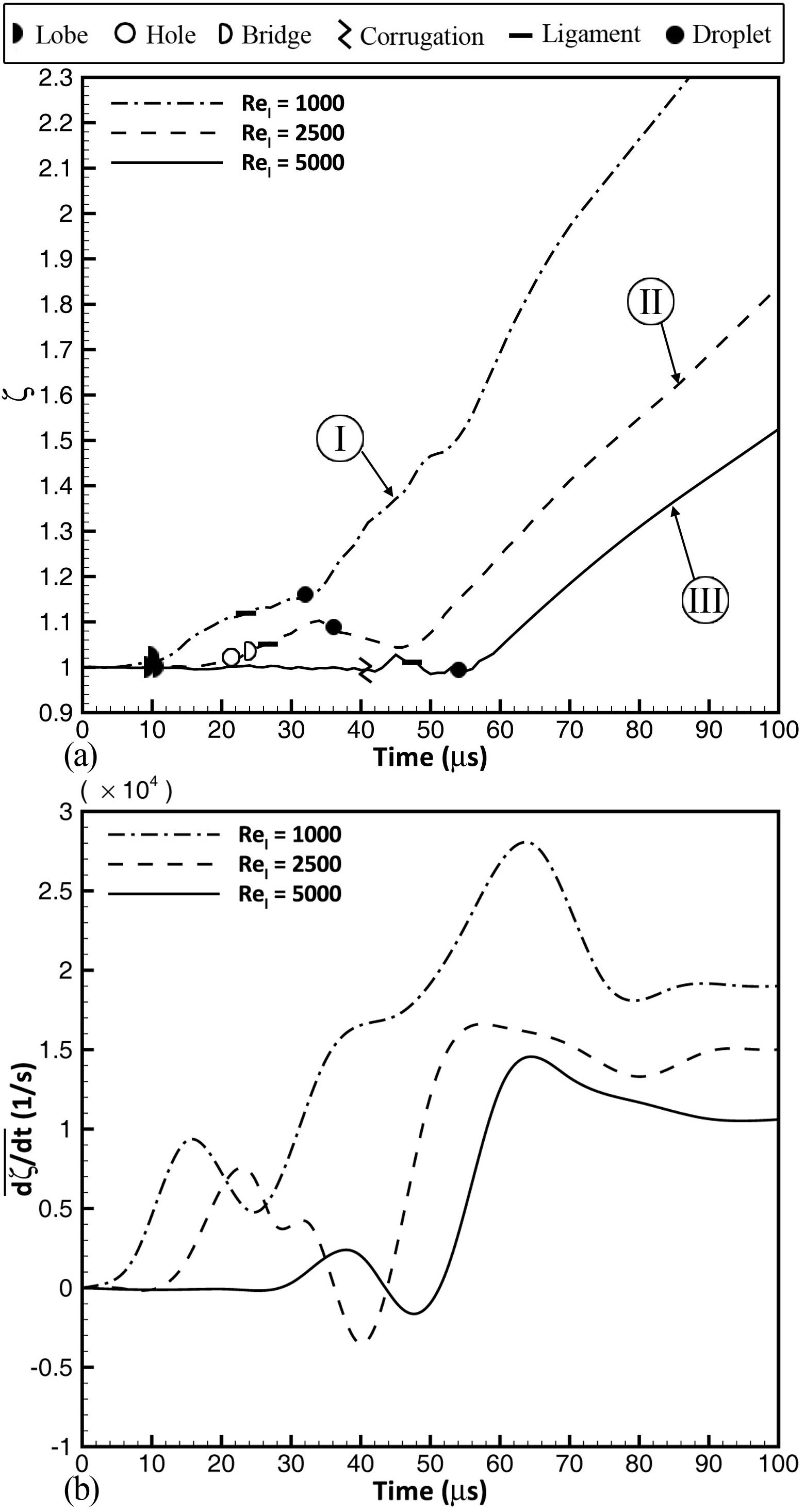}
	\caption{Effect of $Re_l$ on the temporal variation of $\zeta$ (a), and the average spread rate (b); $We_g=7250$, $\hat{\rho}=0.5$, $\hat{\mu}=0.0066$, and $\Lambda = 2.0$.}
	\label{fig:expansion Reynolds}
\end{figure}

Figs.~\ref{fig:expansion Reynolds}(a) and \ref{fig:expansion Reynolds}(b) show the effects of liquid viscosity on $\zeta$ and its growth rate, respectively. Since liquid inertia dominates the viscous effects at higher $Re_l$, the spray is oriented more in the streamwise direction, and $\zeta$ and accordingly the spray angle are smaller at higher $Re_l$. This is consistent with both numerical simulations \citep{Jarrahbashi2} and experimental results \citep{Carvalho, Mansour}. \cite{Mansour} and \cite{Carvalho} showed that the spray angle is reduced by increasing the liquid velocity (or mass flowrate); i.e.~increasing $Re_l$. However, they only reported the final spray angle, but did not address its temporal growth. We show here that the growth rate of the spray width ($\overline{\mathrm{d}\zeta/\mathrm{d}t}$) is lower at higher $Re_l$ (Fig.~\ref{fig:expansion Reynolds}b). The expansion of the spray starts much sooner in Domain I than in Domains II and III, and the asymptotic spray growth rate is lower for higher $Re_l$.

\begin{figure}[b!]
	\centering\includegraphics[width=1.0\linewidth]{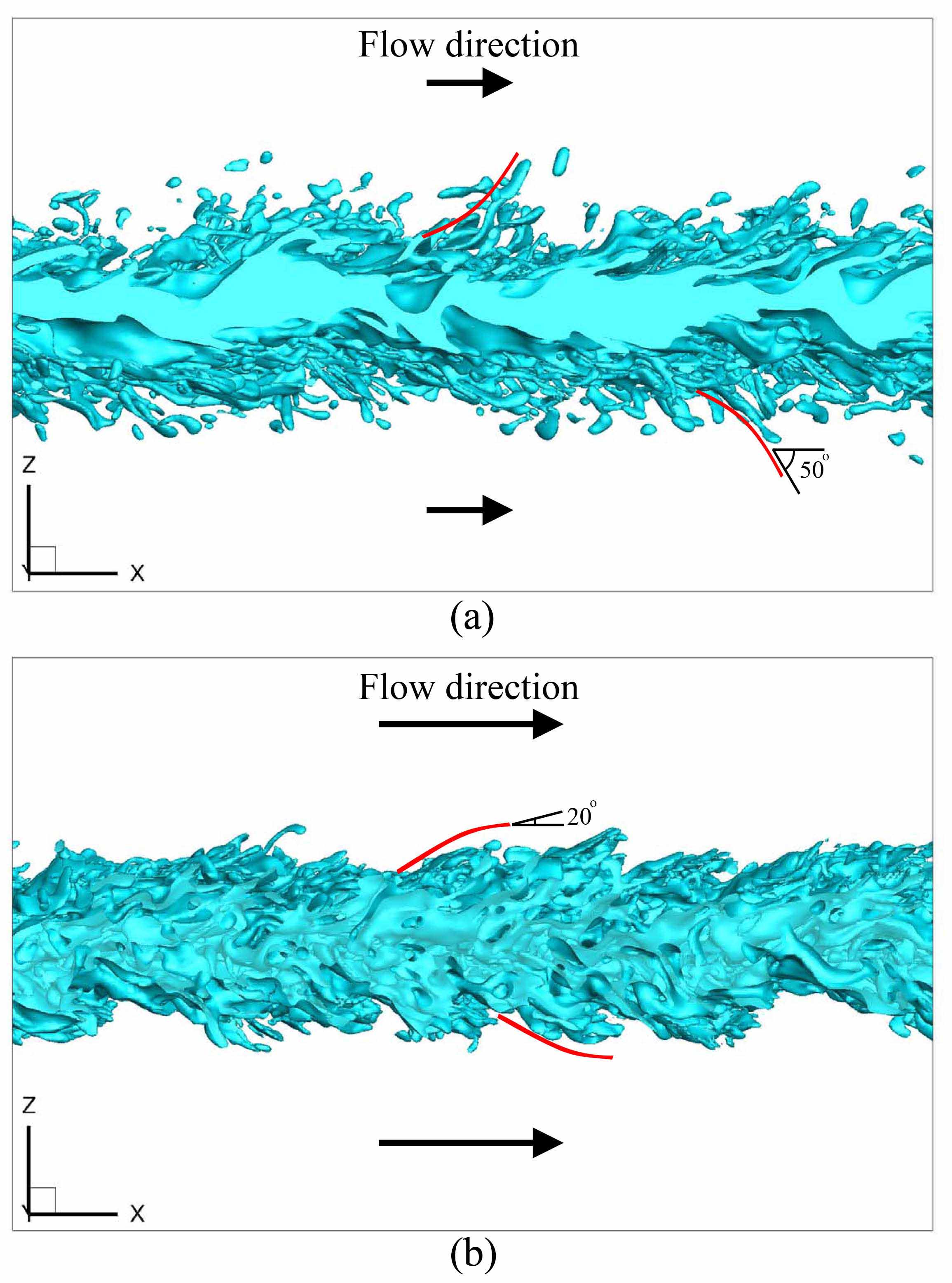}
	\caption{Liquid surface at $t=70~\mu$s from a side view for $Re_l=1000$ (a), and $Re_l=5000$ (b); $We_g=7250$, $\hat{\rho}=0.5$, $\hat{\mu}=0.0066$, and $\Lambda = 2.0$. The red lines indicate the qualitative form and angle of the ligaments.}
	\label{fig:Surface_vs_Re}
\end{figure}

Even though both the highest and the lowest $Re_l$ cases produce similar lobe stretching mechanisms (with and without corrugation formation, respectively), there is a significant difference in their spray expansion. At high $Re_l$, the corrugations on the lobes stretch into streamwise ligaments, resulting in thinner and hence shorter ligaments. The time lapse between ligament formation and ligament breakup is much shorter in Domain III since the ligaments are generally shorter and need less time to thin and break. At low $Re_l$, on the other hand, the lobes directly stretch into ligaments, and the stretching is oriented in the transverse (normal) direction as the viscous forces resist streamwise stretching caused by the inertia. It takes more time for the ligaments to break up into droplets. The ligaments are generally thicker and longer, as also shown for low $Re_l$ by \cite{Marmottant}. This difference originates from the difference in the vortex structures of these two regimes, and shows that these two mechanisms have different causes from vortex dynamics perspective. \cite{Arash3} explain that streamwise vortex stretching is stronger at higher $Re_l$, resulting in streamwise oriented ligaments.

The difference in the jet spread rate at low and high $Re_l$ manifests in the angle of ligaments that grow out of the liquid surface. This is clearly illustrated in Fig.~\ref{fig:Surface_vs_Re}, where the liquid surface at a low $Re_l$ (Fig.~\ref{fig:Surface_vs_Re}a) and high $Re_l$ (Fig.~\ref{fig:Surface_vs_Re}b) are shown at $70~\mu$s. The overall shape of ligaments is denoted by the red lines in this figure, and the average angle of the ligament tips (measured from the streamwise direction) are also indicated. At high $Re_l$, the lower viscous forces on the ligaments are not able to overcome the relatively high gas momentum, and the ligament angle decreases as it penetrates further into the gas. Thus, the transverse velocity is much smaller than the streamwise velocity and the angle is small (about $20^\circ$). On the other hand, the higher liquid viscosity at lower $Re_l$ balances the gas inertia and opposes the streamwise stretching of the ligaments. Consequently, the ligament angle increases as it penetrates further into the gas, resulting in a $50^\circ$ angle at the ligament tip. The ligament shapes can be attributed to the velocity profile -- the boundary layer becomes thinner and the velocity gradient in the $z$-direction becomes steeper as $Re_l$ increases. These results are consistent with the temporal variation of $\zeta$ shown in Fig.~\ref{fig:expansion Reynolds}. Our results correctly predict that the asymptotic jet spread rate is higher at lower $Re_l$. The non-dimensional spread rate is $2.1$, $1.6$, and $1.26~1/$s for $Re_l=1000$, $2500$ and $5000$, respectively (Fig.~\ref{fig:expansion Reynolds}b). This was not the case for jet spread rate versus $We_g$. This confirms that the reason for faster growth of jet width at higher $We_g$ was mainly due to the fact that the droplets break up faster and can be carried away by the vortices near the interface, which is completely different from the cause-and-effect of $Re_l$, explained here.  

\subsection{Density-ratio effects} \label{Density ratio}

\begin{figure}[!t]
	\centering\includegraphics[width=1.0\linewidth]{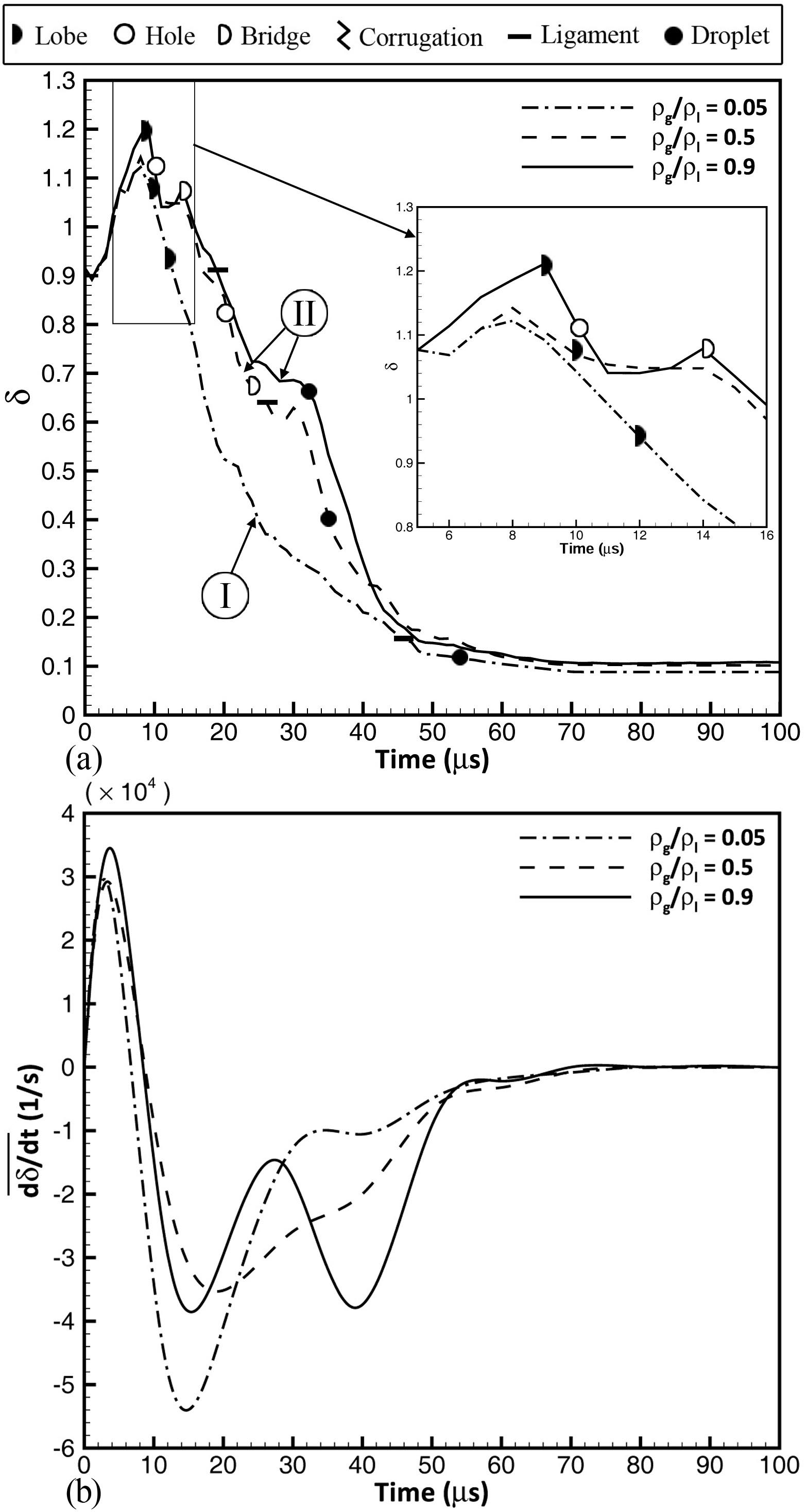}
	\caption{Effect of density ratio on the temporal variation of $\delta$ (a), and the average cascade rate (b); $Re_l=2500$, $We_l=14,500$, $\hat{\mu}=0.0066$, and $\Lambda = 2.0$.}
	\label{fig:scale denratio}
\end{figure}

The effect of density ratio ($\hat{\rho}$) on $\delta$ and its cascade rate are shown in Fig.~\ref{fig:scale denratio}. Three density ratios are studied in a range of $0.05$ (low gas pressure) to $0.9$ (high pressure gas). The liquid Weber number, and thereby the surface tension coefficient, is kept the same for all three cases; thus, $We_g$ is also different for these three cases through gas density. The lowest $\hat{\rho}$ falls in Domain I, where lobes stretch directly into ligaments, and the other two higher density ratios belong to Domain II, and involve hole and bridge formation in their breakup.

$\hat{\rho}$ has only a slight effect on the final length scale. All cases asymptotically reach a nearly similar mean length scale of about $0.1\lambda_0$, with the final length scale being slightly smaller ($\approx0.09\lambda_0$) for the lowest $\hat{\rho}$ (dash-dotted line in Fig.~\ref{fig:scale denratio}a). Besides, $\hat{\rho}$ clearly affects the rate at which the ultimate $\delta$ is achieved. The rate of length-scale cascade is higher at lower gas densities, with the highest rate of approximately $-5.5~\mu$m/$\mu$s occurring between $10~\mu$s and $20~\mu$s for $\hat{\rho}=0.05$. This rate slowly decreases as the lobe stretches into ligaments. Between $30~\mu$s and $45~\mu$s a slower cascade rate ($\approx-1~\mu$m/$\mu$s) is dominant, which corresponds to the formation of first ligaments by the end of this period. As $\hat{\rho}$ increases, the cascade of length scales becomes slower for the early stages ($t<30~\mu$s). With transition into Domain II, a few distinct cascade rates are seen in the process, each corresponding to the formation of a new liquid structure: e.g.~holes, bridges, and ligaments. The cascade rate corresponding to droplet formation for $t>35~\mu$s is larger in Domain II. The reason for this difference is in the mechanisms of droplet formation in these two domains. In Domain I, each lobe results in a single qualitatively large droplet, whereas in Domain II, several smaller droplets are formed from the breakup of ligaments and bridges. Therefore, the average cascade rate is larger for Domain II in those later periods ($t=35$--$45~\mu$s).

The asymptotic length scale is achieved at $64~\mu$s for $\hat{\rho}=0.05$, but at about $68~\mu$s for $\hat{\rho}=0.5$, and at $70~\mu$s for $\hat{\rho}=0.9$. As denoted by the symbols in Fig.~\ref{fig:scale denratio}(a), the rate of ligament and droplet formation is notably affected by $\hat{\rho}$. As $\hat{\rho}$ grows, so that the Domain changes from I to II, the ligament and droplet formation rates undergo a large jump. As $\hat{\rho}$ keeps increasing in the same Domain (II), the ligaments and droplets form sooner, but the difference is not as notable as the jump during the domain transition. \citet{Jarrahbashi2} also showed that the drops form earlier at higher $\hat{\rho}$; however, they did not address the relation between this trend and the change in the breakup process. As gas density increases, the higher gas inertia intensifies the hole formation, therefore expediting the formation of ligaments and droplets. 

$\hat{\rho}$ does not alter the initial length-scale growth rate significantly -- the maximum scale occurs at nearly the same time and the maximum growth rates are also very close (see Fig.~\ref{fig:scale denratio}b) -- which proves it to be correlated with the liquid inertia and not the gas inertia. The asymptotic stage is also driven by the liquid inertia and is almost independent of the gas density in the $\hat{\rho}$ range considered here. As shown in Section \ref{Weber effects}, this stage is also correlated with surface tension. Therefore, the liquid Weber number ($We_l$) and not the gas Weber number ($We_g$) is the key parameter in determining the asymptotic droplet size -- discussed more in this section.

\begin{figure}[!t]
	\centering\includegraphics[width=1.0\linewidth]{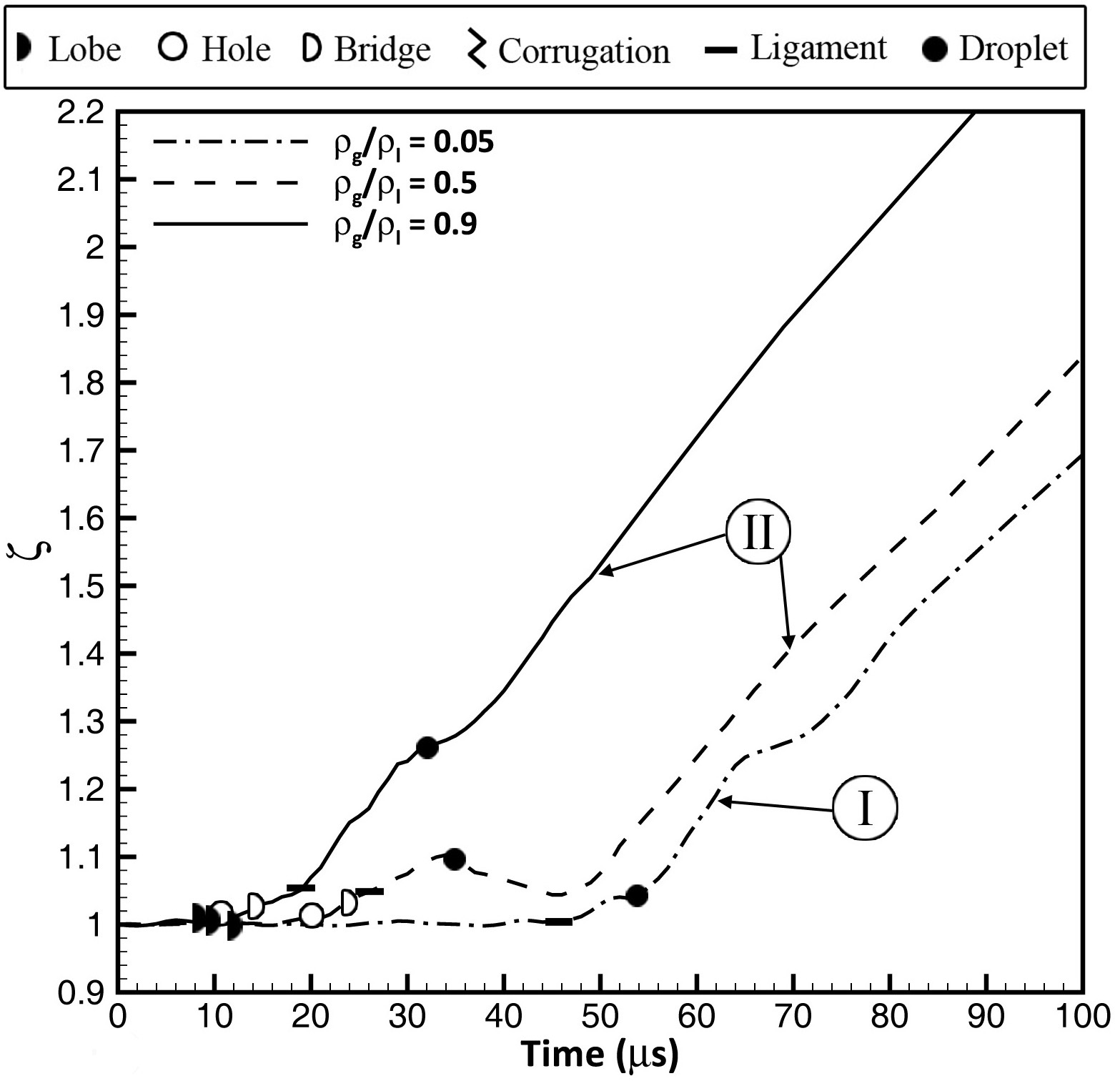}
	\caption{Effect of density ratio on the temporal variation of $\zeta$; $Re_l=2500$, $We_l=14,500$, $\hat{\mu}=0.0066$, and $\Lambda = 2.0$.}
	\label{fig:expansion denratio}
\end{figure}

Fig.~\ref{fig:expansion denratio} shows the temporal variation of $\zeta$ for low, medium, and high density ratios. $\zeta$ increases with increasing $\hat{\rho}$, similar to the findings of \cite{Jarrahbashi2} for round jets. The jet with the highest $\hat{\rho}$ (solid line), which approximates a homogeneous liquid jet, expands much more rapidly than the case with smaller $\hat{\rho}$ (dash-dotted line). The spray width grows from its initial thickness at $10~\mu$s for the highest $\hat{\rho}$, while the expansion is postponed to $46~\mu$s for the lowest $\hat{\rho}$. Some researchers have reported growth of the spray angle with increasing $\hat{\rho}$ or gas-to-liquid momentum ratio \citep{Carvalho, Jarrahbashi2} -- based on the final stage of the fully-developed jet -- but none has shown the temporal growth of the spray to be correlated with $\hat{\rho}$. Our results show that the jet spread rate is also higher at higher $\hat{\rho}$.

Even though the growth rate of $\zeta$ is lower for lower gas densities, the asymptotic growth rate is nearly the same after a long time from the start of the injection, regardless of $\hat{\rho}$. This is seen from the slopes of the curves in Fig.~\ref{fig:expansion denratio}, which become approximately equal near the end of the computations; i.e.~the asymptotic slopes appear to be independent of the gas density. \cite{Jarrahbashi2} also found that the spray spread rate is higher at higher $\hat{\rho}$ for circular jets. However, they used the traditional definition for the jet size, i.e.~distance of the farthest continuous liquid structure from the centerline, which would be ambiguous in some cases, as discussed.

The lower spread rate for low $\hat{\rho}$ is directly related to the vortex dynamics near the interface. The main cause of the lower expansion is the baroclinic effects which are drastically different amongst the range of $\hat{\rho}$ considered here. Due to the larger density gradient, the baroclinic torque is higher at low gas densities. Thus, the vortex cores locate farther from the interface \citep{Arash3}. The induced flow of the vortices away from the interface entrains more gas into the mixing layer and expedites the two-phase mixing \citep{Jarrahbashi2}. However, if the vortices remain closer to the interface, as in higher gas densities, KH roll-up occurs more vigorously, causing a higher growth rate of the instabilities and a faster transverse expansion of the jet. 

\begin{figure}[!t]
	\centering\includegraphics[width=1.0\linewidth]{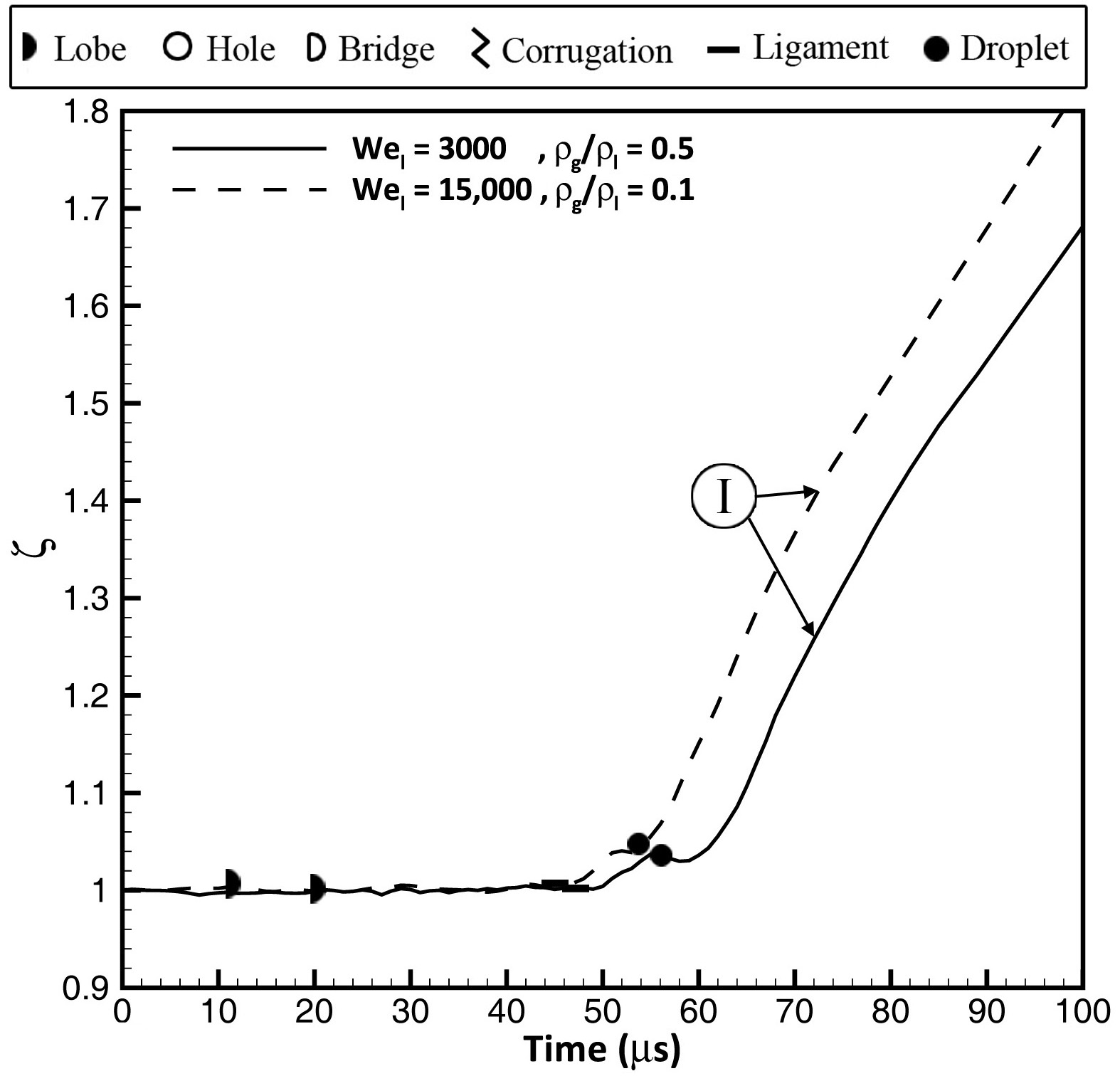}
	\caption{Effects of $\hat{\rho}$ and $We_l$ on the temporal variation of $\zeta$; $Re_l=2500$, $We_g=1500$, $\hat{\mu}=0.0066$, and $\Lambda = 2.0$.}
	\label{fig:overlapping expansion}
\end{figure}

Since an increase in $\hat{\rho}$ increases the spray width and spread rate, and the sheet expansion is also directly proportional to the liquid $We$ (shown in Section \ref{Weber effects}), both $We_l$ and $\hat{\rho}$ affect similarly the jet spread rate. It is interesting now to examine the effects of $We$ and $\hat{\rho}$ combined with the gas-phase Weber number ($We_g$). This is delineated in Fig.~\ref{fig:overlapping expansion}, where the temporal evolution of $\zeta$ for two cases that overlap at the same point in the $We_g$--$Re_l$ map of Fig.~\ref{fig:domains_gas_based} are compared; both cases have the same $We_g=1500$, but different $\hat{\rho}$ and $We_l$. Since $We_g=\hat{\rho}We_l$, $\hat{\rho}$ and $We_l$ should change in opposite directions to keep $We_g$ constant; i.e.~as $\hat{\rho}$ increases (increasing the spread rate), $We_l$ should decrease (decreasing the spread rate). Fig.~\ref{fig:overlapping expansion} shows that the two cases behave very similarly in temporal expansion; both sprays expand at almost the same time and at the same asymptotic rate. The rate of ligament and droplet formation is also comparable in these two cases -- consistent with Eq.~(\ref{eqn:thole}) -- where the ligament stretching time scale $\tau_s$ is inversely proportional to $Re_l$, but independent of the Weber number. Since both cases in Fig.~\ref{fig:overlapping expansion} have equal $Re_l$, their ligament stretching rates are also nearly equal. Therefore, the two parameters, $We_l$ and $\hat{\rho}$, could be combined into a single parameter, $We_g$, for jet expansion analysis. The gas inertia -- and not the liquid inertia -- and liquid surface tension are the key parameters in determining the spray width. This confirms $We_g$ to be the proper choice for categorizing the liquid-jet breakup characteristics, as used in Fig.~\ref{fig:domains_gas_based}.

\begin{figure}[!b]
	\centering\includegraphics[width=1.0\linewidth]{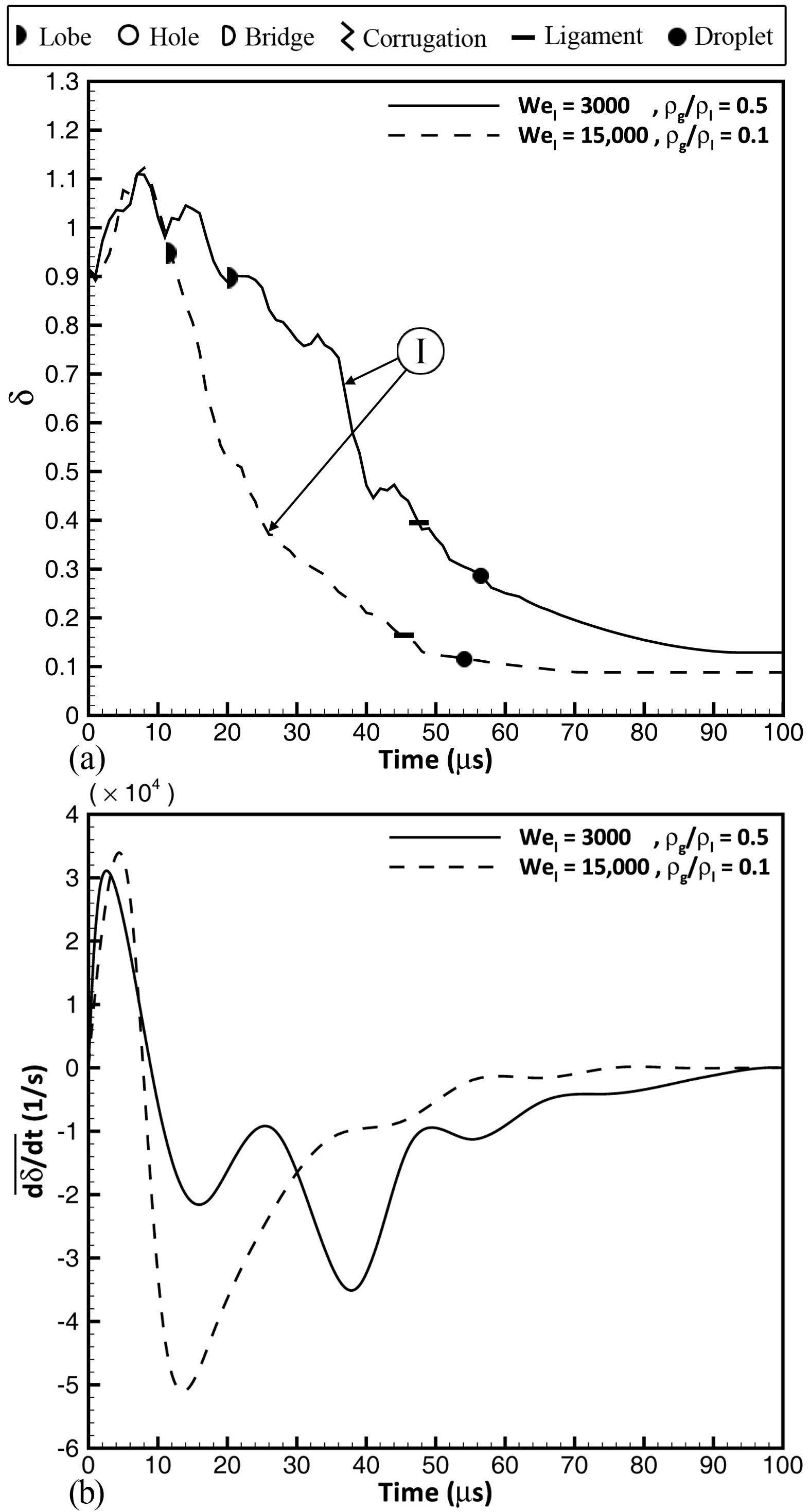}
	\caption{Effects of $\hat{\rho}$ and $We_l$ on the temporal variation of $\delta$ (a), and the average cascade rate (b); $Re_l=2500$, $We_g=1500$, $\hat{\mu}=0.0066$, and $\Lambda = 2.0$.}
	\label{fig:overlapping scale}
\end{figure}

In Section \ref{Weber effects}, it was shown that a decrease in surface tension reduces the asymptotic length scale and $\delta$. $\hat{\rho}$, however, has negligible effect on $\delta$. Thus, $We_l$ is expected to be the key factor in determining the final droplet size (length scale). Fig.~\ref{fig:overlapping scale} confirms this notion and shows that even though the two cases have the same $We_g$, they manifest a significant difference in the cascade process and the asymptotic length scale. Since the case with higher $\hat{\rho}$ (solid line) has a lower $We_l$ -- keeping $We_g$ constant -- it produces larger $\delta$ and has a slower overal cascade. Thus, the liquid inertia is also important for the liquid structure cascade. Even though two jets at the same $Re_l$ and $We_g$ exhibit the same atomization mechanism (both from Domain I), the length scales of the resulting liquid structures depend on $\hat{\rho}$. The lower gas density would result in finer structures. In other words, the atomization domain only determines the breakup quality (the type of process during the cascade), but other factors need to be considered to control the quantitative characteristics of the atomization; e.g.~droplet size and structure length scales. Since surface tension resists lobe formation, the lobes form more slowly at lower $We_l$ (higher $\hat{\rho}$), resulting in a lower cascade rate in the $10~\mu$s -- $30~\mu$s period. Compare the solid and dashed lines in Fig.~\ref{fig:overlapping scale}(b). The rate of lobe stretching and ligament formation, however, is higher for higher $\hat{\rho}$ (solid line) in the $30$--$50~\mu$s period, since the gas inertia assists the stretching of ligaments. With the breakup of ligaments into droplets ($t>50~\mu$s), the cascade rate decreases and approaches zero while a balance is reached between the droplet and ligament formation.

\subsection{Viscosity ratio effects} \label{Viscosity ratio}

\begin{figure}[!t]
	\centering\includegraphics[width=1.0\linewidth]{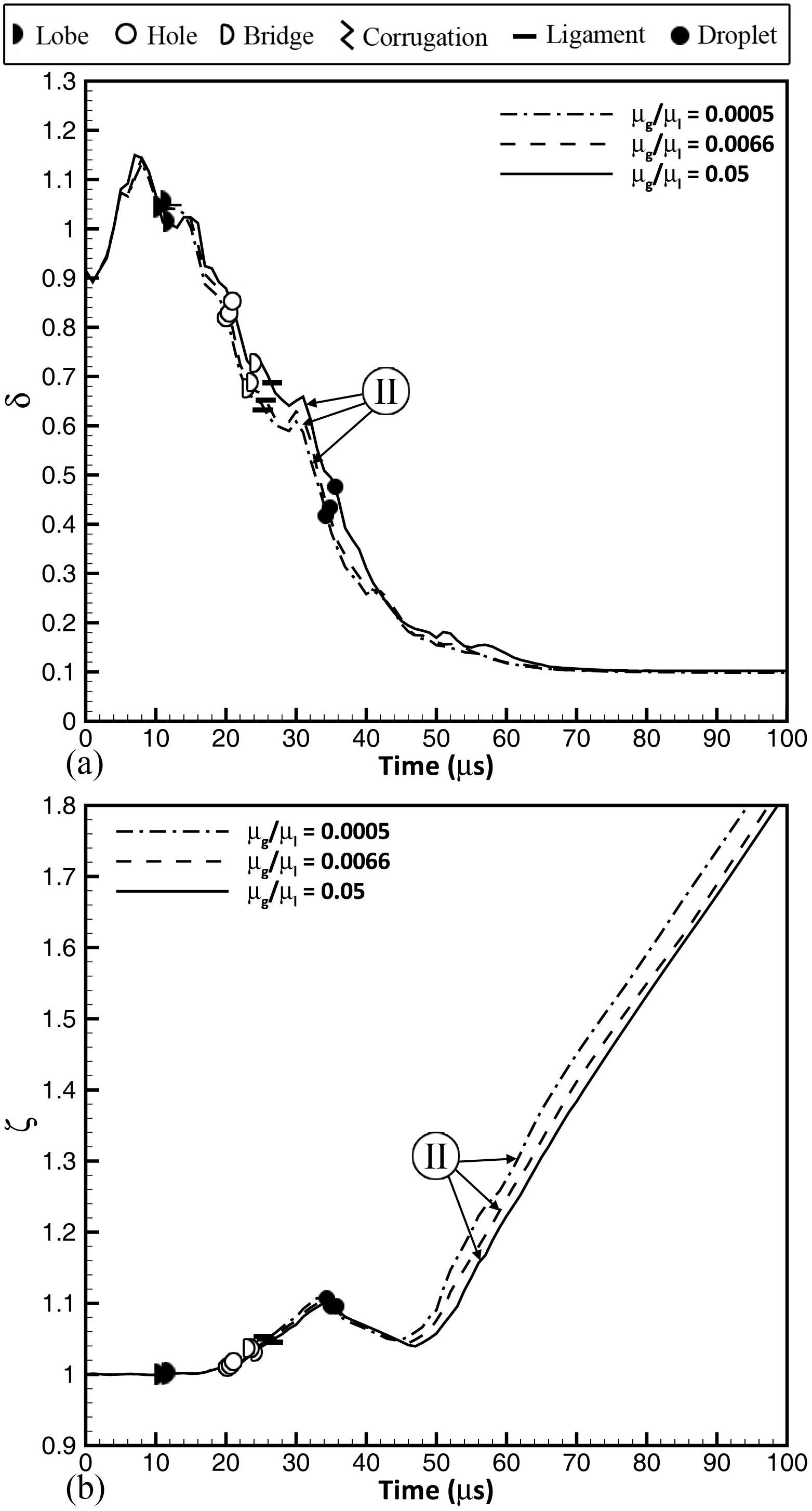}
	\caption{Effect of viscosity ratio on the temporal variation of $\delta$ (a), and $\zeta$ (b); $Re_l=2500$, $We_g=7250$, $\hat{\rho}=0.5$, and $\Lambda = 2.0$.}
	\label{fig:scale expansion visratio}
\end{figure}

\cite{Samuelsen} showed that viscosity ratio ($\hat{\mu}$) has little or no effect on the final droplet size (i.e.~Sauter mean diameter) of liquid jets. Here, the same conclusion is reached, as shown in Fig.~\ref{fig:scale expansion visratio}(a). A wide range of viscosity ratios ($0.0005<\hat{\mu}<0.05$) covering three orders of magnitude are compared here. All cases follow the same cascade with an almost identical rate. The cascade is delayed less than a few microseconds in the period $10$--$40~\mu$s for higher $\hat{\mu}$, but the small difference vanishes at later times. The asymptotic length scale is the same for all viscosity ratios, i.e.~about $0.1\lambda_0 = 10~\mu$m, and the rate of droplet formation also remains the same regardless of $\hat{\mu}$. The gas viscosity is the least important factor in determining the droplet size and has no effect on the structure stretching and length scale growth in the initial stage; i.e.~at $t=0$--$10~\mu$s.

Moreover, $\hat{\mu}$ does not have a notable influence on the jet spread rate either. As plotted in Fig.~\ref{fig:scale expansion visratio}(b), the sheet expands at the same time regardless of the gas viscosity. The spray spread rate, i.e.~the slopes of the curves in Fig.~\ref{fig:scale expansion visratio}(b), is also the same at the end of the process for all viscosity ratios. The only minor difference is that the spray expansion gets delayed a few microseconds as gas viscosity increases. This minor variation over three orders of magnitude $\hat{\mu}$ variation is insignificant. Therefore, the gas viscosity is not important in determining the spray angle and its growth rate, as it was not also for the mean droplet size. The importance of viscosity only manifests through $Re_l$, where an increase in the liquid viscosity, i.e.~lowering $Re_l$, increases the size of droplets and increases the spray angle, as discussed in Section \ref{Reynolds effects}.

\subsection{Sheet thickness effects} \label{Thickness}

\begin{figure}[!t]
	\centering\includegraphics[width=1.0\linewidth]{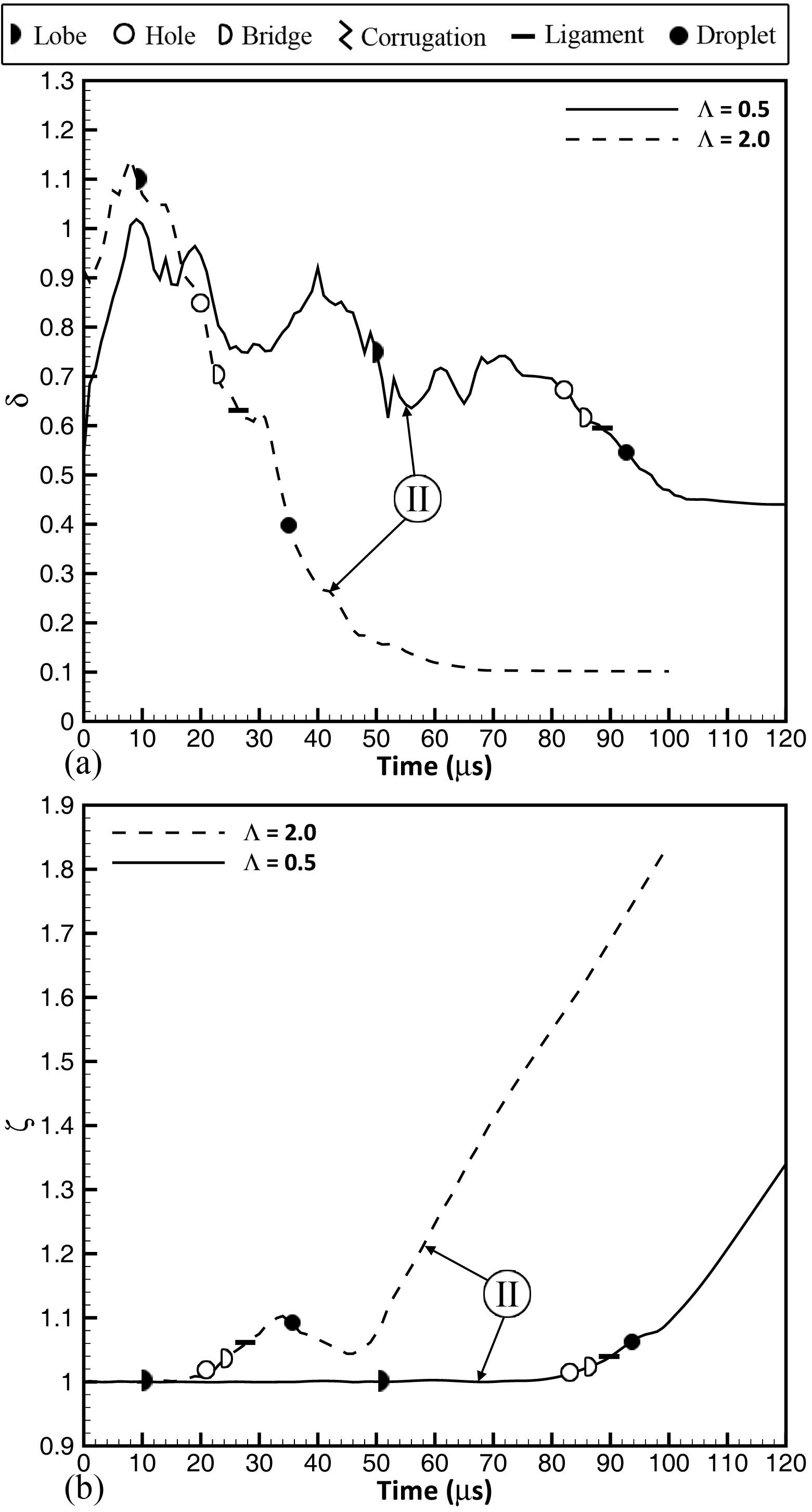}
	\caption{Effect of sheet thickness on the temporal variation of $\delta$ (a), and $\zeta$ (b); $Re_l=2500$, \mbox{$We_g=7250$}, $\hat{\rho}=0.5$, and $\hat{\mu} = 0.0066$.}
	\label{fig:scale expansion thickness}
\end{figure}

\cite{Senecal} showed that ligament diameter is directly proportional to the initial sheet thickness. Our results, illustrated in Fig.~\ref{fig:scale expansion thickness}(a), confirm their findings; the length scales grow as the sheet thickens. For this comparison, two sheets of different widths have been analyzed -- a thin sheet of $50~\mu$m thickness with $\Lambda=2.0$ and a thicker sheet of $200~\mu$m thickness with $\Lambda=0.5$. The initial perturbation wavelength is the same for both cases -- i.e.~$\lambda_0=100~\mu$m.

The cascade occurs much slower and $\delta$ oscillates more in its cascade process for the thicker sheet; see Fig.~\ref{fig:scale expansion thickness}(a). The extra oscillations found in the thicker sheet occur because the initial KH waves take more time to stretch and break into smaller structures as sheet thickness increases. This is explained via vortex dynamics of the interface deformation, where the two vortex layers on top and bottom of the sheet become farther apart and more independent as the sheet becomes thicker. The interaction between the two vortex layers is more intense for the thinner sheets -- a simple consequence of Biot-Savart mutual induction between the vortices in the two layers -- and consequently the cascade occurs faster under the local induction of these two vortex layers \citep{Arash3}. As the initial sinusoidal KH waves stretch in the streamwise direction, the curvature of the interface decreases and $\delta$ grows; as the waves roll over the vortices and the lobes form, $\delta$ decreases. If the waves dampen again, $\delta$ temporarily increases until the next waves start to grow. The process of wave stretching and curling occurs continuously until the lobes stretch enough to cascade into smaller structures; e.g.~ ligaments and droplets. Because of the higher local induction of the KH vortex layers on top and bottom surfaces, the transition towards antisymetry is also faster for thinner sheets, which helps bringing down the length scales more quickly. $\delta$ keeps on decreasing until the asymptotic length scale is achieved. Increasing sheet thickness also significantly delays the formation of liquid structures on the surface. The lobes and ligaments form respectively $40~\mu$s and $60~\mu$s later on the thick sheet than on the thin sheet.

The asymptotic length scales for the thin and thick sheets are $0.1\lambda_0$ and $0.46\lambda_0$, respectively -- agreeing with the findings of \cite{Senecal}. Based on their analytical study, the ligament size should be directly proportional to the initial sheet thickness; thus, as the sheet becomes four times thicker, the mean ligaments size should increase by a factor of four. Since the larger length scales caused by the curvature of the waves are also included in calculation of $\delta$, the ratio of $\delta$ for the thick-to-thin sheets in our case is about $4.6$ -- in fairly good agreement.

The spray expansion is significantly delayed as the sheet thickness increases; see Fig.~\ref{fig:scale expansion thickness}(b). While the thin sheet expands at $20~\mu$s, the expansion of the thicker jet does not start until $80~\mu$s. Even though a thin sheet has higher spread rate at the early stages of spray formation, the thicker sheet achieves a comparable spread rate at its final stage. This is seen from the slopes of the solid and dashed-lines in Fig.~\ref{fig:scale expansion thickness}(b) at the end of the processes, where the slopes are quite equal. This major difference in $\zeta$ is caused by the reduced influence of the vortex layers in the thicker sheet, resulting in a slower shift towards antisymmetry as the vortex layers get farther apart. Since $\zeta$ is normalized by the initial sheet thickness, the absolute spread rate of the thicker sheet in the final phase is much higher than the thin sheet. The instantaneous growth rates become the same after the vortices have grown sufficiently and both sheets have become totally antisymmetric. At this final stage, the vortices in the two layers are equally effective in influencing each other via mutual induction and thus the effect is independent of the sheet thickness.

\subsection{$L_{32}$ calculation}

In liquid-jet atomization, the injector designer is usually interested in the average size based on mass distribution. Sauter mean diameter (SMD, $d_{32}$) is an average of particle size, defined as the diameter of a sphere that has the same volume/surface-area ratio as a particle of interest, normally used in the literature for this purpose. SMD is calculated using the following equation;
\begin{equation}
\mbox{SMD} = \frac{\Sigma_i N_i d_i^3}{\Sigma_i N_i d_i^2},
\label{eqn:SMD}
\end{equation}
where $N_i$ is the number of droplets per unit volume in size class $i$, and $d_i$ is the droplet diameter.

The main assumption in SMD calculation is that all the liquid particles are in the shape of spherical droplets. For the early atomization period considered here, however, a combination of droplets, ligaments and unbroken surfaces exist at the end of computations (see Fig.~\ref{fig:Surface_vs_Weber}). Therefore, more than just the droplet diameter is considered for the measurement of the mean length scale of the spray. Specifically, the Sauter mean diameter is generalized by considering the scales (radius of curvature) of all the droplets, ligaments, waves, and any other liquid structures either broken or still attached to the jet core. In this respect, we present the mean length scale weighted on both spherical droplets ($L_{32}$) and cylindrical ligaments ($L_{21}$). 

Similar to SMD (Eq.~\ref{eqn:SMD}), we calculate a mean based on the length scale $L_i$ and its probability $P(L_i)$. This parameter, called $L_{32}$, is defined as
\begin{equation}
L_{32} = 2\frac{\Sigma_i P(L_i) L_i^3}{\Sigma_i P(L_i) L_i^2}.
\label{eqn:L32}
\end{equation}

Since $L_i$ asymptotes to the droplet radius after all of the jet has broken into droplets, a coefficient of $2$ is multiplied in Eq.~(\ref{eqn:L32}) in order to make it comparable to the mean diameter rather than radius. All the length scales greater than $100~\mu$m are neglected in this analysis since those length scales are much larger than even the largest droplet diameters and instability wavelengths in our computations, and do not represent liquid structures but rather some flat surface on the interface. Consequently, $L_{32}$ is expected to be much larger than SMD in such a parameter range. $L_{32}$ is calculated after the length scales reach a constant value at the end of the computations, and it is expected to reach SMD asymptotically at later time.

\begin{figure}[!t]
	\centering\includegraphics[width=1.0\linewidth]{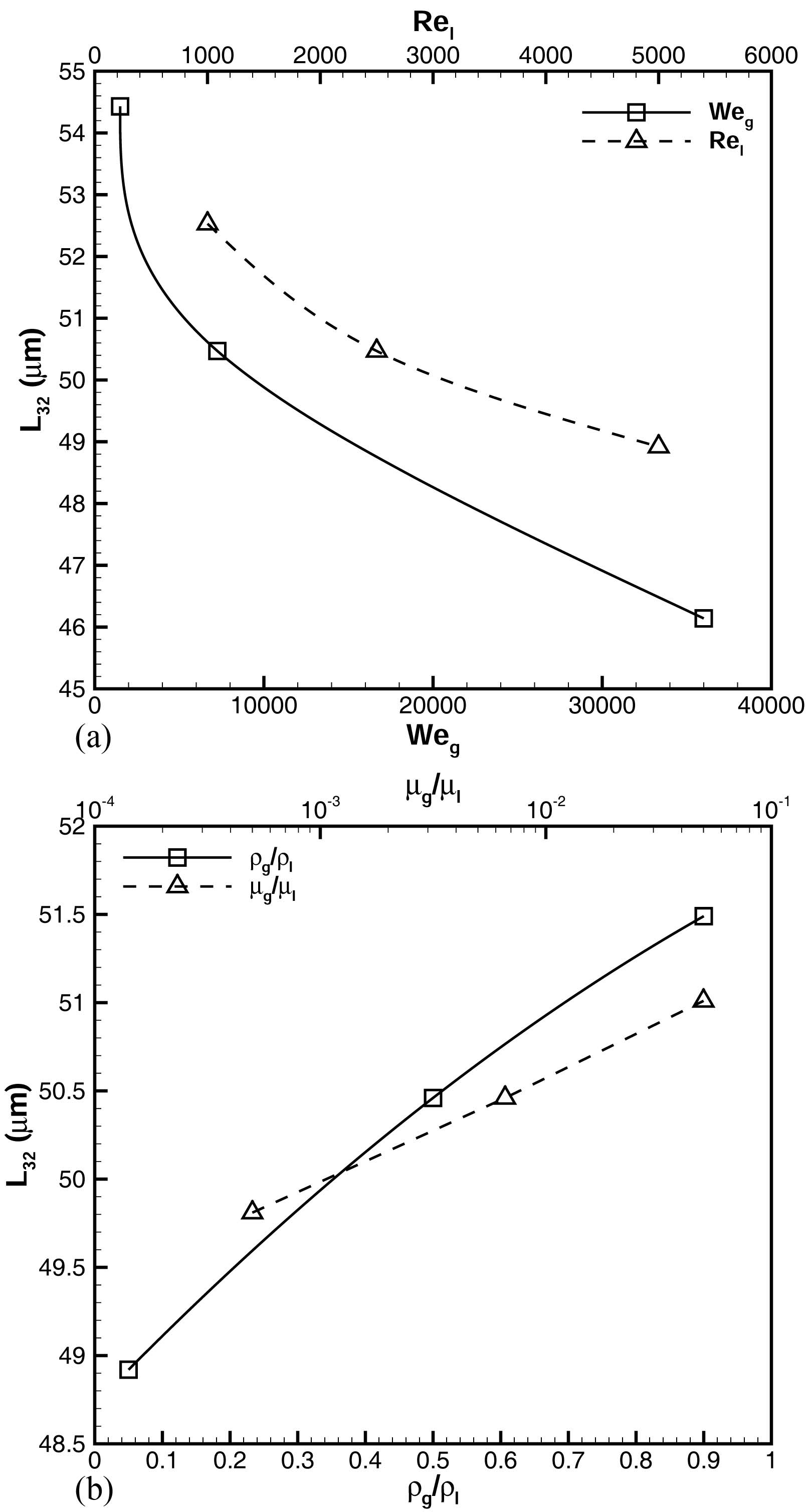}
	\caption{Effects of $We_g$ and $Re_l$ (a), and $\hat{\rho}$ and $\hat{\mu}$ (b) on $L_{32}$; $Re_l=2500$, $We_g=7250$, $\hat{\rho}=0.5$, and $\hat{\mu} = 0.0066$.}
	\label{fig:SMD}
\end{figure}

The effects of $We_g$ and $Re_l$ on $L_{32}$ are shown in Fig.~\ref{fig:SMD}(a). As expected, $We_g$ has the most significant influence on $L_{32}$, where it decreases from $54.5~\mu$m to less than $46~\mu$m as $We_g$ is increased from $1500$ to $36,000$. Even though this decrease is not as huge as predicted by \cite{Varga} for SMD -- since all the length scales are not attributed to droplets only -- its influence is more significant than of the other parameters. The dependence of SMD on $We_g$ as given by \cite{Varga} is limited to a much smaller $We_g$ range than considered in our study. Our study shows that, at the higher range of $We_g$, the effect of $We_g$ is not as pronounced as at lower ranges, though certainly not negligible. 

Increasing $Re_l$ also reduces $L_{32}$, as shown in Fig.~\ref{fig:SMD}(a). The difference in $L_{32}$ is slightly over three microns over the range of $Re_l$ considered in this study; i.e.~\mbox{$1000<Re_l<5000$}. However, $Re_l$ clearly influences $L_{32}$ and the final droplet size, as was discussed in Section~\ref{Reynolds effects}. Both the range of $L_{32}$ and its behavior with respect to $Re_l$ are in fair agreement with results of \cite{Lozano}. The decrease in $L_{32}$ becomes more gradual at high $Re_l$. \cite{Lozano} measured a SMD-to-sheet-thickness ratio of $0.72$ for $Re_l\approx5000$ (based on relative gas-liquid velocity), while we measure $L_{32}/h_0$ of approximately $0.97$ for the same $Re_l$ range. The higher ratio in our case being due to the fact that the jet is not entirely broken into droplets.

Fig.~\ref{fig:SMD}(b) shows the effects of $\hat{\rho}$ and $\hat{\mu}$ on $L_{32}$. Both parameters have minor influence on $L_{32}$ compared to $Re_l$ and $We_g$; even though $\hat{\mu}$ changes over three orders of magnitude and $\hat{\rho}$ ranges from $0.05$ to $0.9$, the difference in $L_{32}$ is only slightly over $1~\mu$m. $L_{32}$ slightly increases with increasing $\hat{\rho}$ and $\hat{\mu}$, where the rate of increase is higher at lower $\hat{\rho}$ ranges. In the ranges covered here, the dependence on $\hat{\rho}$ becomes almost linear at high density ratios ($\hat{\rho}>0.5$), while the dependence on $\hat{\mu}$ is completely linear on the log-scale.

The results presented in Fig.~\ref{fig:SMD} identify the trend in length scale growth or decline for the most important parameters; however, the differences are not very large compared to SMD measurements available in the literature. The reason behind this observation is that, there are still many unbroken liquid structures with large scales in the flow field, which have higher influence on $L_{32}$ and increase its value; thus, the $L_{32}$ measurements end up much closer to each other. 

$L_{32}$ is based on the assumption that the disintegrated elements are spherical droplets, and it gives a volume-to-surface weighting. In the primary atomization period considered here, however, much of the mass will be closer to cylinders (ligaments) rather than spheres. Thus, a comparison between $L_{21}$ (Eq.~\ref{eqn:L21}) might be more suitable, as it gives the volume-to-surface area weighting for a cylinder;
\begin{equation}
L_{21} = 2\frac{\Sigma_i P(L_i) L_i^2}{\Sigma_i P(L_i) L_i}.
\label{eqn:L21}
\end{equation}

\begin{figure}[!t]
	\centering\includegraphics[width=1.0\linewidth]{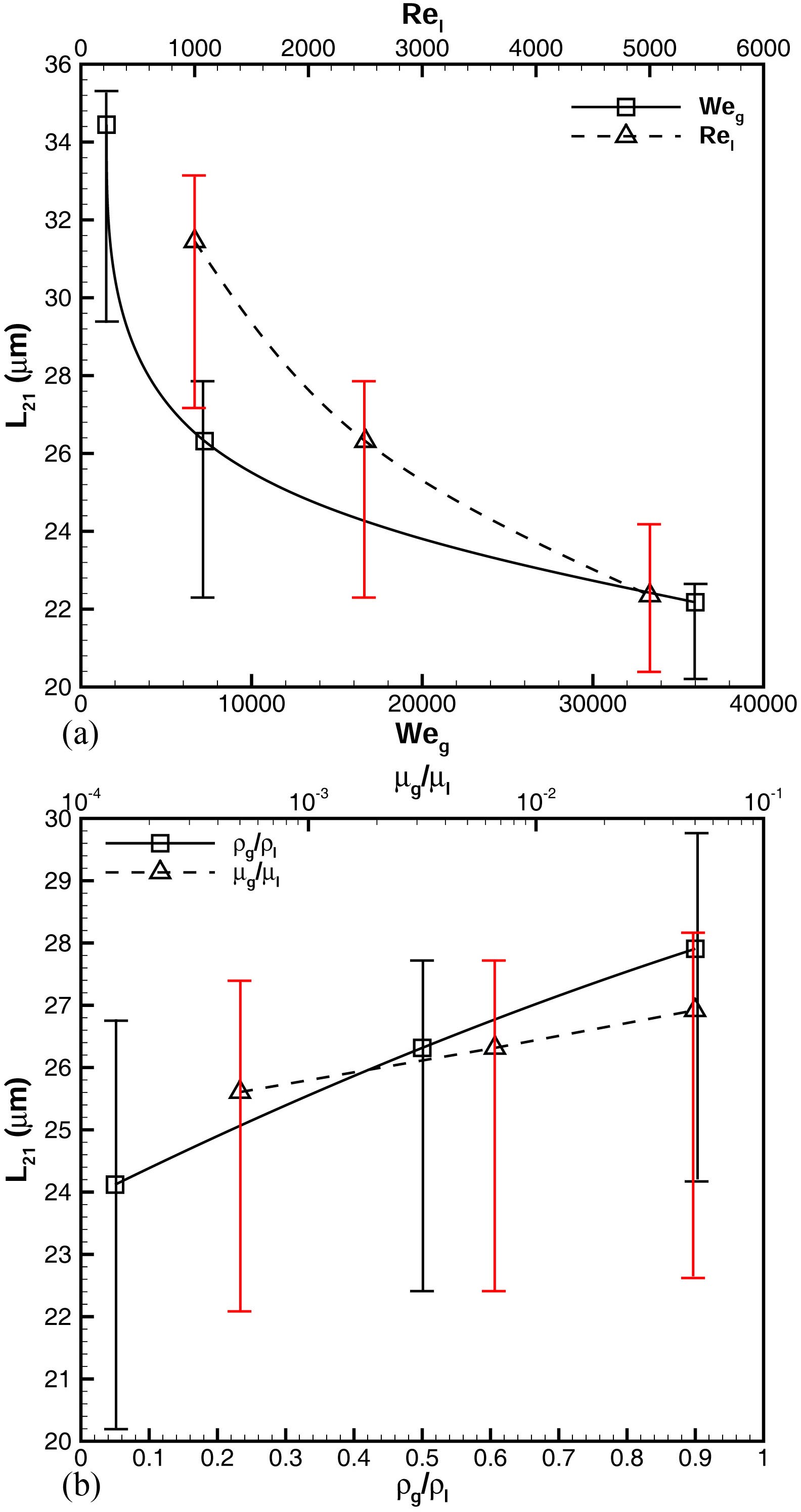}
	\caption{Effects of $We_g$ and $Re_l$ (a), and $\hat{\rho}$ and $\hat{\mu}$ (b) on $L_{21}$; the red and black error bars show the range of ligament diameters. $Re_l=2500$, $We_g=7250$, $\hat{\rho}=0.5$, and $\hat{\mu} = 0.0066$.}
	\label{fig:L21}
\end{figure}

Fig.~\ref{fig:L21}(a) shows that (as expected) $L_{21}$ magnifies the difference between the non-dimensional parameter values. Even though the trends of $L_{21}$ and $L_{32}$ are very similar, more than $12~\mu$m and $9~\mu$m difference in the value of $L_{21}$ is observed over the range of $We_g$ and $Re_l$ considered here, respectively. This indicates that the diameter of ligaments decreases significantly as the atomization domain moves from Domain I to Domain III, following an increase in $Re_l$, or as the atomization moves from Domain I to Domain II, due to an increase in $We_g$. Both these behaviors are in accordance with the characteristics of the atomization domains and the form of ligaments introduced in these three domains; see Fig.~\ref{fig:sketches}. The red and black error bars in Fig.~\ref{fig:L21}(a) indicate the ranges of ligament diameters that are seen in each $Re_l$ and $We_g$ case, respectively. The ligament diameters are measured in post-processing stage by taking a $200~\mu$m$\times200~\mu$m sample on both top and bottom surfaces, similar to what is shown in Fig.~\ref{fig:Surface_vs_Weber}. Fig.~\ref{fig:L21} delineates that the predicted $L_{21}$ is within the range of ligament diameters in all cases. This proves that $L_{21}$ is a justifiable tool for comparison of the ligaments sizes. Both the range and diameter of the ligaments decrease as $We_g$ and $Re_l$ increase. 

Fig.~\ref{fig:L21}(b) shows that in Domain II ($Re_l=2500$, $We_g=7250$), lowering $\hat{\rho}$ reduces the size of ligaments. This is consistent with influence of $\hat{\rho}$ on the KH vortex structures downstream of the waves, as was indicated by \cite{Arash3}. The lobe rims become thicker as $\hat{\rho}$ increases, resulting in thicker liquid bridges and consequently thicker ligaments after bridge breakup. The diversity of ligament diameters also increases as $\hat{\rho}$ is reduced (indicated by the error bars). The effects of $\hat{\mu}$ on the size of ligaments can be neglected compared to the other parameters. $\hat{\mu}$ was also shown to have no effect on the atomization mechanisms and the vortex dynamics near the interface \citep{Arash2, Arash3}. Again, the calculated $L_{21}$ is within the range of ligament diameters denoted by the error bars.

The approximation $R_d\approx 1.67R_1$ yields a new $L_{32}\approx1.67L_{21}$ for droplets formed from these ligaments. This new $L_{32}$ is notably smaller than the values in Fig.~\ref{fig:SMD}.

\subsection{Spray angle}

\begin{figure}[!b]
	\centering\includegraphics[width=1.0\linewidth]{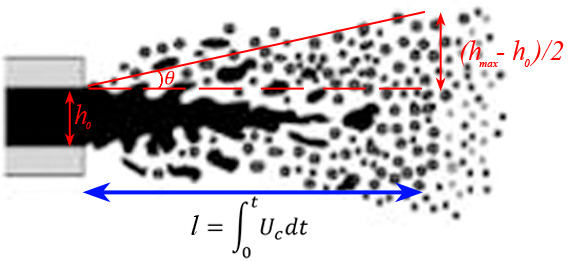}
	\caption{Definition of the spray angle, penetration length, and expanded width.}
	\label{fig:angle}
\end{figure}

Even though it is not possible to directly measure the spray angle in temporal studies like ours, we have calculated the spray angle using the spray width and length in this section. The spray angle ($\alpha$) is twice the half-angle ($\theta$), as shown in Fig.~\ref{fig:angle}. The spray width ($h_{max}$) can be measured at any time from the simulations. The distance that the liquid jet has traveled (penetration length, $l$), however, needs further analysis, since it cannot be measured directly. The penetration length of the jet at any instance can be obtained by integrating the jet convective velocity ($U_c$) from the beginning of the simulation; i.e.~\mbox{$l=\int_{0}^{t}U_c\mathrm{d}t$}. $U_c$ is the liquid-jet convective velocity, also known as Dimotakis velocity \citep{Dimotakis}, $U_c=(U_l+\sqrt{\hat{\rho}}U_g)/(1+\sqrt{\hat{\rho}})$. The convective velocity represents the velocity of the interface at the base of KH waves. Since the liquid velocity $U_l$ grows with time, $U_c$ is not constant and needs to be measured at every time step. The spray angle can then be calculated from the following relation;
\begin{equation}
\alpha=2\theta=2\tan^{-1}\left(\frac{h_{max}-h_0}{2l}\right).
\label{eqn:angle}
\end{equation}

\begin{figure}[!t]
	\centering\includegraphics[width=1.0\linewidth]{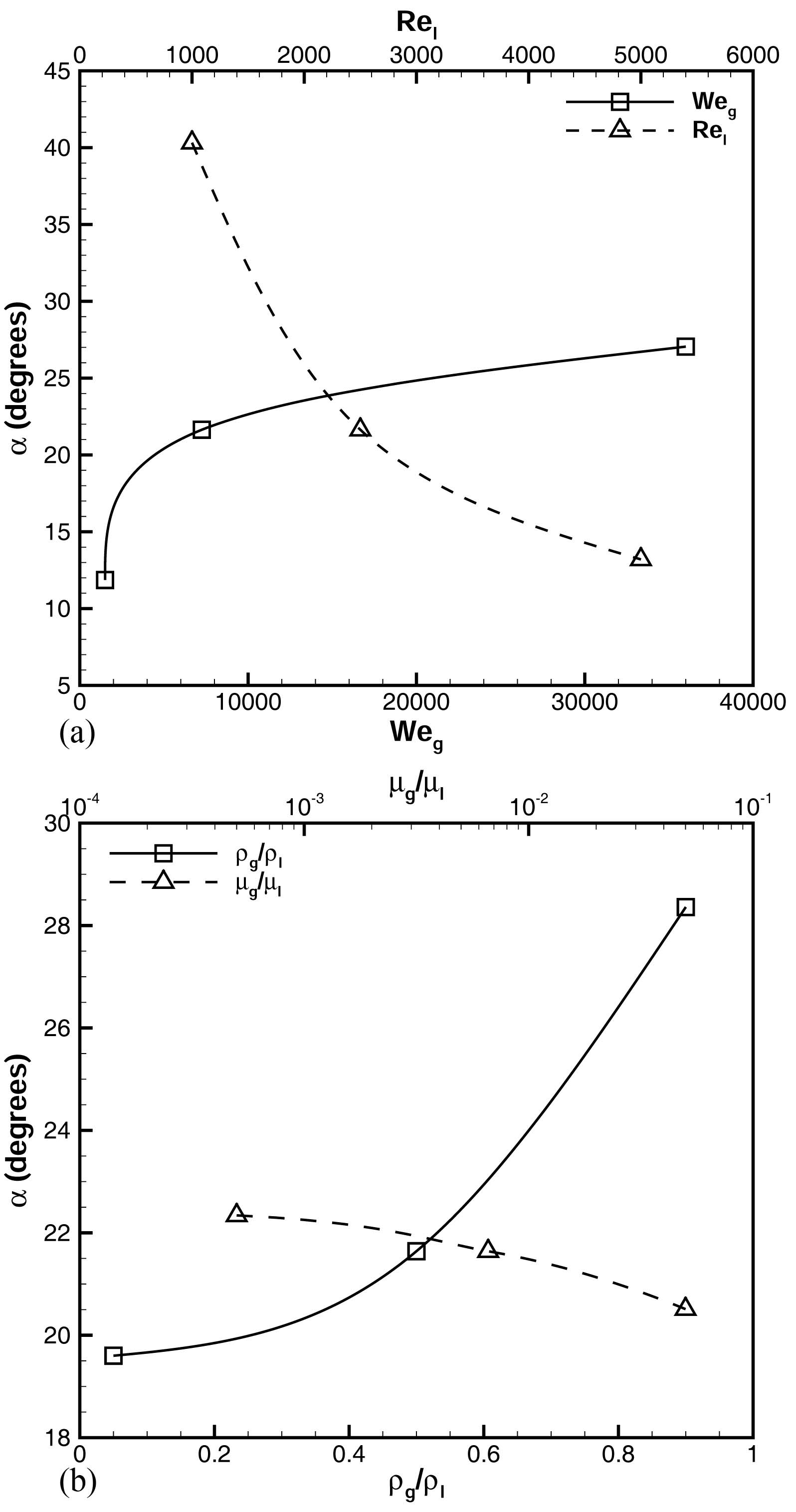}
	\caption{Spray angle $\alpha$ as a function of $We_g$ and $Re_l$ (a), and $\hat{\rho}$ and $\hat{\mu}$ (b); $Re_l=2500$, $We_g=7250$, $\hat{\rho}=0.5$, and $\hat{\mu} = 0.0066$.}
	\label{fig:angle_compare}
\end{figure}

Using Eq.~(\ref{eqn:angle}), $\alpha$ is measured in time and is presented in Fig.~\ref{fig:angle_compare}. In these plots, the values are for the converged value of $\alpha$, after which the angle remains almost constant in time.

As shown in Fig.~\ref{fig:angle_compare}(a), $Re_l$ has the most significant impact on $\alpha$, which decreases with increasing $Re_l$, but less so at higher $Re_l$. Both the range of $\alpha$ and its trend are in good agreement with the experimental results of \citet{Mansour} and \citet{Carvalho} at high liquid mass flow rate. Both of these studies show that $\alpha$ becomes almost independent of the mass flow rate (or Reynolds number) at very high liquid velocity (or $Re_l$). In our results, $\alpha$ decreases from $41^{\circ}$ to about $13^{\circ}$ as $Re_l$ increases from $1000$ to $5000$, because of the reasons that were detailed in discussion of Fig.~\ref{fig:Surface_vs_Re}. As was shown for low $Re_l$, the ligaments bend and grow normally away from the surface, thus increasing the transverse expansion of the spray and $\alpha$. It was shown by \citet{Arash3} that the KH vortices convect away from the interface faster in domain I (low $Re_l$) compared to domains II and III. Thus, as $Re_l$ increases and the breakup regime moves from domain I to domains II and III, the vortices remain closer to the interface and the transverse growth is hindered.

$We_g$ has an opposite effect on $\alpha$ compared to $Re_l$ (Fig.~\ref{fig:angle_compare}a). The increase in $We_g$ increases $\alpha$, but this variation becomes less effective at higher $We_g$. At very low $We_g$, the surface tension is very large and prevents the transverse growth of the instabilities, which results in small $\alpha$. As $We_g$ increases, the instability growth rate increases -- due to the smaller surface tension resistance -- and $\alpha$ increases. Since the breakup of ligaments occurs faster at high $We_g$, the broken ligaments and droplets can be easily carried in the transverse direction by the gas flow; this increases the spray angle. The far-field gas stream bounds the transverse growth of the spray, and prevents $\alpha$ from growing indefinitely with increasing $We_g$; thus, $\alpha$ grows very gradually at very large $We_g$. Also, as $We_g$ increases, the breakup regime of the liquid jet moves from Domains I/III to Domain II (Fig.~\ref{fig:domains_gas_based}), where the movement of vortical structures are mostly in the normal direction \citep{Arash3}. This contributes to increase of $\alpha$.

Fig.~\ref{fig:angle_compare}(b) shows that $\hat{\rho}$ also has a significant effect on $\alpha$, but $\hat{\mu}$ effects are almost negligible. The effect of $\hat{\rho}$ on the growth of instabilities and $\alpha$ is consistent with vortex dynamics analyses of \citet{Hoepffner} and \citet{Arash3}. Both theses studies show that at higher $\hat{\rho}$, the KH vortex remains closely downstream of the KH waves and rolls up the KH waves and increases the growth of instabilities. At low $\hat{\rho}$ on the other hand, the downstream KH vortex fails to roll gas and liquid together, but takes the form of a gas vortex sheltered from the main stream by the liquid body of the wave \citep{Hoepffner}. The wave grows a tongue which undergoes flapping. Liquid drops are torn from the wave through this flapping motion, and sent partly off to the gas stream and partly into the vortex core. The droplets are mostly convected in the streamwise direction; thus, the spray width and $\alpha$ become smaller than in the higher $\hat{\rho}$ cases. As $\hat{\rho}$ decreases further, the vortices become smaller and $\alpha$ decreases. Since the vortices cannot decrease indefinitely, the change in $\alpha$ becomes almost negligible at very low $\hat{\rho}$. The effect of $\hat{\rho}$ on $\alpha$, as given in Fig.~\ref{fig:angle_compare}(b), agrees with experimental results of \citet{Mansour}, where they showed that increasing the gas pressure (gas density) for a fixed liquid flow rate increased the spray angle.

\section{Conclusions} \label{Conclusion}

Two PDFs were formed for the liquid-structure length scale and the spray width from the numerical data that was obtained from a transient 3D DNS on a liquid-sheet segment. The PDFs provided statistical information about the length-scale distribution and the qualitative number density of ligaments/droplets during early liquid-jet atomization. The temporal variation of the mean of the PDFs gave the rate of cascade of liquid structures in different atomization domains. The mean and PDF of the spray width also showed the first instance of lobe and ligament breakup. The effects of gas Weber number ($We_g$), liquid Reynolds number ($Re_l$), density ratio ($\hat{\rho}$), viscosity ratio ($\hat{\mu}$), and wavelength-to-sheet- thickness ratio ($\Lambda$) on the mean length scale, the cascade rate, and the spray angle are quantified and summarized in Table~\ref{tab:Summary1}. The size of arrows in this table indicates the relative significance of the change.

\begin{table}[!t]
	\begin{center} 
		\caption{\label{tab:Summary1}Summarized effects of non-dimensional parameters on quantities of interest.}
		\begin{tabular}{c|c|c|c}
			\specialrule{.2em}{.1em}{.1em}
			Quantity  & $L_{32}$ \& $L_{21}$ & Cascade rate & Spray angle\\
			\specialrule{.1em}{.05em}{.05em}
			$We_g \uparrow$ & \huge{$\downarrow$} & \huge{$\uparrow$} & \Large{$\uparrow$}\\
			\midrule
			$Re_l \uparrow$ & \Large{$\downarrow$} & \LARGE{$\uparrow$} & \huge{$\downarrow$}\\
			\midrule
			$\hat{\rho} \uparrow$ & \small{$\uparrow$} & $\downarrow$ & \Large{$\uparrow$}\\
			\midrule
			$\hat{\mu} \uparrow$ & \small{$\uparrow$} & -- & --\\ 
			\midrule
			$\Lambda \uparrow$ & \huge{$\downarrow$} & \huge{$\uparrow$} & \large{$\uparrow$}\\ 
			\specialrule{.2em}{.1em}{.1em}
		\end{tabular}
	\end{center}
\end{table}

As the resistance of surface tension forces against surface deformation decreases by increasing $We_g$, the droplet size decreases, the cascade of structures and ligament breakup occur faster, and the spray width as well as the liquid surface area grow at higher rates. The initial growth of the length scales due to the stretching of the waves and lobes is affected more by liquid inertia than by the surface tension, as higher inertia results in a more vigorous and prolonged stretching and more flat surfaces. The asymptotic stage of length scale cascade, on the other hand, is affected mostly by surface tension and liquid inertia, but less by liquid viscosity. 

The liquid-structure cascade rate is significantly increased by increasing $Re_l$ as the viscous resistance against surface deformation decreases. The spray width is larger at lower $Re_l$, and the spray angle and the spray spread rate decrease as $Re_l$ increases -- attributed to the change in the angle of ligaments that stretch out of the sheet surface. Gas-to-liquid density ratio has minor influence on the final length scale, but the cascade occurs slower as density ratio increases. Gas inertia and liquid surface tension are the key parameters affecting the spray width, as it grows significantly with increasing gas density. Viscosity ratio has negligible effect on both the spray width and the final droplet size. Increasing the sheet thickness, however, decreases both the normalized spray width and its growth rate, while decreasing the structure cascade rate and producing larger droplets.

\begin{table}[!t]
	\begin{center}
		\caption{\label{tab:Summary2}Summarized effects of Domain change on quantities of interest.}
		\begin{tabular}{c|c|c|c}
			\specialrule{.2em}{.1em}{.1em}
			Domains  & $L_{32}$ \& $L_{21}$ & Cascade rate & Spray angle\\
			\specialrule{.1em}{.05em}{.05em}
			I $\rightarrow$ II & \huge{$\downarrow$} & \huge{$\uparrow$} & \large{$\uparrow$}\\
			\midrule
			II $\rightarrow$ III & \large{$\downarrow$} & \large{$\uparrow$} & \huge{$\downarrow$}\\
			\midrule
			I $\rightarrow$ III & \LARGE{$\downarrow$} & \LARGE{$\uparrow$} & \huge{$\downarrow$}\\
			\specialrule{.2em}{.1em}{.1em}
		\end{tabular}
	\end{center}
\end{table}

The cascade process and the spray expansion rate are decoupled for different atomization domains, and the trend of these quantities following the transition between the atomization domains is summarized in Table \ref{tab:Summary2}. Differences were notable for the length-scale distribution and spray expansion, which were correlated with the vortex structures at each domain. The times of length-scale cascade and sheet expansion were related to the formation of various liquid structures, showing that the ligament and droplet formation occurs faster at higher density ratios.

\section*{Acknowledgments}

Access to the XSEDE supercomputer resources under Allocation CTS170036 and to the UCI HPC cluster were very valuable in performing our high resolution computations.


\bibliographystyle{elsarticle-harv} 
\bibliography{IJMF}




\end{document}